\documentclass[floatfix,reprint,nofootinbib,amsmath,amssymb,aps,prd,superscriptaddress
]{revtex4-1}

\usepackage{aas_macros}
\usepackage{graphicx}
\usepackage{dcolumn}
\usepackage{bm}
\usepackage{diagbox}
\usepackage{mathtools}
\usepackage[normalem]{ulem}
\usepackage{xcolor}
\definecolor{colorLink}{rgb}{0.9,0,0} 
\definecolor{colorCite}{rgb}{0,0.7,0} 
\definecolor{colorURL} {rgb}{0,0,0.8} 
\usepackage[colorlinks=true,linktocpage=true,linkcolor=colorLink,citecolor=colorCite,urlcolor=colorURL]{hyperref}

\newcommand{\beq}{\begin{equation}}
\newcommand{\eeq}{\end{equation}}
\newcommand{\beqa}{\begin{eqnarray}}
\newcommand{\eeqa}{\end{eqnarray}}

\def\dng{\delta_{\textsc{n}_\textup{g}}}

\def\d12{\delta_{\textup{I}_{12}}}

\def\Nm{\textup{N}_\textup{m}}

\def\d12{\delta_{\textup{I}_{12}}}
\def\C{\textbf{C}}
\def\L{\mathcal{L}}
\def\k{{\bf k}}

\def\x{{\bf x}}

\def\I{\textup{I}}

\def\fun#1#2{\lower3.6pt\vbox{\baselineskip0pt\lineskip.9pt
        \ialign{$\mathsurround=0pt#1\hfill##\hfil$\crcr#2\crcr\sim\crcr}}}

\def\kMpc{\, h \, {\rm Mpc}^{-1}}

\defcitealias{WadSco19}{WS19}
\defcitealias{IvaSimZal19}{Iva19}
\defcitealias{DodSch1309}{DS13}
\defcitealias{PerRosSan1404}{Per14}

\newcommand{\be}{\begin{equation}}
\newcommand{\ee}{\end{equation}}

\renewcommand\a{\alpha}
\renewcommand\b{\beta}

\def\d{\partial}
\newcommand{\bseq}{\begin{subequations}}
\newcommand{\eseq}{\end{subequations}}

\begin{document}

\title{Cosmological constraints from BOSS with analytic covariance matrices}

\author{Digvijay Wadekar}\email{jay.wadekar@nyu.edu}\affiliation{Center for Cosmology and Particle Physics, Department of Physics, New York University, New York, NY 10003, USA}
\author{Mikhail M. Ivanov}
\affiliation{Center for Cosmology and Particle Physics, Department of Physics, New York University, New York, NY 10003, USA}
\affiliation{ Institute for Nuclear Research of the Russian Academy of Sciences,\\ 60th October Anniversary Prospect, 7a, 117312 Moscow, Russia
}
\author{Roman Scoccimarro}
\affiliation{Center for Cosmology and Particle Physics, Department of Physics, New York University, New York, NY 10003, USA}

\date{September 1, 2020}

\begin{abstract}

We use analytic covariance matrices to carry out a full-shape analysis of the
galaxy power spectrum multipoles from the Baryon  Oscillation Spectroscopic Survey (BOSS). We obtain parameter estimates 
that agree well with those based on the sample covariance from two thousand galaxy mock catalogs, thus validating the analytic approach 
and providing substantial reduction in computational cost. 
We also highlight a number of additional advantages of analytic covariances.  
First, the analysis does not suffer from sampling noise,  
which biases the constraints
and typically requires inflating parameter error bars. 
Second, it allows us to study convergence of the cosmological constraints when recomputing the  
analytic covariances to match the best-fit power spectrum, which can be done at a negligible computational cost, unlike when using mock catalogs. These effects reduce the systematic error budget of cosmological constraints, which suggests that the analytic approach may be an important tool for upcoming high-precision  galaxy redshift surveys such as DESI and Euclid. Finally, we study the impact of various ingredients in the power spectrum covariance matrix and show that the non-Gaussian part, which includes the regular trispectrum and super-sample covariance, has a marginal effect ($\lesssim 10 \%$) on the cosmological parameter error bars. We also suggest improvements to analytic covariances that are commonly used in Fisher forecasts.

\end{abstract}
\maketitle
\section{Introduction}


The analytic calculation of 
the power spectrum
covariance matrices
developed in~\cite{WadSco19} (hereafter \citetalias{WadSco19})
is a potential alternative 
to the covariance matrix estimation
from mock catalogs. 
Moreover, the analytic  approach allows one 
to address at least two issues that cannot be easily tackled  with mock catalog based covariances:
(a) sampling noise due to a finite number of the mocks,
(b) the difference between the extracted cosmology and the one used to create the mocks.

The construction of the covariance matrix from mock catalogs is a standard tool 
in redshift survey analyses. 
The main advantage of this approach is that 
the mock simulations capture
the effects 
of the survey geometry and gravitational clustering beyond
perturbation theory.
However, running many full N-body 
simulations for $\mathcal{O} (\textup{Gpc}^3)$ volumes is  computationally expensive, 
and 
in order to produce a large amount of mock catalogs to reduce sampling noise typically a number of 
 approximations are used instead \cite{BonMye9603,ScoShe02,ManScoPer1301,TasZalEis1306,WhiTinMcB1401,KitYepPra14,ChuKitPra15,IzaCroFos1607,LipSanCol1810,BloCroSef1905,ColSefMon1811}.
Even with these simplifications, however, mock catalog production for real surveys can take millions of CPU hours~\cite{KitRodChu1603,Zha20}. 
There are also proposed semi-analytic methods for calculating the covariance from small-volume simulations or from data \cite{2017MNRAS.472.4935H,2018MNRAS.478.4602K,OcoEisVar1611,OcoEis18,2020MNRAS.491.3290P}, but such methods do not clearly take into account all the physical effects that affect galaxy clustering.


The difference between the fiducial cosmology used to generate the mocks and the actual cosmology inferred from the data is a potential source of systematic error.
To eliminate this error, 
in principle, one should iterate the analysis using the new covariance matrix re-evaluted for the output cosmology \cite{Teg9711,Tegmark:1996qt,EifSchHar09,WhiPad15,MorSch13}.
However, this cannot be done with the mock covariances 
because of their prohibitive computational cost.

Another uncertainty present in the covariance matrix estimated from a finite sample of mocks is sampling noise.
Its impact on parameter constraints has been thoroughly studied in the literature~\cite{HarSimSch0703,TayJoaKit1306, DodSch1309,PerRosSan1404,TayJoa14,SelHea1602}. 
Sampling noise is typically taken into account by the inflation of parameter variances.
However, due to the stochastic nature of this noise, even the inflation of error bars, in principle, does
not guarantee 
that the parameter estimation 
is accurate and unbiased when only one
realization of the sample covariance matrix is used.

The danger of sampling noise has already stimulated a broad line of research
devoted to noise reduction techniques e.g. tapering \citep{2015MNRAS.454.4326P}, shrinkage \citep{2017MNRAS.466L..83J},  sparsity-based methods~\citep{2016MNRAS.460.1567P}, singular-value decompositions \cite{GazSco0508,PhiIva20inprep,Sco0012,EisZal0101} or Taylor expanding the precision matrix about a smooth fiducial model \citep{2018MNRAS.473.4150F}.
The systematic errors 
generated by the sampling noise 
and the difference between fiducial and best-fit cosmologies cannot be reliably estimated using a single sample covariance constructed from a finite number of mocks,
which may compromise 
cosmological results from a galaxy survey. 
In this paper, we show that both of the aforementioned problems can be circumvented with the use of analytic covariance matrices. 
We will follow the approach of~\citetalias{WadSco19}, who has recently put forward
the first analytic model for the full (diagonal and non-diagonal) covariance of the 
redshift-space 
galaxy power spectrum multipoles. Their model is based on
perturbation theory (PT) and
includes the effects of radial redshift space distortions, arbitrary survey window, shot noise, nonlinear gravitational evolution and the effect of super-survey modes. 
Because of the dominance of shot noise and the super-survey modes over non-linearities in the covariance at small scales, 
their PT-based model was shown to have an excellent agreement with mock simulations up to quasi-linear scales ($k\simeq 0.6 \kMpc$). 
However, in the case when the range of eigenvalues of a matrix is large (as is typically the case for the covariance matrices of the power spectrum), very similar looking matrices can have completely different inverses.
Consequently, the likelihood for cosmological parameter estimation, which depends on the inverse of the multipole covariance matrix, could be affected.

It is therefore important to validate the analytic covariance approach in the 
likelihood analysis of the actual galaxy survey data, which is the main goal of our work.
To that end we will redo the full-shape analysis of the BOSS galaxy power spectrum \cite{IvaSimZal19,DAmico:2019fhj,Ivanov:2019hqk} (see also \cite{SanScoCro1701,GilPerVer1702,BeuSeoSai1704,GriSanSal1705,AlaAtaBai1709}), following the methodology of \cite{IvaSimZal19} (hereafter \citetalias{IvaSimZal19}), using the analytic covariance matrices of~\citetalias{WadSco19}  instead of those from mock catalogs~\cite{KitRodChu1603}. 
The main result of our study is that the cosmological constraints obtained in the previous analyses using the mock catalog covariances agree with the results based on the analytic covariances to $\sim 0.1\sigma$.
We will show that this residual shift is produced by sampling noise in the mock covariance, and it cannot be fully captured by the standard approach of inflating the error bars.
Finally, we will show that our cosmological constraints are also stable with respect to updating the
covariance matrix to match the best-fit output cosmology.

The stability of the constraints under variations of the covariance matrix is an important test, 
which is usually done in the case of the CMB \cite{Abbott:2017wau,Aghanim:2019ame,Spe03} and for some weak lensing
data analyses~\cite{HikOguHam1903,JoaLimAsg20}.
Thus, our work also validates the results of the previous full-shape analyses of the galaxy power 
spectrum (\cite{IvaSimZal19,DAmico:2019fhj,Ivanov:2019hqk}) and removes  the uncertainty associated with the choice of covariance matrices.

Our analytic approach also provides insight into the relevance of various physical effects that form the parameter constraints. 
The non-Gaussian contribution to the covariance, especially the super-sample covariance (SSC), has been the subject of extensive studies (see \cite{ScoZalHui9912,HamRimSco0609,SefSco0503,SefCroPue0607,PutWagMen1204,TakHu1306,LiHuTak1404,LiHuTak1411,AkiTakLi1704,LiSchSel1802} for some examples).
In this work we explicitly
demonstrate the effect of the non-Gaussian covariance on parameter constraints from a realistic spectroscopic survey.
Our results suggest that the bulk of the constraints is coming from the Gaussian part of the covariance;
the non-Gaussian
contributions affect 
the parameters error bars at $\lesssim 10\%$ level up to $k=0.25$~$h~\text{Mpc}^{-1}$, with roughly equal contributions from the regular trispectrum
and SSC. This should be contrasted with the commonly used signal-to-noise ratio, which is significantly affected by the non-Gaussian covariance \cite{WadSco19,ChaBlo17,TarNisJeo20,HikTakKoy20, SugSaiBeu1908}. 
Thus, our analysis demonstrates that 
the signal-to-noise may be a misleading 
metric to illustrate the effects of the covariance matrix.

It is useful to compare our
work to
the previous literature~\cite{Yamamoto:2005dz,Grieb:2015bia,BlaCarKod1810,2019MNRAS.490.5931P,LiSinYu1811} focused on the Gaussian/disconnected part of the power spectrum covariance. 
Some of these works have already pointed out three generic advantages of the analytic covariance over the mock simulations:  
(i) negligible 
computation cost, 
(ii) absence of sampling noise
that requires the inflation of error bars, 
(iii)  the possibility to use the best-fit power spectrum as an input.
In our work we demonstrate 
these points
on the example of the \textit{full} covariance 
matrix calculation 
that includes 
the connected part as well. 
Moreover, we will explicitly study the impact of the analytic covariance on cosmological 
constraints, 
focusing on the  
parameter inference from the  BOSS data.

The outline of the paper is as follows. We first discuss our analysis method in Section~\ref{sec:AnalysisMethod}. We give a brief overview of the analytic covariance in Section~\ref{sec:AnalyticCovariance} where we also compare the effect of its non-Gaussian component on parameter constraints using a Fisher forecast. We describe the mock catalogs we use and discuss the effects of sampling noise on parameter constraints in Section~\ref{sec:Mocks}.
We present our BOSS analysis results in Section~\ref{sec:Results}. We discuss and conclude in Section~\ref{sec:Discussions}. Appendix~\ref{AppA} provides more details on analytic covariance matrices including suggestions for improving Fisher forecasts, while Appendix~\ref{AppB} characterizes the noise in covariance matrices from mock catalogs.
Finally, Appendix~\ref{sec:bg3}
contains further tests of our bias treatment.

\section{Data and Methodology}
\label{sec:AnalysisMethod}

In what follows we will analyse
monopole and quadrupole moments of the BOSS galaxy power spectra
using the full-shape method 
of \citetalias{IvaSimZal19}.
We measure the power spectrum multipoles
from the publicly available catalogs\footnote{\href{https://data.sdss.org/sas/dr12/boss/lss/}{
\textcolor{blue}{https://data.sdss.org/sas/dr12/boss/lss/}}
} of BOSS DR12 \cite{AlaAtaBai1709}
using the estimator in~\cite{Sco1510}.
As a cross-check, we analyzed the same catalogs with the \texttt{nbodykit}
code \cite{Hand:2017pqn} and found excellent agreement. The catalogs 
contain four different data samples: high-z (effective redshift $z_{\rm eff}=0.61$) and low-z ($z_{\rm eff}=0.38$),
north and south galactic cap (NGC and SGC) data  chunks. 
Note that for each particular chunk we use the same data vector across all the analyses. This means that only the covariance matrices are varied, the data vectors are fixed for every chunk.

We fit the power spectrum data with
the one-loop IR-resummed perturbation
theory prediction \cite{Blas:2016sfa,Ivanov:2018gjr,Ivanov:2019hqk}.
This model was verified in a 
blind challenge to measure  cosmological parameters
from the 
BOSS-like mock catalogs, whose cumulative volume is $\sim 100$ times bigger than the actual BOSS volume~\cite{Nishimichi:2020tvu}.
Our theoretical 
predictions are
evaluated with the \texttt{CLASS-PT}
code~\cite{ChuIvaSim20} (based on the FFTLog method \cite{Simonovic:2017mhp}),
and then convolved with the survey window function as described in \cite{Ivanov:2019hqk}.
We use the data in the wavenumber range $[0.01,~0.25]$~$h~\text{Mpc}^{-1}$,
which is robust w.r.t.
survey systematics 
and two-loop corrections omitted 
in our theoretical calculation
\cite{IvaSimZal19}.

We assume the minimal flat  $\Lambda$CDM model that is characterized by $\omega_{\rm b},\omega_\textup{cdm},H_0,A_s,n_s$\footnote{$\omega_{\rm b} =\Omega_{\rm b} h^2$ and $\omega_\textup{cdm}=\Omega_\textup{cdm} h^2$ are the physical densities of baryons and DM respectively; $A_s$ and $n_s$ are the amplitude and tilt of the primordial spectrum of scalar perturbations; and $H_0$ is the Hubble parameter.}. The neutrino sector is approximated by a single massive state with $m_\nu=0.06$ eV.
This choice is made only for simplicity and does not affect the conclusions of our paper.
In general, we believe that it is more appropriate to treat $m_\nu$ as a free parameter in the fit, as it is done 
in \citetalias{IvaSimZal19}.
We also fix $\omega_{\rm b}$ and $n_s$ to the Planck best-fit values~\cite{Aghanim:2018eyx},
\begin{equation}
n_s=0.9649\,,\quad 
\omega_{\rm b}=0.02237\,.
\end{equation}
Using these fixed values or imposing the tight Planck Gaussian priors on these parameters produce identical results. 
For convenience, we will use the primordial power spectrum amplitude 
normalized to the Planck best-fit value,
\begin{equation}
    A\equiv \frac{A_s}{A_{s,\,{\rm Planck}}}\,.
\end{equation}
All in all, our MCMC chains will sample the following set of cosmological parameters,
\begin{equation}
\{ \omega_\textup{cdm},~H_0,~A\}\, ,
\label{eq:CosmoParameters}\end{equation}
for which we do not assume any priors.

As far as the nuisance parameters are concerned, we will use the following physical priors,\footnote{We denote flat priors as $p \in (1,4)$ and Gaussian priors with mean $\mu$ and
variance $\sigma^2$ as $p\sim \mathcal{N}(\mu,\sigma^2)$. }
\begin{equation}
\begin{split}
  & b_1A^{1/2} \in (1,4)\,,\quad   b_2A^{1/2}\sim \mathcal{N}(0,1)\,,\quad 
  b_{{\cal G}_2}A^{1/2}\sim   \mathcal{N}(0,1)\,,\\
  & c_0, c_2 \sim 
   \mathcal{N}(0,30^2)~[\text{Mpc}/h]^2\,,\\
   & \tilde{c}\sim  
    \mathcal{N}(500,500^2)~[\text{Mpc}/h]^4\,, \\
&  P_{\rm shot} \sim  \mathcal{N}(0,~5000^2)~[\text{Mpc}/h]^3\,,
     \end{split}\label{eq:NuisanceParameters}
   \end{equation}
where $b_1$ denotes linear bias, $b_2$ \& $b_{{\cal G}_2}$ denote quadratic bias parameters. 
As far as the quadratic and cubic tidal biases $b_{\mathcal{G}_2}$ and 
$b_{\Gamma_3}$ are concerned,
the power spectrum 
data cannot measure them
separately. It can
only constrain
their following combination:
\be 
\label{eq:bG2}
b'_{\mathcal{G}_2}=b_{\mathcal{G}_2} + 0.4b_{\Gamma_3}\,.
\ee 
If we scan over $b_{\mathcal{G}_2} $
and $b_{\Gamma_3}$ separately, 
we find that they are fully degenerate, 
and their 1D
the marginalized distributions are flat even within very wide priors. 
This makes the sampling of these posterior distributions significantly time-consuming. 
In order to facilitate 
the convergence of our MCMC chains,
we follow \cite{Ivanov:2019hqk,Nishimichi:2020tvu} and scan only over the principal component \eqref{eq:bG2}. 
This can be done by
keeping $b_{\mathcal{G}_2}$
in the fit and
setting the cubic bias to zero,
\be 
b_{\Gamma_3}=0\,.
\ee
This choice does not affect the constraints for the cosmological parameter~\cite{Ivanov:2019hqk,Nishimichi:2020tvu}, see Appendix~\ref{sec:bg3} for more detail.
It should be kept in mind that the constraints on $b_{\mathcal{G}_2}$ presented in our paper are, in fact, the constraints on $~b'_{\mathcal{G}_2}$.

In addition, $c_0,c_2,\tilde{c}$ are counter-terms that account for the impact of small-scale velocity dispersion on the redshift-space power spectrum. Note that even though the standard Poissonian pair-counting shot noise contribution is already 
removed by the power spectrum estimator \cite{FelKaiPea9405}, 
one still needs to marginalize over a constant offset, $P_{\rm shot}$
to account for the residual 
contribution produced by fiber collisions and exclusion effects.
Note also that the priors 
on the cosmological and nuisance parameters used in this paper are somewhat different from the ones used in \cite{Ivanov:2019hqk}; they have been used in the analysis presented in Ref.~\cite{ChuIvaSim20}.

Our MCMC analysis is carried out 
using the \texttt{Montepython} code \cite{Audren:2012vy,Brinckmann:2018cvx}. 
We consider the chains to be converged if they satisfy the Gelman-Rubin convergence criterion \mbox{$R-1<0.01$} \cite{Gelman:1992zz,Brooks:1997me}.
All plots with posterior densities are produced with the \texttt{getdist} package\footnote{\href{https://getdist.readthedocs.io/en/latest/}{
\textcolor{blue}{https://getdist.readthedocs.io/en/latest/}}
}~\cite{Lewis:2019xzd}.


\section{Analytic covariances}
\label{sec:AnalyticCovariance}


In this section, we highlight some important aspects of the perturbation theory (PT) - based covariance matrix of  \citetalias{WadSco19}.
 We closely follow their notation and refer
the readers to \citetalias{WadSco19} for further details. 
We will use their \textsc{Cova-PT} code\footnote{\label{covapt}\textsc{Cova-PT} is publicly available at \href{https://github.com/JayWadekar/CovaPT}{\textcolor{blue}{https://github.com/JayWadekar/CovaPT}}.} for computing the analytic covariance matrices corresponding to different BOSS samples.
In most of the paper we will focus on the high-$z$ north galactic cap (NGC) sample ($0.5<z<0.75$) of the BOSS data  (with the exception of Sec.~\ref{sec:FullBossResults} which contains the results for the full BOSS survey).
The power spectrum covariance can be represented as a combination of two components: the Gaussian/disconnected part
that
corresponds to the product of two power spectra\footnote{
Note that the Gaussian covariance does not assume that the density field itself is Gaussian, but it includes contributions from discreteness and non-linearity.}, and the non-Gaussian part
described by the 
connected four-point function, or trispectrum in Fourier space.
The latter also contains  contributions from the super-survey modes.
Let us now discuss these two parts in more detail.

\subsection{Gaussian covariance}
\label{sec:GaussCova}
In order to understand the important aspects of the Gaussian covariance, let us rewrite here the final results from Eqs.~(57) and (88) of \citetalias{WadSco19} for the case of the continuous gaussian covariance (G) and the shot noise contribution to the Gaussian covariance (SN-G):

\begin{equation}
\begin{split}
 \textbf{C}^\textup{G}_{\ell_1\ell_2} (k_1,k_2)\simeq& \sum_{\ell'_1,\ell'_2} P_{\ell'_1}(k_1)\, P_{\ell'_2}(k_2) \ \mathcal{W}^{(1)}_{\ell_1,\ell_2,\ell'_1,\ell'_2} (k_1,k_2)\\
  \textbf{C}^\textup{SN-G}_{\ell_1\ell_2} (k_1,k_2)\simeq&\Big[ \sum_{\ell'} P_{\ell'}(k_1)\mathcal{W}^{(2)}_{\ell_1,\ell_2,\ell'} (k_1,k_2) +(k_1 \leftrightarrow k_2)\Big]\\
&+\mathcal{W}^{(3)}_{\ell_1,\ell_2}(k_1,k_2)\,,
\label{eq:GaussCova}\end{split}
\end{equation}

where $P_\ell$ corresponds to the power spectrum multipoles and $\mathcal{W}$ corresponds to different window kernels and depends on the width of the $k$-bins; it includes the effect of the survey geometry and of the changing line-of-sight (LOS) over the volume of the survey.
The explicit expressions for $\mathcal{W}$ are given in Eq.~(\ref{eq:W_kernel}) and are evaluated using FFTs of the survey random catalog (see \cite{LiSinYu1811,OcoEisVar1611,OcoEis18,PhiEis19} for alternate methods using correlation functions).
One aspect of the Gaussian covariance in Eq.~\ref{eq:GaussCova}, which makes it 
computationally very efficient, is the factorization of the clustering and the survey geometry terms. Such factorization is based on the assumption that the convolution of the power spectrum with the survey window can be ignored, as the convolution only affects the scales close to the size of the survey window\footnote{We have explicitly 
taken into account the leading corrections beyond this approximation and checked that they do not affect the parameter constraints in our analysis.}.

One can immediately see from Eq.~\ref{eq:GaussCova} that calculating the Gaussian covariance for multiple sets of cosmology and galaxy bias parameters is computationally cheap, as it amounts to simply computing the power spectrum multipoles for that set of parameters. The factorization of geometry and clustering also allows one to include  velocity dispersion effects and loop corrections in the Gaussian covariance quite easily through the theoretical model for $P_\ell(k)$. Therefore, the analytic model for the Gaussian covariance is accurate up to  $k_\textup{max}$ where the theoretical power spectrum model is accurate.

The Gaussian covariance in Eq.~\ref{eq:GaussCova} is not exactly 
diagonal because a few neighboring $k$-bins
get correlated due to the survey mask. In order to make forecasts for future surveys in the literature, the Gaussian covariance is typically assumed to be diagonal and the effects of FKP weights, survey geometry and changing LOS are ignored. We discuss these approximations in detail and their impact on cosmological parameter constraints in Appendix~\ref{apx:GaussCova}.

\subsection{Non-Gaussian covariance and Fisher forecast}
\label{sec:NGcovariance}



The non-Gaussian (NG) part of the covariance is composed of two main contributions: the regular trispectrum \cite{HarPen12,HarPen13,Bertolini:2015fya,MohSelVla1704, TarNisJeo20} and the contribution due to super-survey modes called beat-coupling \cite{HamRimSco0609, SefCroPue0607} or super sample covariance (SSC) \cite{TakHu1306,LiHuTak1404,LiSchSel1802} (see Appendix~\ref{apx:NGcova} for more detail on the NG expressions we use).
\citetalias{WadSco19} calculated the NG covariance at tree-order (which is formally of the same order as one-loop in the Gaussian covariance).
One of the important conclusions of their analysis was that using the FKP estimator leads to the impact of super-survey modes on the covariance being stronger than was previously assumed in the literature.
In this paper we will explicitly demonstrate that even the stronger SSC effect only leads to a marginal degradation of parameter constraints.

One of the motivations to study the NG covariance has been an observed degradation of the signal-to-noise ratio (S/N),
\beq
 ({\rm S/N})^2=\sum_{ij}^{k_\textup{max}} P(k_i)\, \C^{-1}(k_i,k_j)\, P(k_j)
\label{eq:S/N}\eeq
compared to the Gaussian case, both in real space \cite{ChaBlo17,TarNisJeo20} and redshift space \cite{WadSco19,HikTakKoy20, SugSaiBeu1908}.
In our joint (monopole + quadrupole) analysis, we also find that the S/N reduces by $\sim 33$\% for $k_\textup{max}=0.25 \kMpc$ upon including the NG covariance, which is consistent with previous results of \citetalias{WadSco19}.
However, the S/N ratio 
is not directly related to parameter constraints.
Indeed, S/N is inversely proportional to the variance of the \textit{real space} power spectrum amplitude in linear theory, provided that all other parameters are fixed.
However, even in redshift space linear theory,
the velocity fluctuation amplitude $f\sigma_8$ measurements
 result from the breaking of the degeneracy 
with the linear galaxy bias $b_1$ in the joint analysis of
the monopole and quadupole moments, and hence are not directly related to S/N (see for e.g. \cite{KobNisTak20}).
As we go to the non-linear level, the measurement of cosmological parameters becomes sensitive to the marginalization over nuisance parameters, which obscures the interpretation of S/N even further.

\begin{figure}
\centering
\includegraphics[scale=0.8,keepaspectratio=true]{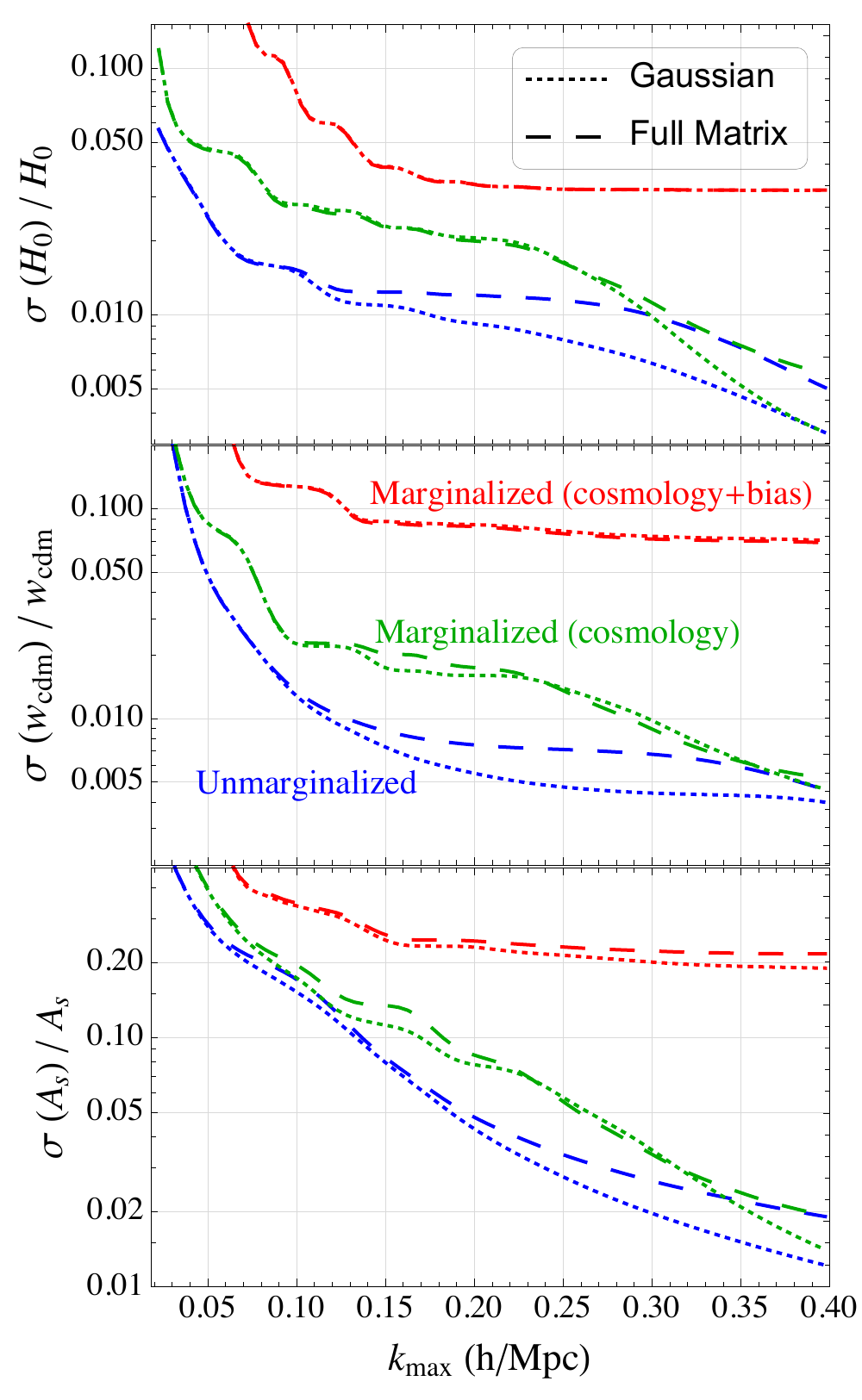}
\caption{Forecasted error on the cosmological parameters upon using two different cases of the covariance matrix. In the unmarginalized scenario for a particular parameter in blue, all other cosmology and bias parameters are fixed. 
The green curves display the effect of marginalization over the cosmological parameters $\{\omega_{cdm},H_0,A_s\}$ only.
The red curves show the effect of marginalizing over all 10 parameters given in Eq.~\ref{eq:10paramset}. The NG covariance affects the unmarginalized constraints significantly (similar to what a naive S/N analysis would suggest), but the effect of the NG covariance on the marginalized parameters is much weaker.
Note that our actual MCMC analysis will be limited to $k_{\rm max}=0.25~h/$Mpc; the results for higher $k_{\rm max}$ are displayed only for illustration purposes.}
\label{fig:NG_marginalization}
\end{figure}

To gain some intuition into 
the effect of the NG covariance on cosmological constraints, we perform a simple 
Fisher forecast using the full non-linear model and two choices of the analytic covariance: the full case and the Gaussian part only. We focus on the cosmological parameters $\omega_{cdm},H_0,A_s$. 
For each parameter and each choice of the covariance matrix we do two different analyses: compute the errorbars with other parameters fixed (``unmarginalized'') or 
marginalize over full set of
other parameters.
We calculate the Fisher matrix given by \cite{Teg9711},
 \beq
F_{\a \b}=\sum_{ij}^{k_\textup{max}} \frac{\partial P(k_i)}{\partial p_\a}\, \C^{-1}(k_i,k_j)\, \frac{\partial P(k_i)}{\partial p_\b}\,, \label{eq:Fisher}\eeq
for the same parameter set that we use in our BOSS analysis,
\be
\begin{split}
p_\alpha=\{\omega_{cdm},H_0,A_s,b_1A^{1/2},&b_2A^{1/2},b_{\mathcal{G}_2}A^{1/2}\\
&,c_0,c_2,\tilde{c},P_{\rm shot}\}\,. 
\end{split}
\label{eq:10paramset}
\ee
and also include the priors mentioned in Eq.~\ref{eq:NuisanceParameters}.
We show the resulting parameter errorbars as a function of the data cut $k_{\rm max}$ in Fig.~\ref{fig:NG_marginalization}. The unmarginalized case is similar to the naive S/N analysis and we indeed find that the NG covariance affects the errorbars significantly. However, the effect of the NG part becomes much weaker after marginalization over other parameters:
the resulting constraints for $H_0$ and $\omega_\textup{cdm}$ are nearly unaffected by the NG part and there is a marginal change in the resulting $A_s$ constraints. To see the relative effect of the nuisance parameters on the constraints, we also show the case when the parameter marginalization is carried out without including the nuisance parameters. Remarkably, the marginalization over the cosmological parameters already suppresses the effect of the NG covariance,
even if the 
nuisance parameters are fixed. 

In our actual data analysis, we will use
$k_{\rm max}=0.25 \kMpc$.
However, it is instructive to push to higher $k_{\rm max}$ as the ratio of the amplitude of the NG covariance to that of the G covariance increases with $k$~\cite{WadSco19}, and we indeed find that the effect of NG part on the $A_s$ errorbar also increases (the errorbar degradation in the marginalized case increases from $\sim 8$\% for $k_\textup{max}=0.25 \kMpc$ to $\sim 15$\% for $k_\textup{max}=0.4 \kMpc$). 
However, in a realistic case, one would need to include two-loop corrections with additional 
nuisance parameters in this case, and hence the marginalization effects can become stronger on these scales.
Thus, going to higher $k_{\rm max}$ with our one-loop theory model  is, of course, overly optimistic.

One other insight gained from the above discussion is that S/N is a misleading metric to quantify the impact of the NG covariance. 
We will also see in Sec.~\ref{sec:Results} that our Fisher results are in good agreement with the full MCMC analysis for $k_\textup{max}=0.25 \kMpc$. See also \cite{YuSel20inprep} for an analysis with a different methodology and performed at a higher $k_\textup{max}$.

 Finally, it is worth briefly discussing the effect of higher-order non-linearities like loop and fingers-of-god (FoG) corrections, which have not been included in NG covariance model of \citetalias{WadSco19}.
For the auto-covariance of monopole and quadrupole, the contribution from super survey modes and shot noise is expected to dominate over the loops and fingers-of-god even at high-$k$ \citepalias{WadSco19}. For the cross-covariance, however, the effect of long modes and shot noise is sub-dominant, while the effect of FoG is particularly important. The NG cross-covariance is thus expected to be smaller than our calculation at high-$k$ and likewise its effect on the parameter constraints, so our analysis can be treated as a 
first conservative approximation.




\section{Covariance from mock catalogs}
\label{sec:Mocks}

In this section we provide details of the mocks that we use and a brief overview of the sampling noise due to a finite number of mocks. We use the V6C MultiDark-Patchy mock galaxy catalogs~\cite{KitRodChu1603} (hereafter referred to as Patchy mocks), which were also used in SDSS-BOSS parameter constraints papers by the collaboration~\cite{SanScoCro1701,GilPerVer1702,BeuSeoSai1704,GriSanSal1705,AlaAtaBai1709}. These catalogs were generated by using the \textsc{patchy} code \cite{KitYepPra14} and calibrated using the BigMultiDark $N$-body simulation \cite{KlyYep1604, RodChuPra1610}. The work in~\cite{KitRodChu1603} gives a rough estimate of the computations involved in generating the 2048 mocks to be 0.5 million CPU hours (for the same volume, full $N$-body simulations would have required $\sim$ 9 billion CPU hours). For comparison, the analytic covariance of \citetalias{WadSco19} requires $\sim$100 CPU hours to compute the Monte-Carlo integrals for evaluating the window kernels in Eq.~\ref{eq:W_kernel} for a given survey random catalog; the rest of the steps are computationally trivial.

We measure the power spectrum from mocks using the  estimator of~\cite{Sco1510}; in particular, we use the best (lower variance) of the two estimators presented  for the quadrupole. We have included the
veto mask which excludes the un-observable regions on the sky, e.g. near bright stars. We choose not to use the standard fiber collision weights in the Patchy catalogs because 
they are based on the nearest neighbor approximation, which is not entirely accurate \cite{HahScoBla17}. 
A more accurate way of accounting for fiber collisions is the effective window method, supplemented with the marginalization of appropriate nuisance parameters ($P_{\rm shot}$ and $\tilde c$ in our case). 
However, it has been shown in Refs.~\cite{IvaSimZal19,DAmico:2019fhj}
that this marginalization accounts for fiber
collisions even if the effective window method
is not applied, i.e. the whole effect of the effective window can be absorbed into the nuisance parameters. Given this reason, 
we do not explicitly implement the 
effective window in our analysis, but allow for wide priors on the nuisance parameters instead
(we also checked that including the fiber collision weights in the Patchy mocks does not affect our results). 

To estimate the multipoles power spectrum covariance from a sample of $\textup{N}_\textup{m}$ mocks, we use the standard empiric estimator,
\begin{equation}\begin{split}
&\hat{\textbf{C}}_{\ell_1\ell_2}(k_i,k_j)\\
&\equiv \frac{1}{\textup{N}_\textup{m}-1} \left[\sum_n^{\textup{N}_\textup{m}}\, [P_{\ell_1}^{(n)}(k_i)- \bar{P}_{\ell_1}(k_i)] [P_{\ell_2}^{(n)}(k_j)- \bar{P}_{\ell_2}(k_{j})] \right]\, ,
\end{split}\label{eq:CovEstimator}\end{equation}
where the sample mean power spectrum is given by $\bar{P}_\ell(k_i)=\sum_n^{\textup{N}_\textup{m}} P_\ell^{(n)}(k_i)/\textup{N}_\textup{m}$.
We will use $\textup{N}_\textup{m}=2048$ throughout this paper, just like the previous BOSS analyses~\cite{SanScoCro1701,GilPerVer1702,BeuSeoSai1704,GriSanSal1705,AlaAtaBai1709}.
In order to get
the unbiased inverse covariance in the 
ensemble-averaged limit, the estimator of the inverse covariance (precision) matrix needs to be rescaled by a pre-factor as shown in the following equation,
\beq
\hat{\C}^{-1}=\frac{\textup{N}_\textup{m}-n_b-2}{\textup{N}_\textup{m}-1}(\hat{\C}^{-1})_\textup{measured}
\eeq
which is called the ``Hartlap factor'' \cite{HarSimSch0703,And03} (originally derived by Wishart in Ref.~\cite{Wishart28}) and has an almost negligible difference from unity (= 0.95) in our case of $n_b=96$ bins and $\Nm=2048$.

\subsection{Sampling noise}
\label{sec:SamplingNoise}

\setlength{\unitlength}{1cm}
\begin{figure}
\centering
\includegraphics[scale=0.75,keepaspectratio=true]{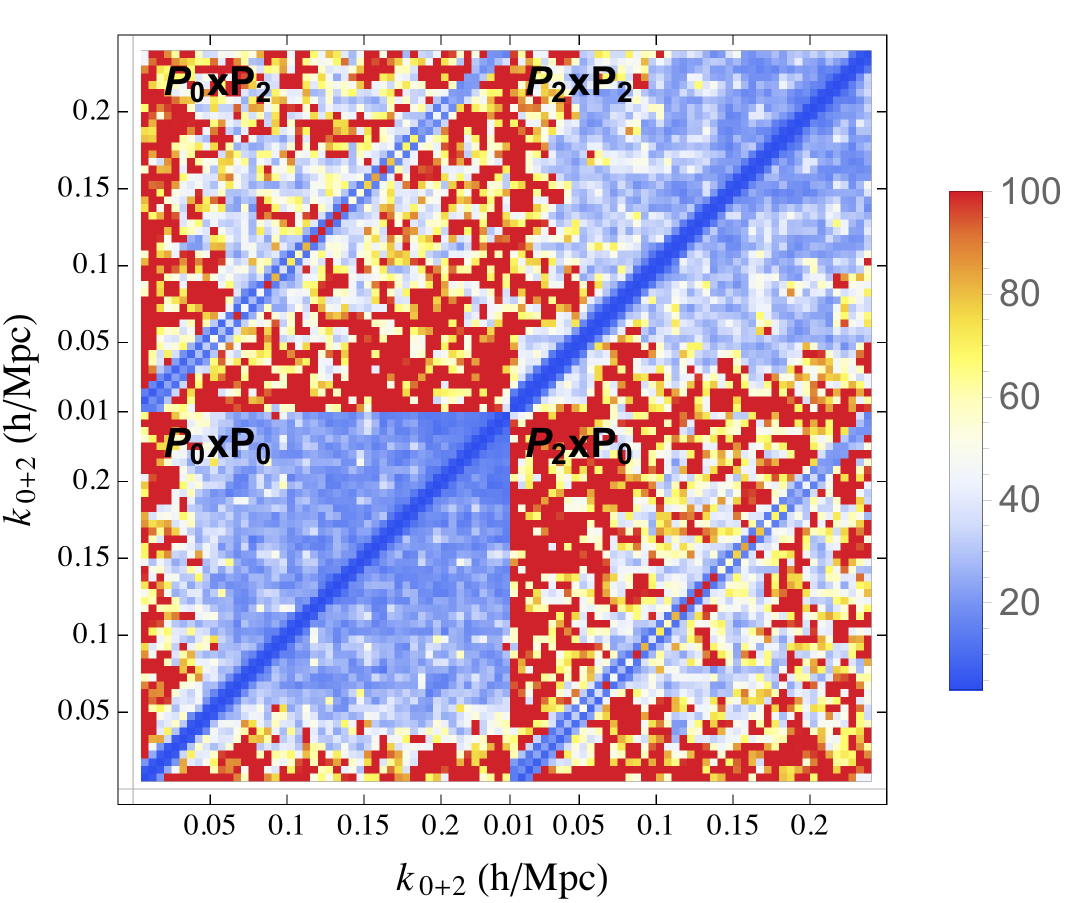}
\put(-0.05,4.2){\rotatebox{270}{\% error}}
\caption{Percentage error in the elements of the power spectrum multipoles covariance matrix from 2048 Patchy mocks due to sampling noise, calculated using Eq.~(\ref{eq:CovaTheoryError}).}
\label{fig:Error2D}
\end{figure}

When a finite sample of mocks is used to estimate the covariance matrix, sampling noise is introduced in the matrix. If one assumes that the variations in the power spectra measured from the mock catalogs are Gaussian distributed,
the error in the elements of the estimated covariance matrix is given by\footnote{See Appendix E of \citetalias{WadSco19} for a justification of the non-Gaussian contribution to the error on the covariance being sub-dominant for $k\lesssim 10 \kMpc$.} \cite{TayJoaKit1306,And03}
\begin{equation}
\begin{split}
\Delta \C_{\ell_1 \ell_2}(k_i,k_j)  = & \frac{1}{\sqrt{\Nm-1}}\Big[\C_{\ell_1 \ell_2}^2(k_i,k_{j})\\
&+ \C_{\ell_1 \ell_1}(k_i,k_i)\C_{\ell_2 \ell_2}(k_j,k_j)\Big]^{1/2}\, .
\label{eq:CovaTheoryError}\end{split}
\end{equation}
Substituting the values of the matrix elements obtained from Patchy in the RHS of the above equation, we show in Fig.~\ref{fig:Error2D} the resultant error on individual elements of the Patchy multipoles covariance matrix (we also checked that the estimates are roughly consistent with the error estimated using bootstrapping). 
The effect of sampling noise from Fig.~\ref{fig:Error2D} is quite significant, especially for the cross-covariance elements. 
The elements for which error is larger than their absolute value are labelled as 100\%.
The relative error 
in the diagonal 
and off-diagonal 
elements is $\sqrt{2/2047}\simeq$ 3.1\%  and $\gtrsim 30$\%, respectively. This happens because
the diagonal elements are larger than the 
off-diagonal ones,
and dominate the variance of the covariance in Eq.~(\ref{eq:CovaTheoryError}) for all elements,
\be
\frac{\Delta \C_{\ell_1 \ell_2}(k_i,k_j)}{ \C_{\ell_1 \ell_2}(k_i,k_j)}
 \sim \bigg[ \frac{1}{\Nm} \frac{\C_{\ell_1 \ell_1}(k_i,k_i)\C_{\ell_2 \ell_2}(k_j,k_j)}{\C_{\ell_1 \ell_2}^2(k_i,k_{j})} \bigg]^{1/2}
\ee
We can see that the relative noise has a quite weak scaling with the number of mocks, $\propto \Nm^{-1/2}$,
and hence reducing the noise on the off-diagonal elements from, e.g., $\sim 50\%$ down to a level of $10\%$ requires 
increasing the number 
of mocks by more than an order of magnitude.


Let us now consider the effects of sampling noise on parameter constraints; it leads to: (a) stochastic inflation/deflation of parameter errorbars and (b) stochastic shifts of the best-fit values of parameters. 
Averaging over ensembles of the estimated covariance matrices, Ref.~\cite{DodSch1309} derived a general formula for the RMS value of the shifts in the best-fit parameters as
 \beq
 \langle \Delta p^2\rangle^{1/2}  = \sigma_p \sqrt{B(n_b-n_p)}\,, \label{eq:RMSshift}
 \eeq
where $\sigma_p$ is the (unknown) true parameter error. Using $B$ from Eq.~\ref{eq:KernelsAB}, the RMS in our case is $0.2 \sigma_p$. 
The stochasticity due to the shifts can be included in the total error budget by adding the RMS value in quadrature to the usual statistical error. This, along with including the effect of (a), leads to rescaling the parameter errorbars by the widely-used $M_1 $ factor \cite{PerRosSan1404},
\beq\label{eq:m1}
    M_1 \equiv  \sqrt{m_1}= \sqrt{\frac{1+B(n_b-n_p)}{1+A+B(n_p+1)}}\, ,
\eeq
where the embedded terms are given by \cite{TayJoaKit1306},
\be
\begin{split}
    & A = \frac{2}{(\Nm-n_b-1)(\Nm-n_b-4)},\\
    &B = \frac{\Nm-n_b-2}{(\Nm-n_b-1)(\Nm-n_b-4)}\,.
    \end{split}
\label{eq:KernelsAB}\ee

Multiplying by $M_1$ is a common practice in the analyses based
on the sample covariance (see e.g.~\cite{BeuSeoRos1701}). 
In our setup $n_p=10$ , $n_b=96$ and $\Nm=2048$, which gives $M_1=1.02$,
and hence, this factor is only a small correction to the typically reported $\sim 10\%$ precision on the errorbars. 
Naively, given a small difference of $M_1$ from unity, one might conclude that the sampling noise
can be ignored.
But re-scaling by $M_1$ only ensures that the constraints are unbiased if we could average the covariance matrix  over an ensemble of noise realizations. 
It does not guarantee 
that the constraints obtained with a single realization of the sample covariance are unbiased\footnote{To test this, we performed a MCMC analysis using a set of 50 synthetic sample covariance matrices (see Appendix~\ref{apx:GeneratingSamples} for details). We indeed found that the errorbars on cosmological parameters in some cases shrank by a much larger amount than what the $M_1 (=1.02)$ factor corrects for (see Fig.~\ref{fig:SamplingEffect} where the $H_0$ errorbars shrink by $\sim$18\% for `Sample 1').}.
It is also important to note that even when the bias of the best-fits due to sampling noise is taken into account by the $M_1$ factor at the level of the errorbars, the mean 
of the posterior distribution from a single realization of the covariance could still have 
 $\sim 0.2\sigma$ shifts w.r.t. true values obtained in the absence of sampling noise. 
This will be important for the interpretation of the results from our BOSS analysis in the next section.

There is another important caveat in using the $M_1$ factor and in Eq.~(\ref{eq:RMSshift}), the derivation of both is based on the assumption of Gaussian parameter likelihood in a Fisher analysis. This assumption can be inaccurate in practice; a classic example is the sum of neutrino masses, whose distribution is peaked at the boundary of the sharp prior $\sum m_\nu >0$, as is found in the analysis of the CMB Planck data and also of spectroscopic surveys \cite{Ala20}.
Other examples of highly non-Gaussian distributions are given by the posteriors for the nuisance parameters (like $b_2$) \cite{IvaSimZal19}.
Such deviation from Gaussianity is expected to exacerbate the effects of sampling noise as compared to a naive Fisher analysis. Therefore, it is imperative to validate the cosmological constraints with different choices of the covariance matrix, which is one of the goals of this paper.

Apart from sampling noise, the covariance matrix is also sensitive to noise due to various numerical approximations involved in its computation. One way to roughly estimate the impact of such numerical noise on the inversion of a matrix is to calculate its conditional number (which is defined as the ratio of the largest eigenvalue to the smallest eigenvalue). The conditional number for the Patchy multipole covariance matrix is quite large ($1.2 \times 10^6$), which implies that the matrix is sensitive to numerical instabilities during inversion. Various approximate techniques used in the generation of mock simulations could therefore further degrade the parameter constraints.

\section{Results}
\label{sec:Results}

In this section we compare the parameter constraints obtained using the covariance matrix from Patchy mocks with those using the analytic covariance. We will first focus on the NGC high-z data chunk ($z_{\rm eff}=0.61$)
and discuss the results for the other samples in the end of this section.

\subsection{Case study: high-z NGC BOSS sample}
\label{sec:HighZNgcResults}

Let us discuss the constraints obtained from the high-$z$ NGC sample, which 
has the largest volume among all the BOSS samples.
We will use this case to illustrate three key aspects of our analytic covariance analyses:
the comparison with the mock covariance
for the same fiducial cosmology, effect of the update to match the best-fit output cosmology,
and the contribution of the non-Gaussian covariance.


\begin{figure*}
    \centering
        \includegraphics[scale=0.5,keepaspectratio=true]{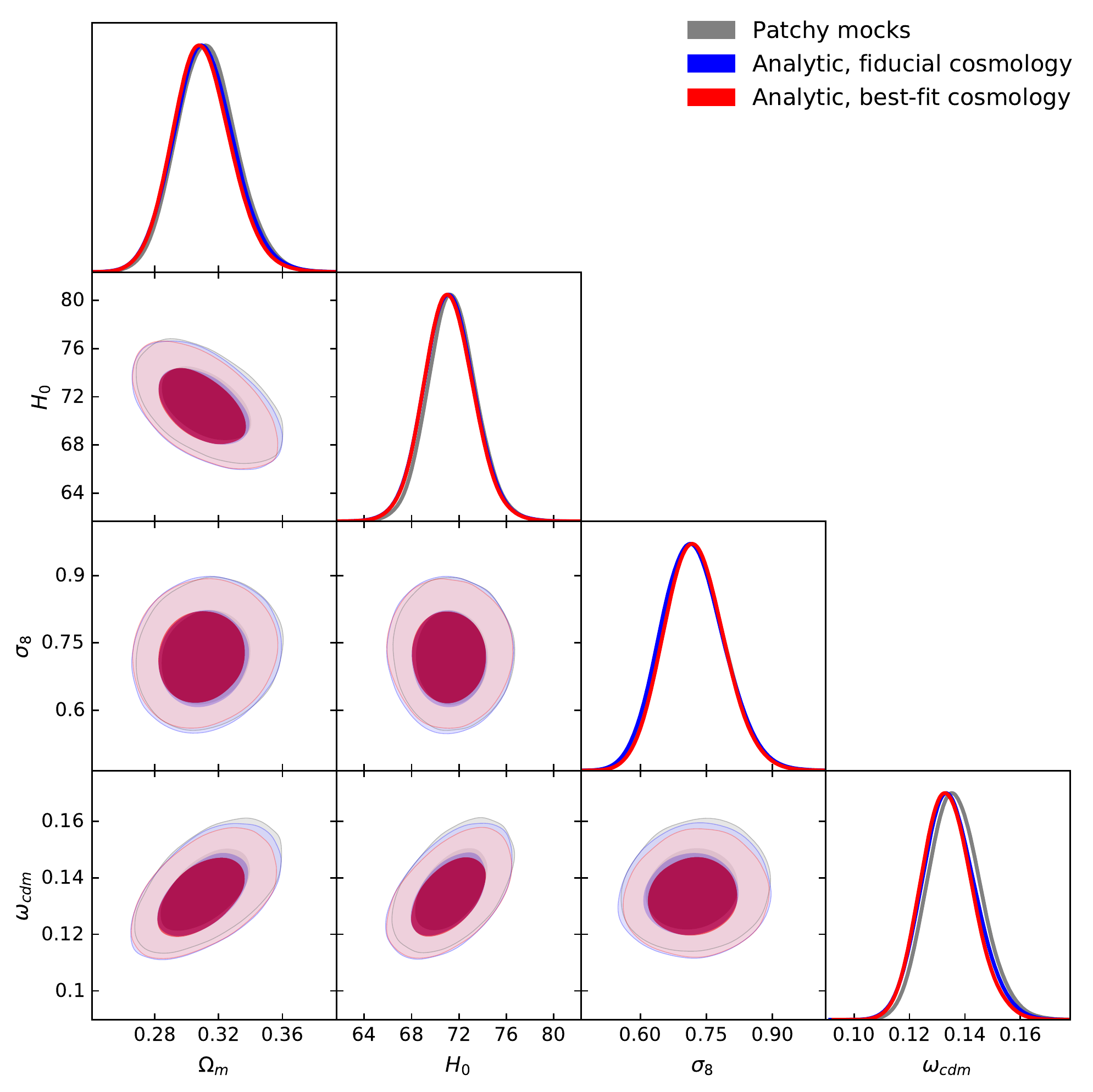}
\caption{Cosmological parameters inferred from the BOSS NGC high-$z$ data chunk using different covariance matrices; corresponding tabulated values are in Table~\ref{tab:ngcz3}. The constraints from the analytic covariance of \citetalias{WadSco19} are quite similar to the ones from 2048 Patchy mocks, except for small $\sim 0.2 \sigma$ shifts which are generated by sampling noise (see the cross-checks shown in Figs. \ref{fig:SamplingEffect} \& \ref{fig:denoise}). The constraints are stable under the change of cosmology of covariance matrix by comparing the cases of the fiducial cosmology used in Patchy and the best-fit output cosmology.
}
    \label{fig:covar_ngc3}
\end{figure*}

\begin{table*}[t!]
  \begin{tabular}{|c||c|c|c|} \hline
    \diagbox{Parameter}{Covariance}   &  ~~~~~~~~Patchy mocks~~~~~~~~ & Analytic (fiducial cosmo)  &  Analytic (best-fit cosmo)   
      \\ [0.2cm]
      \hline 
$H_0$ (km/s/Mpc)   & $71.44_{-2.2}^{+2.0}$ 
& $71.19_{-2.3}^{+2.1}$
& $71.15_{-2.3}^{+2.2}$ \\ \hline
$A^{1/2}$   & $0.8135_{-0.093}^{+0.077}$ 
& $0.8194_{-0.098}^{+0.08}$
& $0.8276_{-0.094}^{+0.078}$
\\ 
\hline
     $\omega_\textup{cdm}$  & $0.1364_{-0.01}^{+0.0091}$
  & $0.1345_{-0.011}^{+0.0092}$
  & $0.1336_{-0.01}^{+0.0089}$ 
  \\ \hline
   $b_1 A^{1/2}$ & $1.905_{-0.056}^{+0.063}$ & $1.915_{-0.056}^{+0.062}$& 
   $1.916_{-0.055}^{+0.062}$
   \\ \hline
   $b_{\mathcal{G}_2} A^{1/2}$ 
   & $0.1627_{-0.23}^{+0.2}$
   &  $0.1468_{-0.23}^{+0.2}$
   & 
  $0.1345_{-0.23}^{+0.19}$
 \\  \hline \hline
$\Omega_m$   & $0.3126_{-0.02}^{+0.018}$
& $0.3111_{-0.021}^{+0.019}$
& $0.3098_{-0.02}^{+0.018}$  \\ \hline
$\sigma_8$   & $0.721_{-0.074}^{+0.064}$
&$0.719_{-0.076}^{+0.065}$
& $0.724_{-0.072}^{+0.063}$ \\ 
\hline
\end{tabular}
\caption{Mean values and 68\% CL minimum credible
intervals for the parameters of the base $\Lambda$CDM model fitted to
the high-$z$ NGC chunks of the 
BOSS data. 
The upper part of the table displays the parameters that were sampled directly.
The lower group lists derived parameters. We show only those nuisance parameters whose posteriors are
noticeably narrower than the priors. 
}
\label{tab:ngcz3}
\end{table*}

\begin{figure*}[ht!]
    \centering
        \includegraphics[scale=0.5,keepaspectratio=true]{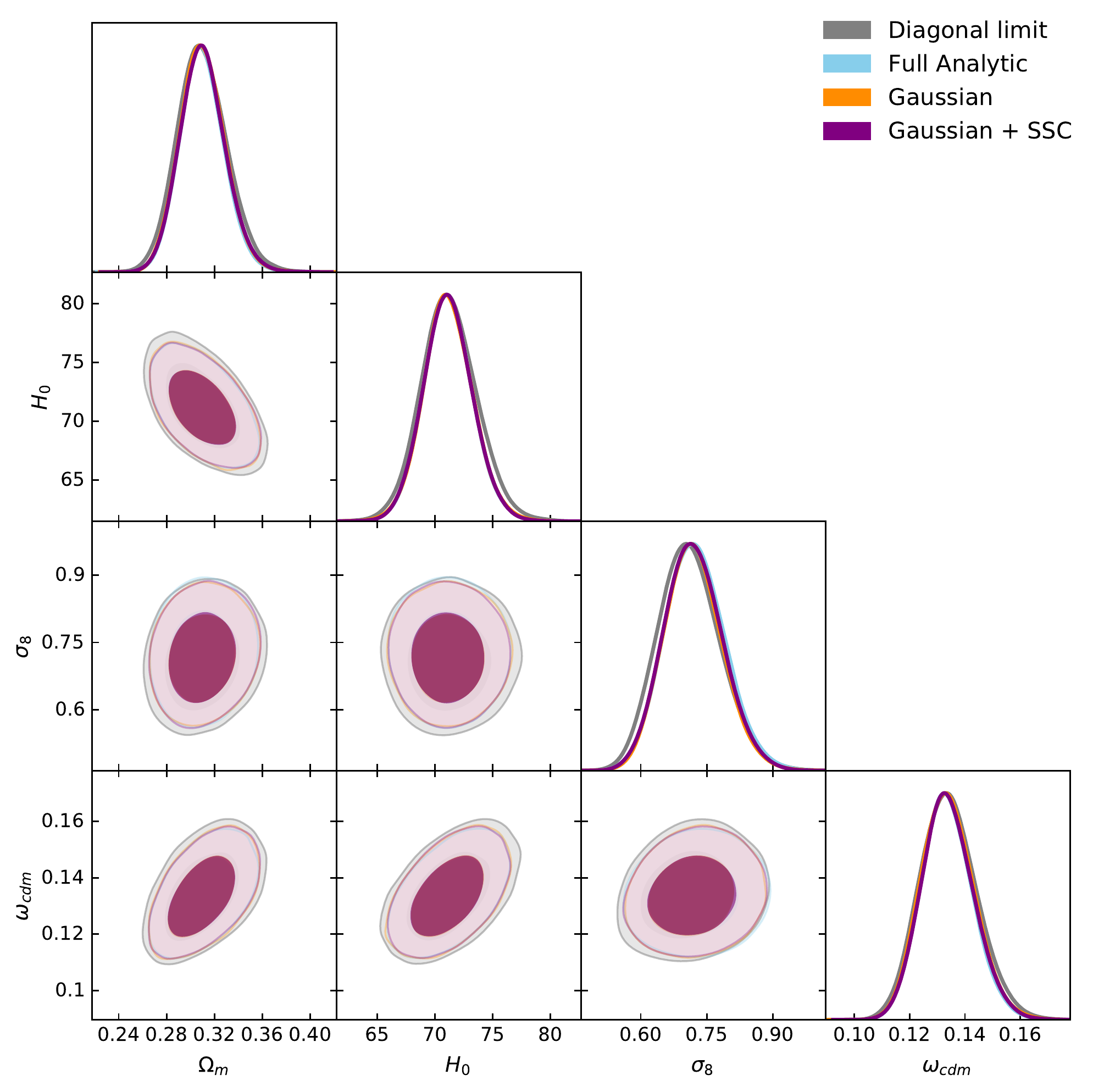}
\caption{Same as Fig.~\ref{fig:covar_ngc3} but including different variations of the analytic covariance matrix:
using only its Gaussian/disconnected part and using the Gaussian part supplemented with the super-sample covariance (SSC);
corresponding tabulated values are in Table~\ref{tab:ngcz3_break}.
We also show the constraints obtained with the approximate diagonal limit version of the Gaussian covariance, which is discussed in detail in
Appendix~\ref{apx:GaussCova} and further compared in Fig.~\ref{fig:ElementsCov_DiagLim}. All contours other than the gray ones are very similar to each other, which shows that the non-Gaussian contributions to the covariance affect the parameter constraints marginally. 
}
\label{fig:breakdown_covar}
\end{figure*}
\begin{table*}[th!]
  \begin{tabular}{|c||c|c|c|c|c|} \hline
    \diagbox{Parameter}{Covariance}   
     &  
     ~~~
     Full analytic 
         ~~~
      &  
          ~~~
      Gaussian 
          ~~~
     &     ~~
     Gaussian + SSC 
         ~~
      &  
          ~~ Diagonal limit     ~~
          &     ~~ Forecast approx.
              ~~
      \\ [0.2cm]
      \hline 
$H_0$    
& $71.15_{-2.3}^{+2.2}$
& $71.16_{-2.2}^{+2.1}$ 
& $71.15_{-2.3}^{+2.1}$ 
& $71.25_{-2.7}^{+2.4}$
& $71.18_{-2.3}^{+2.1}$
\\ \hline
$A^{1/2}$   
& $0.8276_{-0.094}^{+0.078}$
& $0.8218_{-0.089}^{+0.074}$
& $0.8223_{-0.091}^{+0.075}$ 
& $0.8115_{-0.097}^{+0.076}$
& $0.8244_{-0.086}^{+0.071}$
\\ 
\hline
     $\omega_\textup{cdm}$  
  & $0.1336_{-0.01}^{+0.0089}$  
  & $0.1339_{-0.011}^{+0.009}$
  & $0.1339_{-0.01}^{+0.009}$  
  & $0.1341_{-0.011}^{+0.0096}$ 
  & $0.1338_{-0.01}^{+0.0088}$ \\ \hline
    $b_1 A^{1/2}$ 
    & $1.916_{-0.055}^{+0.062}$ 
    &$1.914_{-0.057}^{+0.065}$
    & 
   $1.915_{-0.057}^{+0.065}$
   & $1.914_{-0.064}^{+0.072}$
   & $1.913_{-0.056}^{+0.065}$
   \\ \hline
   $b_{\mathcal{G}_2} A^{1/2}$ 
   & $0.1345_{-0.23}^{+0.19}$ 
   &  $0.1295_{-0.22}^{+0.2}$
   & $0.1291_{-0.22}^{+0.2}$
   & $0.1513_{-0.24}^{+0.22}$
   & $0.1338_{-0.22}^{+0.2}$
   \\ \hline \hline 
$\Omega_m$  
& $0.3098_{-0.02}^{+0.018}$
& $0.3101_{-0.02}^{+0.018}$
& $0.3102_{-0.02}^{+0.018}$ 
& $0.31_{-0.023}^{+0.02}$
& $0.3098_{-0.02}^{+0.018}$ 
\\ \hline
$\sigma_8$   
& $0.724_{-0.072}^{+0.063}$
& $0.719_{-0.069}^{+0.060}$ & $0.720_{-0.070}^{+0.062}$ &$0.711_{-0.076}^{+0.064}$ & 
$0.722_{-0.066}^{+0.057}$\\ 
\hline
\end{tabular}
\caption{Same as Table~\ref{tab:ngcz3} but for different variations of the analytic covariance matrix. See the text for more detail.}
\label{tab:ngcz3_break}
\end{table*}

\subsubsection{Comparison for the same fiducial cosmology}

As a first step, we analyze the data with the analytic covariance evaluated
for the fiducial cosmology used in the Patchy mocks. 
This direct comparison will later allow us to isolate the effects due to
cosmology-dependence of the covariance matrix, which will be discussed in the next sub-section. 
We show the posterior 
distribution of the inferred cosmological parameters in
Fig.~\ref{fig:covar_ngc3}; the corresponding 1d marginalized limits are in Table~\ref{tab:ngcz3}. Note that we display only the bias parameters $b_1A^{1/2}$ and $b_{\mathcal{G}_2}A^{1/2}$ because their limits are substantially narrower
than the priors, unlike the posteriors for other nuisance parameters. 
The only noticeable difference 
between the two cases is a
$0.2\sigma$ shift in $\omega_\textup{cdm}$.
The shift also 
propagates into
a derived parameter $\Omega_m$.
Apart from this, the means in the posterior distributions are very similar and the sizes of error bars agree to $5\%$. Even though the difference is
quite insignificant, it is
somewhat larger than the ensemble-averaged expectation value $M_1=1.02$.

The observed shift of $0.2\sigma$
in $\omega_\textup{cdm}$
is consistent with the theoretical estimate of the shift due to sampling noise in Eq.~(\ref{eq:RMSshift}). Nevertheless, we perform two additional tests to show that the shift resulted due to sampling noise. We give an overview of the tests below and leave the details to Appendices \ref{apx:GeneratingSamples} and \ref{apx:Denoising}. 
In the first test, we constructed synthetic sample covariance matrices using the analytic covariance as reference.
Having analyzed the data with these synthetic covariances, we found very similar $\sim 0.2\sigma$ shifts in different parameters, which indicates that these shifts are indeed to be expected due to sampling noise from $\sim 2000$ mocks.

Our second test is based on the singular value decomposition technique (SVD) to reduce the sampling noise in a sample covariance matrix (see Appendix \ref{apx:Denoising} for details). We find that the slight tension between the analytic and Patchy results is removed once we apply the denoising method to the Patchy covariance.


\subsubsection{Updating the covariance matrix for the best-fit cosmology}
\label{sec:Results_BestFitupdate}


The cosmological and bias parameters found in our likelihood analysis turned out to be different from the ones used to generate the Patchy covariance (see Fig.~\ref{fig:Ckk_fiducial} for the difference between the power spectra obtained using the best-fit parameters and the mean of the Patchy mocks). 
In such a case, one should redo the analysis with the updated covariance evaluated for the best-fit cosmological and nuisance parameters. 
In principle, one should iterate this procedure until the obtained best-fit cosmology matches the one used to generate the covariance. The analytic covariance method allows one to follow this procedure at a negligible computational cost, unlike the case of mock-derived covariance matrices. 

We redo our analysis using the covariance matrix recomputed using the best-fit output cosmology obtained from the first run; we substitute the best-fit power spectrum in Eq.~\ref{eq:GaussCova} to derive the new Gaussian covariance and use the posterior cosmological and bias parameters to resimulate the non-Gaussian terms in Eq.~\ref{eq:A12}. The results are shown in Fig.~\ref{fig:covar_ngc3}.
The updated more accurate covariance causes some changes to constraints as seen in Table~\ref{tab:ngcz3}, but the changes are quite minor: the errorbars agree 
within $10\%$ and means are shifted by $\lesssim 0.1\sigma$. As the changes are not statistically significant enough to warrant another run, our iterative procedure has converged already after one step.
\subsubsection{Impact of the non-Gaussian covariance}
\label{sec:Results_CovaComponents}

\begin{figure*}
\centering
\includegraphics[scale=0.6,keepaspectratio=true]{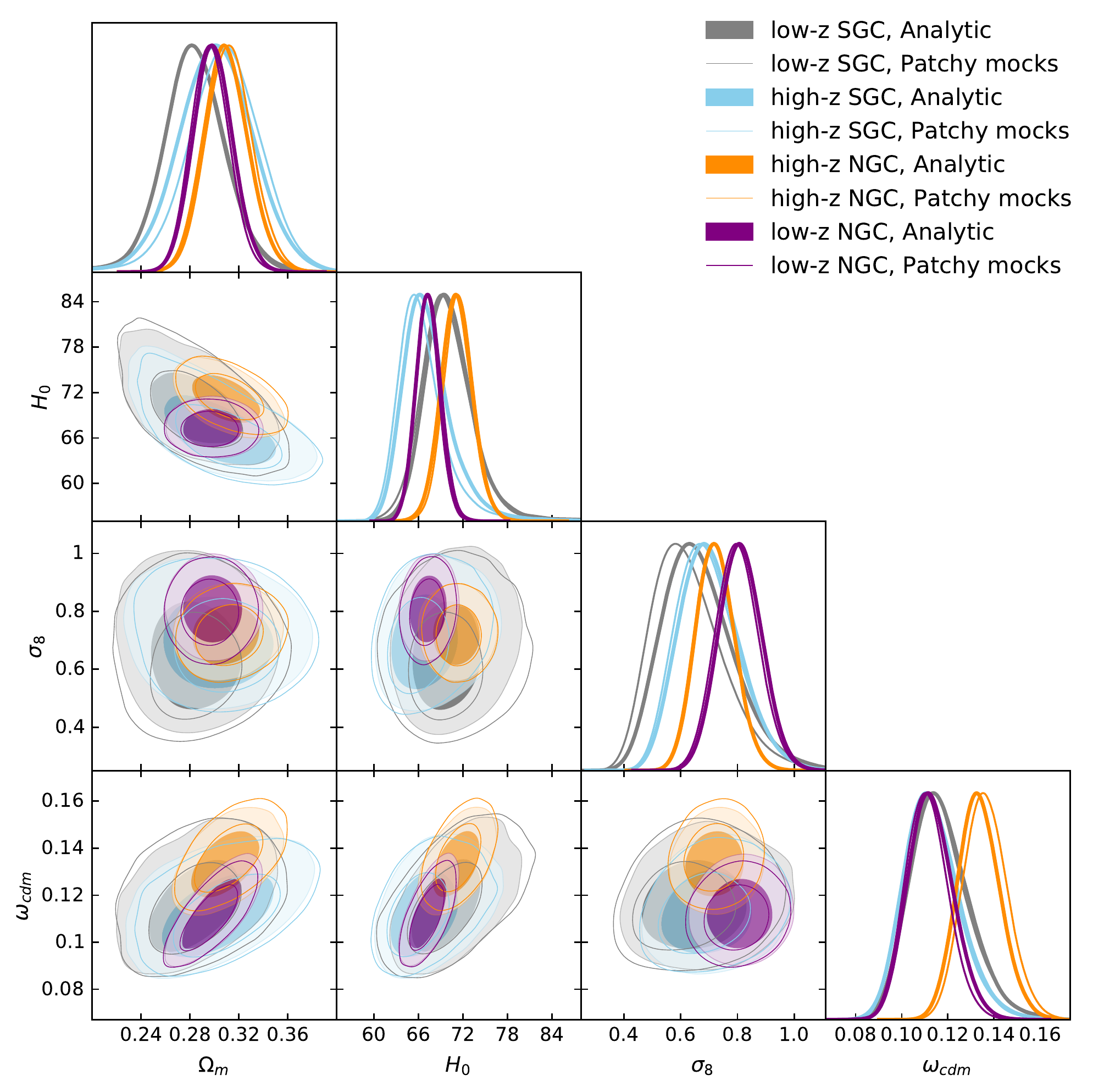}
\caption{Same as Fig.~\ref{fig:covar_ngc3} but for the different BOSS data chunks; the lined (filled) contours are for the Patchy (analytic) covariance. The behavior for each chunk is similar to that in Fig.~\ref{fig:covar_ngc3}: the parameter constraints are almost identical modulo small stochastic parameter shifts
at the $0.1\sigma -0.2\sigma$ level.}
    \label{fig:all_chinks}
\end{figure*}
\begin{table*}[th!]
  \begin{tabular}{|c||c|c||c|c||c|c|} \hline
  Chunk (Cov.)
     &  high-z SGC (M)
      &   high-z SGC (A)
     &   low-z NGC (M)
      &   low-z NGC (A)
          &    low-z SGC (P)
          & low-z SGC (A)
      \\ [0.2cm]
      \hline 
$H_0$ (km/s/Mpc)   
& $66.68_{-4.0}^{+2.2}$
& $67.57_{-4.3}^{+2.3}$ 
& $67.22_{-1.7}^{+1.5}$
& $67.36_{-1.8}^{+1.6}$
& $70.55_{-4.7}^{+2.6}$
& $70.51_{-4.2}^{+2.6}$ 
\\ \hline
$A^{1/2}$   
& $0.899_{-0.16}^{+0.12}$
& $0.9067_{-0.17}^{+0.12}$
& $1.041_{-0.12}^{+0.11}$
& $1.047_{-0.12}^{+0.11}$
& $0.7881_{-0.19}^{+0.12}$
& $0.8343_{-0.19}^{+0.13}$
\\ 
\hline
     $\omega_\textup{cdm}$  
  & $0.1133_{-0.014}^{+0.01}$ 
  & $0.1136_{-0.014}^{+0.01}$ 
  & $0.1111_{-0.01}^{+0.0087}$ 
  & $0.1127_{-0.011}^{+0.0092}$
    &  $0.1165_{-0.016}^{+0.011}$ 
  &  $0.1168_{-0.015}^{+0.011}$
   \\ \hline  
    $b_1A^{1/2}$  
  & $2.064_{-0.093}^{+0.13}$
  & $2.04_{-0.093}^{+0.13}$
  & $1.864_{-0.058}^{+0.061}$
  & $1.865_{-0.056}^{+0.059}$
  & $1.854_{-0.093}^{+0.13}$
  & $1.848_{-0.088}^{+0.12}$
  \\ \hline  
    $b_{\mathcal{G}_2}A^{1/2}$  
  & $0.02241_{-0.33}^{+0.34}$
  & $-0.02387_{-0.34}^{+0.35}$
  & $-0.215_{-0.16}^{+0.17}$
  & $-0.1775_{-0.15}^{+0.16}$
  & $0.2599_{-0.35}^{+0.29}$
  & $0.2535_{-0.33}^{+0.25}$
   \\ \hline \hline 
$\Omega_m$  
& $0.3079_{-0.03}^{+0.031}$
&$0.3007_{-0.031}^{+0.032}$
& $0.2966_{-0.017}^{+0.015}$
& $0.299_{-0.017}^{+0.016}$
& $0.2822_{-0.027}^{+0.028}$
& $0.2824_{-0.026}^{+0.027}$
\\ \hline
$\sigma_8$   
& $0.698_{-0.13}^{+0.087}$
& $0.705_{-0.13}^{+0.089}$ 
& $0.799_{-0.075}^{+0.075}$ &$0.811_{-0.076}^{+0.076}$ 
& $0.630_{-0.15}^{+0.09}$
& $0.667_{-0.15}^{+0.1}$\\ 
\hline
\end{tabular}
\caption{Same as Table~\ref{tab:ngcz3} but for different BOSS data chunks and two choices of the covariance matrix: Patchy mocks (M) and analytic (A) computed for the best-fit cosmologies from the Patchy covariance runs.}
\label{tab:chunks_break}
\end{table*}

\begin{figure*}
    \centering
\includegraphics[scale=0.5,keepaspectratio=true]{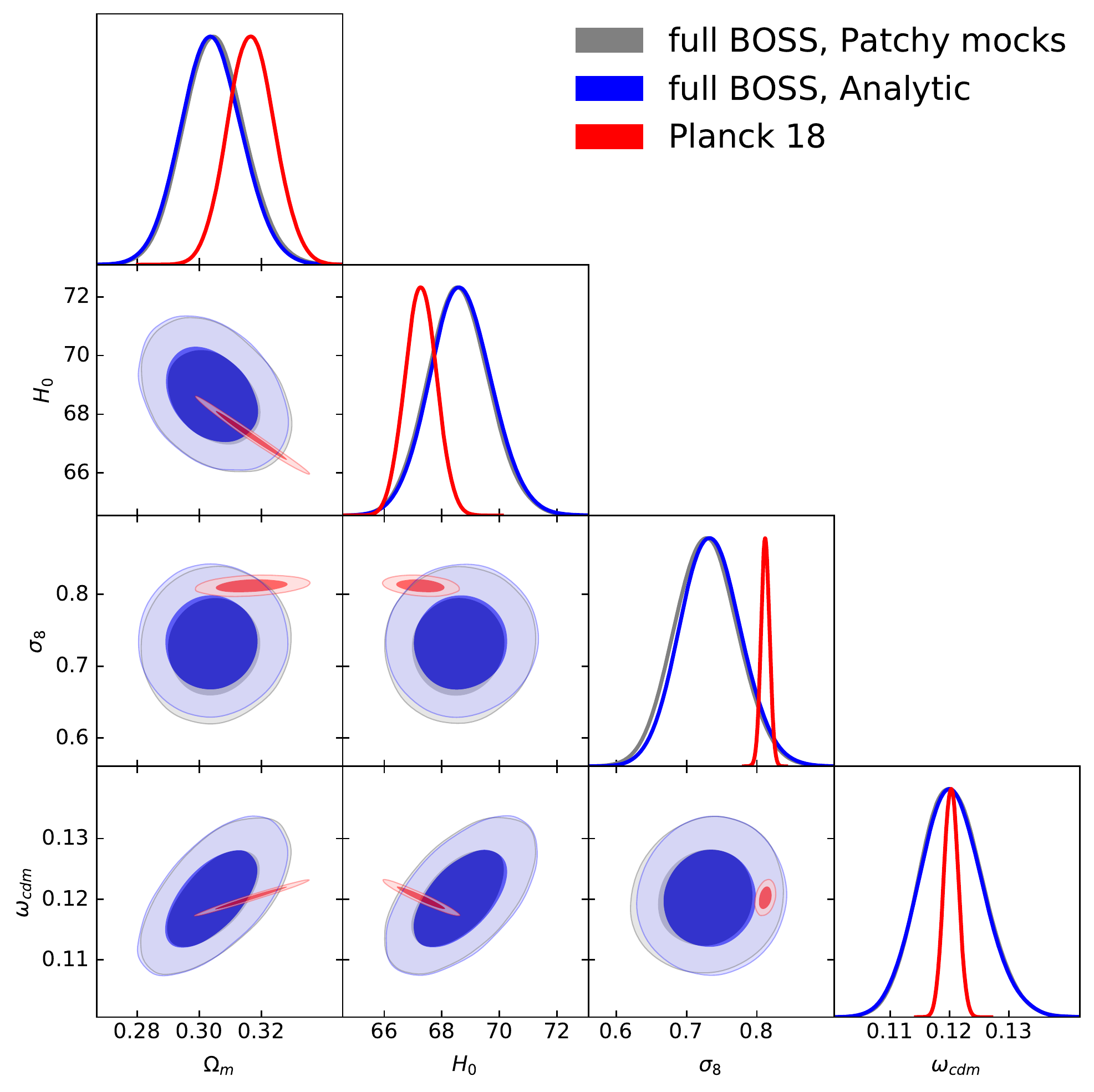}
\caption{Cosmological parameters inferred from the combination of all BOSS data chunks using different covariance matrices; corresponding tabulated values in Table~\ref{tab:full}. The combination of different chunks leads to cancellation of the stochastic shifts seen in Fig.~\ref{fig:all_chinks}, and the resulting constraints are in excellent agreement.
We also show the Planck 2018 results for comparison.}
    \label{fig:all}
\end{figure*}

\begin{table*}[t!]
  \begin{tabular}{|c||c|c|} \hline
    \diagbox{Parameter}{Covariance}   &  ~~~~~~~~Patchy mocks~~~~~~~~ &  Analytic (best-fit cosmo)   
      \\ [0.2cm]
      \hline 
$H_0$ (km/s/Mpc)   & $68.59_{-1.1}^{+1.1}$
& $68.66_{-1.1}^{+1.1}$  \\ \hline
$A^{1/2}$   & $0.8942_{-0.063}^{+0.057}$ 
& $0.9018_{-0.064}^{+0.057}$ 
\\ 
\hline
     $\omega_\textup{cdm}$  & $0.1204_{-0.0057}^{+0.0051}$
  &$0.1203_{-0.0056}^{+0.0052}$
   \\ \hline \hline 
$\Omega_m$   & $0.3048_{-0.011}^{+0.0096}$
& $0.3039_{-0.01}^{+0.0097}$\\ \hline
$\sigma_8$   & $0.728_{-0.044}^{+0.044}$
&$0.734_{-0.043}^{+0.043}$ \\ 
\hline
\end{tabular}
\caption{Same as Table~\ref{tab:ngcz3} but for  BOSS data combined from all chunks.}
\label{tab:full}
\end{table*}

An analytic calculation of the covariance matrix enables one to investigate the effects of various physical contributions on the parameter constraints.
To illustrate this, we rerun our analysis for two more cases: using only the disconnected (Gaussian) part, and including the super-sample covariance (SSC). 
Fig.~\ref{fig:breakdown_covar} compares the parameter constraints with ones from the full (Gaussian + non-Gaussian (NG)) analytic covariance.
We see that the NG covariance (both with and without SSC) affects the parameter constraints marginally \footnote{Note that, contrary to what one might naively expect, there are cases where including the NG covariance can improve constraints (for e.g. $b_1A^{1/2}$ and $b_{\mathcal{G}_2}A^{1/2}$ in Table~\ref{tab:ngcz3_break} and sections of the green curves in Fig.~\ref{fig:NG_marginalization}). This is similar to the behavior seen for bispectrum constraints in \cite{SefCroPue0607}; see their Sec.~5.C for an explanation.} and this effect is also consistent with the Fisher analysis estimates in Fig.~\ref{fig:NG_marginalization}. 
It is worthwhile to note that the effect of the NG covariance increases as the shot noise in the survey decreases and also as we go to smaller scales \cite{WadSco19}.
Because the analytic formalism of \citetalias{WadSco19} currently allows for a cheap computation of the NG part to tree-order,
it should be included in the 
analysis in order to keep the systematic error budget within 
the desired $0.1\sigma$ limit.

We have only considered spectroscopic surveys in this work but it is worth contrasting the effect of NG covariance on parameter constraints from the photometric surveys, for which the analysis is typically done to much smaller scales at which the NG contribution to the total covariance is larger. More importantly, there is a damping of long modes in spectroscopic surveys, because the FKP estimator normalizes the density fluctuations by the total number of galaxies in the survey \cite{WadSco19}. There is no analogous damping in the photometric surveys and therefore the NG covariance is expected to have a relatively larger effect on the parameter constraints \cite{BarKraSch1806,LacGra19}.

Finally, we also repeated our analysis using approximate diagonal versions of the Gaussian covariance matrix used in the literature (see Appendix~\ref{apx:GaussCova} for details), where the effects of the survey geometry and changing LOS are neglected. We first use the diagonal version of the Gaussian covariance written in Eq.~(\ref{eq:DiagonalLimit}), which was referred to as ``diagonal limit'' by \cite{LiSinYu1811}. We also use an even cruder form written in Eq.~(\ref{eq:ForecastApprox}), which is typically used in Fisher forecasts and we label it as ``forecast approximation''. 
Quite surprisingly, these very crude choices yielded parameter constraints that are quantitatively very similar to the ones 
obtained in the full analysis as seen in Fig.~\ref{fig:breakdown_covar} and Table~\ref{tab:ngcz3_break}.
It is important to note that the two cases mentioned above are not at all controlled assumptions and produce drastically different precision matrices as seen in Fig.~\ref{fig:ElementsCov_DiagLim}. For these reasons we do not recommend these for any practical application to data. 
We leave the further discussion to Appendix~\ref{apx:GaussCova}.

\subsection{Results for the full BOSS survey}
\label{sec:FullBossResults}

We use the same procedure as the previous section to analyze each of the BOSS data chunks: we first analyze the data using the Patchy covariance matrices and find the best-fit cosmological and nuisance parameters. We use those best-fit parameters to compute the analytic covariance and use it to reanalyze the BOSS data. Our results, displayed in Fig.~\ref{fig:all_chinks} and in Table~\ref{tab:chunks_break}, are qualitatively similar to the ones obtained for the NGC high-$z$ case: the parameter constraints are almost identical modulo some insignificant parameter shifts.


Importantly, we do not see any significant difference in the behavior of low-$z$ and high-$z$ bins. This suggests that the higher-order
perturbation theory corrections 
omitted in the calculation of the NG covariance
have no sizeable impact on parameter
inference, in agreement with the 
arguments given in \citetalias{WadSco19}.
Indeed, if these corrections had impact on the final result, it would be more pronounced in the low-$z$ bin for two reasons: the non-linear clustering becomes stronger at lower $z$, and the impact of shot noise for the low-$z$ bin is less dominant\footnote{The low-$z$ bin has nearly twice the galaxy number density as the high-$z$ bin.}, which further increases the relative importance of non-linearities.


Overall, we do not see any pattern that would hint at 
a systematic error induced by the analytic covariance.
The parameters from
the four chunks are in good agreement
with each other, and therefore can be combined. Since different chunks represent non-overlapping patches of
the sky and
redshift bins, their data are independent 
and we can simply multiply the corresponding likelihoods.\footnote{It should be mentioned that the contiguous galaxies located at the boundaries of the redshift bins are, in fact, correlated, but the number of these galaxies is small compared to the full samples, such that the independence of different $z$-bins is a reasonable approximation.}
The results are shown in Fig.~\ref{fig:all}, 
where, for reference, we also display the baseline Planck 2018 TTTEEE+low E+lensing
results obtained for the same base $\Lambda$CDM model~\cite{Aghanim:2018eyx}; the 1d marginalized limits are given in Table~\ref{tab:full}.
The posteriors for any of the
parameters are not significantly 
affected by the choice of the 
covariance matrix.
The difference between the constraints is $0.14\sigma$ 
for $\sigma_8$
and less than $0.1\sigma$ 
for the other parameters, in particular, for $\omega_{cdm}$. 
Recall that the individual chunks exhibited somewhat bigger
$\sim 0.2\sigma$ tensions between the analytic and Patchy results. 
The observed reduction of these tensions
is yet
another confirmation that they were produced by stochastic fluctuations in the 
Patchy covariance, which average out when independent samples 
are combined. 

Finally, it is worth commenting on the 
differences of our constraints w.r.t. the previous analysis of Ref.~\cite{ChuIvaSim20},
which used similar priors on cosmological and nuisance parameters. 
This analysis was based on the publicly available measurements of the BOSS power spectra and Patchy covariance matrices\footnote{
\href{https://fbeutler.github.io/hub/hub.html}{\textcolor{blue}{https://fbeutler.github.io/hub/hub.html}}},
and yielded $H_0$ lower by $0.5\sigma$
and $\omega_\textup{cdm}$ lower by $1\sigma$
compared to our present results for the Patchy covariance. 
Using the same public data products,
we have found that these shifts are generated by the difference between the public power spectra and our  measurements from the BOSS galaxy catalogs.
As a cross-check, we performed our power spectra measurements with two independent codes (see Sec.~\ref{sec:AnalysisMethod}) and found identical results, which still slightly differ from the public spectra. 
Since the difference in the eventual constraints is not very significant and it is not caused by the covariance matrix, its thorough investigation goes
beyond the scope of this paper.




\section{Discussion and Conclusions}
\label{sec:Discussions}

In this paper we have 
presented a new analysis of the BOSS full-shape data 
using the perturbation-theory (PT)
based covariance 
matrices of \citetalias{WadSco19}.
This approach is a well-controlled and an extremely cheap complement to the usually adopted way of estimating the sample 
covariance from large sets of mock catalogs. 
The key advantages of the analytic approach over the mock simulations
are that the analytic covariance matrix (i)
can be easily recomputed 
for any input 
cosmology
and (ii) it does not 
contain any finite 
sampling noise, 
which is the main source of bias in the likelihood analyses based on the mock covariances.


The noise in the covariance 
matrix constructed from mock catalogs
biases means and 
variances of inferred 
cosmological parameters. 
Most of studies account for these effects by inflating the 
resulting errorbars by the so-called factor $M_1$~\cite{PerRosSan1404}. 
This practice, however, is not
perfectly accurate: 
it does not guarantee that the constraints obtained with a particular realization of the sample covariance matrix would be unbiased. 
Besides, it assumes 
that the likelihood in the space of parameters
is Gaussian,
which is not the case for realistic 
large-scale structure
likelihoods. 

To illustrate these arguments, 
we have performed a detailed 
study of the sampling noise 
in the covariance matrix
and its impact on the actual BOSS likelihood, presented in Appendix~\ref{AppB}.
We have found that 
indeed the sampling noise
can produce bias on the means 
and variances 
of cosmological parameters,
which cannot be fully taken into account by multiplying the errors with a factor $M_1$ (equal to $1.02$ in our case).
This result can be contrasted with
the previous studies based on toy Gaussian likelihoods~\cite{PerRosSan1404,DodSch1309}, which do not fully capture all features of the real data.

We have explicitly demonstrated 
that the analytic covariance 
matrices give accurate and unbiased constraints,
and moreover, took advantage of the
property that they can 
be easily updated  
to match the output cosmology.
In particular, 
we have reanalyzed the BOSS data 
using the analytic covariance matrices based on the cosmology 
preferred by the data itself.
Modulo small shifts discussed above, 
we found constraints statistically 
consistent 
with the ones based the Patchy mock covariance.
This test also validates 
the previous full-shape results obtained
in the literature 
and removes the 
uncertainty of the BOSS
cosmological constraints based on the covariance matrix. 

Finally, we have discussed the effect of various components of the covariance matrix on the parameter constraints. 
Specifically, we have found that the non-Gaussian covariance, which includes the regular trispectrum and super-sample covariance, affects the parameter errorbars at a marginal level ($\lesssim 10\%$)
and the effect is expected to only mildly increase at a higher $k_\textup{max}$. This is welcome news for a perturbative calculation of the covariance because the treatment of non-linearities in the trispectrum becomes difficult at high-$k$.
Additionally, we have suggested an improvement to Fisher forecasts, which typically use a very crude version of the Gaussian covariance matrix.
Namely, we have provided explicit expressions for the covariance matrix that take into account the non-trivial radial selection function of the survey, see Appendix~\ref{sec:FisherApprox}.

Overall, our results demonstrate the utility of the perturbation theory approach to covariance matrices. We believe that it is an important tool for the upcoming high-precision galaxy redshift surveys such as Euclid~\cite{Laureijs:2011gra,Amendola:2016saw} 
and DESI~\cite{Aghamousa:2016zmz}.



The analytic approach to covariance can be extended in various ways.
The current calculation of the analytic non-Gaussian covariance
does not include higher-order non-linearities (i.e. loop corrections and  ``fingers-of-God'').
Although we did not find evidence that these corrections are important for the BOSS parameter constraints,
they can be consistently included 
within the PT framework along the lines of Refs.~\cite{Bertolini:2015fya,MohSelVla1704, TarNisJeo20}. The impact of non-linearities relative to shot noise can be roughly compared for upcoming surveys like DESI and Euclid using the signal to noise ratio ($\bar{n}P$) at the BAO scale (see Fig.~2 of \cite{FonMcDMos1405}). The value of $\bar{n}P$ is largely similar to that of BOSS (the largest deviation being for the BGS sample of DESI where the value is only $\lesssim 2.5$ larger), so we expect roughly similar impact of non-linearities as seen in BOSS.
Also, including a better treatment of the window convolution in the analytic covariance at low-$k$ would be required in an analysis aimed at constraining primordial non-Gaussianity of local type. 
Another interesting research direction is the calculation and consequence validation 
of the covariance matrices between the power spectrum multipoles and the anisotropic BAO parameters extracted from the post-reconstructed power spectra,
along the lines of~\cite{Philcox:2020vvt}. We leave these questions for future investigation.






\section*{Acknowledgements}

We thank Florian Beutler, Chang Hahn, Yin Li, Sukhdeep Singh, Uros Seljak, Atsushi Taruya, 
Masahiro Takada,
Marcel Schmittful,
Martin White,
Byeonghee Yu and
Matias Zaldarriaga for useful discussions and Alex Eggemeier for his help with the Compass code.
We are indebted to Marko Simonovi\'c, Oliver Philcox and Martin White for their comments on the draft of this paper.
We also thank Florian Beutler for making his measurements publicly available. We have used the \textsc{Cova-PT}$^{\ref{covapt}}$ code to calculate the analytic covariance matrices.

MI is partially supported by the Simons Foundation's \textit{Origins of the Universe} program.


\appendix

\section{More details on analytic covariances}
\label{AppA}

In Sec.~\ref{sec:Results_CovaComponents} we discussed the effects of different components of the covariance matrix and of different approximations on the parameter constraints.
In this appendix we present some of the theoretical expressions we used for analytic covariance and describe the connections to various versions of the covariance used in the literature.  

We adopt the following notation from \citetalias{WadSco19}:
\begin{subequations}
\begin{align}
W_{ij}(\x) \equiv& \bar{n}^i(\x) w^j(\x)\, ,    \\
\I_{ij} \equiv& \int _{\x} \bar{n}^i(\x)w^j(\x)\, ,
\label{eq:Iij} \end{align}\label{WandI}
\end{subequations}where $i,j$
are some (integer) power exponents,
$\bar{n}(\x)$ is the redshift distribution of objects in the survey (also called the radial selection function) and
\beq
 w(\x)\equiv\frac{1}{1+\bar{n}(\x)P_0},
\eeq
 is the well-known FKP weight \cite{FelKaiPea9405} and we adopt $P_0=10000\,  h^{-3}$ Mpc$^{-3}$ \cite{BeuSeoSai1704}.

\subsection{Gaussian covariance and its diagonal limit}
\label{apx:GaussCova}

\begin{figure*}
\centering
\includegraphics[scale=0.45,keepaspectratio=true]{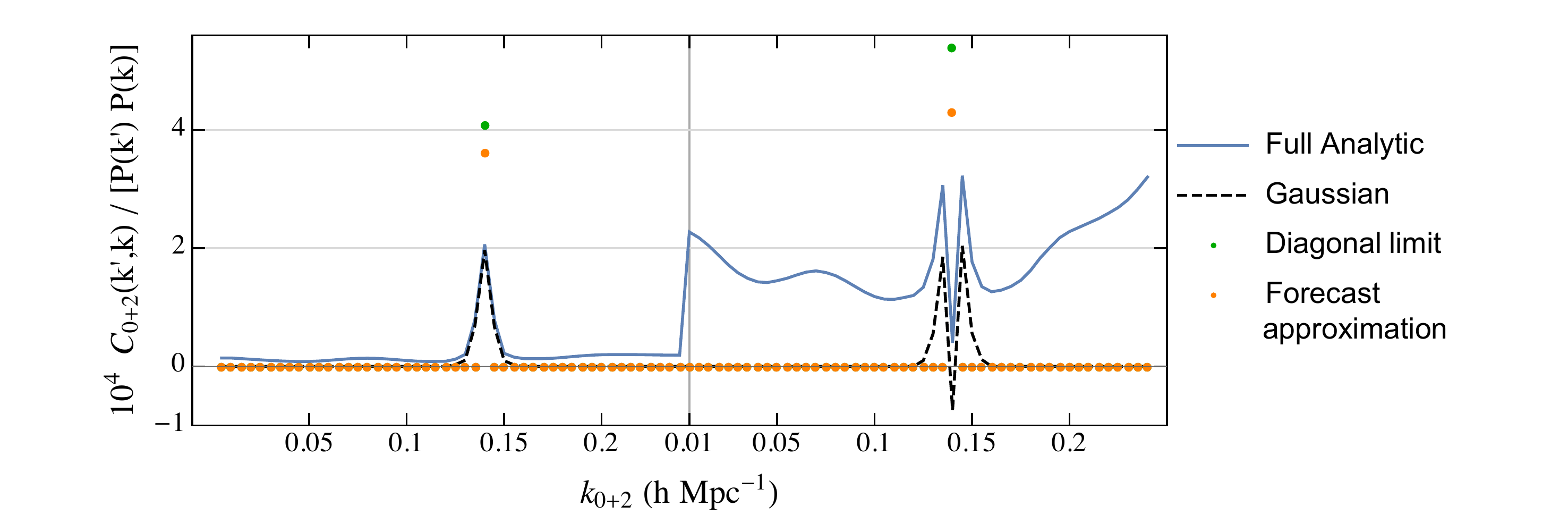}
\includegraphics[scale=0.45,keepaspectratio=true]{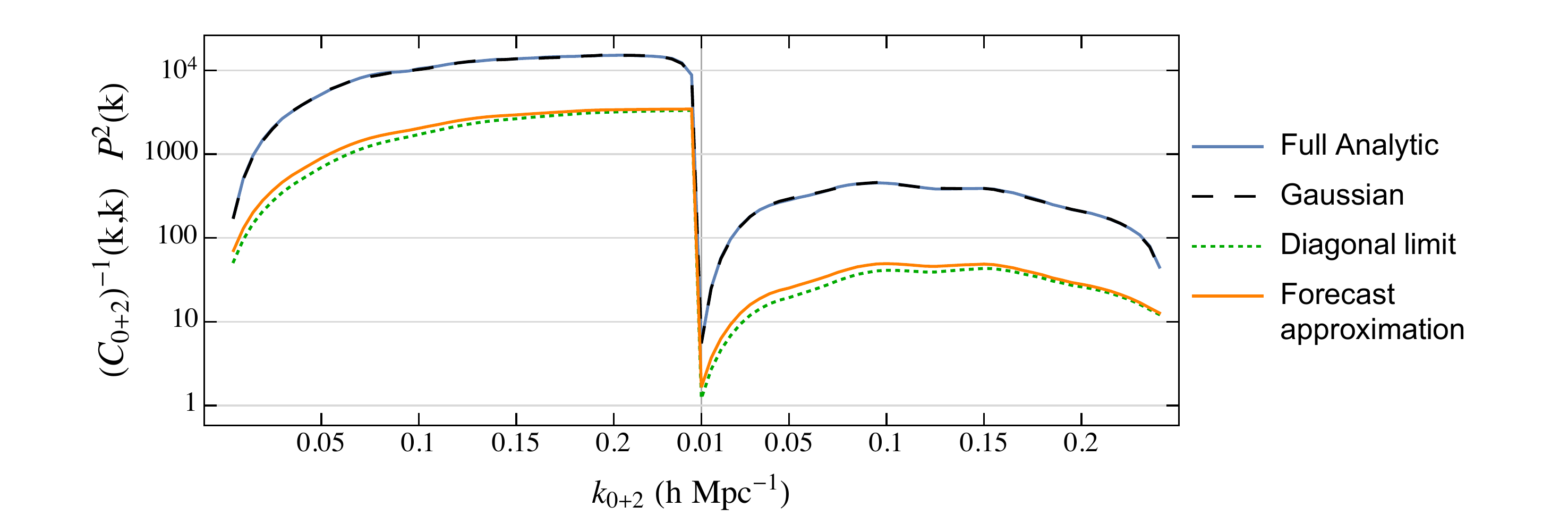}
\caption{Top: A row of the multipole covariance matrix corresponding to $k'_0$=0.147 $\kMpc$. Bottom: Diagonals of the multipole precision matrix. Along with the full analytic covariance and its Gaussian part, we also compare two crude approximations of the Gaussian covariance: the diagonal limit form of the covariance given in Eq.~\ref{eq:DiagonalLimit}, and an even cruder form which is typically used in Fisher forecasts of surveys given in Eq.~\ref{eq:FisherCova}. Even though the two crude approximations give comparable constraints on cosmological parameters in our case, they are not controlled approximations as suggested by the comparison in the bottom panel.}

\label{fig:ElementsCov_DiagLim}
\end{figure*}

Let us start with the Gaussian covariance and write the explicit expressions of the window kernels mentioned in Eq.~(\ref{eq:GaussCova}) \citepalias{WadSco19}:
\begin{widetext}
\beq\begin{split}
\mathcal{W}^{(1)}_{\ell_1,\ell_2,\ell'_1,\ell'_2} (k_1,k_2)\equiv&
 \frac{(2\ell_1+1)(2\ell_2+1)}{\I_{22}^2}\int_{\hat{\k}_1,\hat{\k}_2,\x_1,\x_2} W_{22}(\x_1) W_{22}(\x_2)\, e^{-i(\x_1-\x_2)\cdot(\k_1-\k_2)}\\
& \qquad \times \mathcal{L}_{\ell_1}(\hat{\x}_1 \cdot \hat{\k}_1) \mathcal{L}_{\ell'_1}(\hat{\x}_2 \cdot \hat{\k}_1) \mathcal{L}_{\ell'_2}(\hat{\x}_1 \cdot \hat{\k}_2) \Big[ \mathcal{L}_{\ell_2}(\hat{\x}_2 \cdot \hat{\k}_2) + \mathcal{L}_{\ell_2}(\hat{\x}_1 \cdot \hat{\k}_2) \Big]\, ,\\
\mathcal{W}^{(2)}_{\ell_1,\ell_2,\ell'} (k_1,k_2)\equiv&\frac{(1+\bar{\alpha})}{2} \frac{(2\ell_1+1)(2\ell_2+1)}{\I_{22}^2}\int_{\hat{\k}_1,\hat{\k}_2,\x_1,\x_2}
e^{-i(\k_1-\k_2)\cdot (\x_1-\x_2)} W_{22}(\x_1)W_{12}(\x_2) \L_{\ell'}(\hat{\x}_1\cdot\hat{\k}_1)\\
 \times \Big[
\L_{\ell_1}(\hat{\x}_1\cdot\hat{\k}_1)&
\L_{\ell_2}(\hat{\x}_2\cdot\hat{\k}_2)+\L_{\ell_1}(\hat{\x}_2\cdot\hat{\k}_1)
\L_{\ell_2}(\hat{\x}_1\cdot\hat{\k}_2)+\L_{\ell_1}(\hat{\x}_1\cdot\hat{\k}_1)
\L_{\ell_2}(\hat{\x}_1\cdot\hat{\k}_2)+\L_{\ell_1}(\hat{\x}_2\cdot\hat{\k}_1)
\L_{\ell_2}(\hat{\x}_2\cdot\hat{\k}_2)\Big] \bigg \}\\
\mathcal{W}^{(3)}_{\ell_1,\ell_2}(k_1,k_2)\equiv&(1+\bar{\alpha})^2 \frac{(2\ell_1+1)(2\ell_2+1)}{\I_{22}^2}\int_{\hat{\k}_1,\hat{\k}_2,\x_1,\x_2} W_{12}(\x_1)W_{12}(\x_2) e^{-i(\k_1-\k_2)\cdot(\x_1-\x_2)}\\
&\qquad \qquad \qquad \times \L_{\ell_1}(\hat{\x}_1\cdot\hat{\k}_1)\Big[\L_{\ell_2}(\hat{\x}_1\cdot\hat{\k}_2)+\L_{\ell_2}(\hat{\x}_2\cdot\hat{\k}_2)\Big]\\
\end{split}\label{eq:W_kernel}\eeq
where $\bar{\alpha}$ is the ratio of number of objects in the galaxy and random catalogs ($\equiv \textup{N}_\textup{g}/\textup{N}_\textup{r}\ll 1$) and $\int_{\hat{\k}_i}$ represents an integral over the volume of the $k_i$ bin.
Note that the Gaussian covariance in Eq.~(\ref{eq:GaussCova}) is not exactly diagonal due to the leakage into the neighboring bins introduced by the survey window which has a finite width in $k$-space.
If one uses the approximations that the width of the survey window in $k$-space is much smaller than the width of the $k$-bins then the Gaussian covariance becomes diagonal. If one further assumes that
the LOS along the survey volume is fixed to a particular direction $\hat{\textbf{n}}$, i.e $\hat{\k}\cdot \hat{\x}_i \rightarrow \hat{\k}\cdot \hat{\textbf{n}}$, one gets vast simplifications in the kernels in Eq.~(\ref{eq:W_kernel}). By using identities like $\int_{\k} W_{22}(\k) W_{22}(-\k)=\int_{\x} W^2_{22}(\x)= \I_{44} $, the Gaussian covariance in Eq.~(\ref{eq:GaussCova}) simplifies as:

\beq
\begin{split}
\C_{00}(k_i,k_j) =&\frac{2}{V_k} \frac{\delta^\textup{K}_{ij}}{\I^2_{22}} \bigg[\I_{44}\bigg(P^2_0+\frac{1}{5} P^2_2+\frac{1}{9} P^2_4\bigg) + 2\, \I_{34}P_0+\I_{24}\bigg]\\
\C_{02}(k_i,k_j)=&\frac{2}{V_k} \frac{\delta^\textup{K}_{ij}}{\I^2_{22}}\bigg[ \I_{44}\bigg(2\, P_0 P_2+ \frac{2}{7}P^2_2 +\frac{4}{7}P_2P_4+\frac{100}{693} P^2_{4}\bigg)+2\, \I_{34}\, P_2\bigg]\\
\C_{22}(k_i,k_j)=&\frac{2}{V_k}  \frac{\delta^\textup{K}_{ij}}{\I^2_{22}} \bigg[\I_{44} \bigg( 5P_0^2+\frac{20}{7}P_0P_2+\frac{20}{7}P_0P_4+\frac{15}{7}P_2^2 +\frac{120}{77}P_2P_4+\frac{8945}{9000}P^2_4\bigg)\\
& \ \ \ \ \ \ \ \ \ + \I_{34} \bigg(10 P_0+\frac{20}{7}P_2 +\frac{20}{7}P_4 \bigg)+ 5\, \I_{24} \bigg]\\
\C_{04}(k_i,k_j)=&\frac{2}{V_k} \frac{\delta^\textup{K}_{ij}}{\I^2_{22}}\bigg[ \I_{44}\bigg(\frac{18}{35} P^2_2+ 2\, P_0P_4 +\frac{40}{77}P_2P_4+\frac{162}{1001} P^2_4\bigg)+2\, \I_{34}\, P_4\bigg]\\
\C_{24}(k_i,k_j)=&\frac{2}{V_k} \frac{\delta^\textup{K}_{ij}}{\I^2_{22}}\bigg[ \I_{44}\bigg(\frac{36}{7} P_0P_2+\frac{108}{77} P^2_2+ \frac{200}{77} P_0P_4 +\frac{3578}{1001}P_2P_4+\frac{900}{1001} P^2_4\bigg)+\I_{34} \bigg(\frac{36}{7} P_2+\frac{200}{77} P_4\bigg)\bigg]\\
\C_{44}(k_i,k_j)=&\frac{2}{V_k}  \frac{\delta^\textup{K}_{ij}}{\I^2_{22}} \bigg[\I_{44} \bigg( 9P_0^2+\frac{360}{77}P_0P_2+\frac{16101}{5005}P_2^2+\frac{2916}{1001}P_0P_4 +\frac{3240}{1001}P_2P_4+\frac{42849}{17017}P^2_4\bigg)\\
& \ \ \ \ \ \ \ \ \ + \I_{34} \bigg(18 P_0+\frac{360}{77}P_2 +\frac{2916}{1001}P_4 \bigg)+ 9\, \I_{24} \bigg]
\end{split}\label{eq:DiagonalLimit}
\eeq

\end{widetext}

where $V_k \equiv \frac{4 \pi k^2dk}{(2\pi)^3} $ is proportional to the volume of the $k$-bin with width $dk$ and $P_\ell(k)$ are the typical Kaiser multipoles given by
$P_\ell(k)\equiv \int \frac{d\mu}{2} (1+\beta \mu^2)^2 P_{\rm lin}(k)$ and $\mu\equiv\hat{\k}\cdot \hat{\textbf{n}}$.
We have not used the hexadecapole in our BOSS analysis but we write its expressions here for completeness.

We compare the parameter constraints obtained using the diagonal limit covariance in Fig.~\ref{fig:breakdown_covar}.
We also compare elements of the covariance and the inverse covariance matrix for different cases in Fig.~\ref{fig:ElementsCov_DiagLim}. Although the diagonal limit covariance is seen to produce qualitatively similar constraints in Fig.~\ref{fig:breakdown_covar}, it is clear from the comparison of diagonals of the precision matrix that the diagonal limit is not a controlled approximation as the elements differ by almost an order of magnitude. Further checks are needed to quantify the cases where the approximations involved in the diagonal limit are controlled. 
Note also in Fig.~\ref{fig:ElementsCov_DiagLim} that the cross-covariance for the Gaussian case has a non-trivial shape because the survey window is not isotropic but itself has a quadrupole (see also Fig.~2 of \citetalias{WadSco19}) but such effects due to shape of the survey window are neglected in the diagonal limit case.
Furthermore, the effects of the changing LOS along the survey volume are also neglected in the diagonal limit case but these should affect low redshift bins the most. The full Gaussian covariance in Eq.~(\ref{eq:GaussCova}) therefore should be preferred.

\subsubsection{Approximations used in Fisher forecasts}
\label{sec:FisherApprox}
Let us now compare the diagonal limit case in Eq.~\ref{eq:DiagonalLimit} to the often-used expression of Gaussian covariance in Fisher and MCMC forecasts for future surveys, e.g.~~\cite{Audren:2012vy,Aghamousa:2016zmz,Chudaykin:2019ock,Blanchard:2019oqi,Nis20,Ivanov:2020ril} (labelled as `forecast approximation' hereafter) which is given by

\beq
\begin{split}
\C_{\ell_1\ell_2} (k_i,k_j) &= \delta^\textup{K}_{ij} \frac{2}{N_{k_i}} (2\ell_1+1) (2\ell_2+1)\\
\times& \int_{-1}^1 \frac{d\mu}{2} \L_{\ell_1}(\mu) \L_{\ell_2}(\mu)\\
&\times \bigg[\sum_{\ell'_1}P_{\ell'_1}(k_i) \L_{\ell'_1}(\mu) +\frac{1}{\bar{n}_\textup{survey}}\delta^\textup{K}_{\ell'_1 0}\bigg]\\
&\times \bigg[\sum_{\ell'_2}P_{\ell'_2}(k_j)\L_{\ell'_2}(\mu)+\frac{1}{\bar{n}_\textup{survey}}\delta^\textup{K}_{\ell'_2 0}\bigg]
\end{split}
\label{eq:FisherCova}\eeq
where $N_{k} = \textup{V}_\textup{survey} V_k$
is interpreted as total number of $k$-modes in a bin of width $dk$ and $\bar{n}_\textup{survey}$ is calculated as the ratio of total number of galaxies to the volume of the survey ($ \equiv \textup{N}_\textup{survey}/ \textup{V}_\textup{survey}$).

\begin{figure}
\centering
\includegraphics[scale=0.8,keepaspectratio=true]{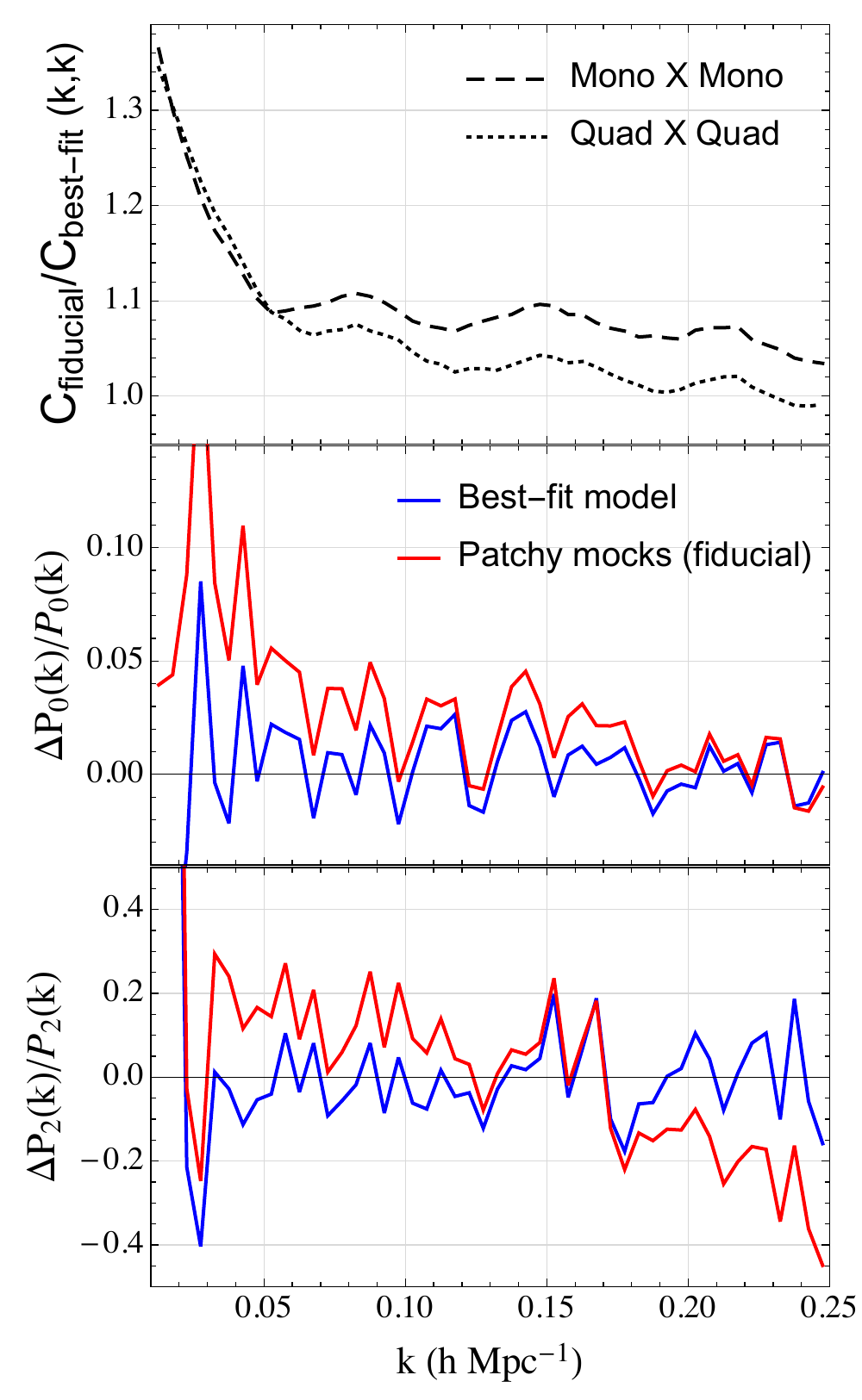}
\caption{Top panel shows the change in the diagonal elements of the analytic auto-covariance matrix on using our best-fit output cosmology as compared to the fiducial cosmology used in Patchy mocks \cite{KitRodChu1603}. The relative difference in the monopole (quadrupole) power spectrum compared to the BOSS NGC high-$z$ measurements is shown in the middle (bottom) panel; the mean of the power spectra from Patchy mocks is in red and our best-fit power spectrum model is in blue. 
}
\label{fig:Ckk_fiducial}
\end{figure}

If one neglects the FKP weights (which amounts to ignoring the radial selection function and assuming that the survey has a uniform number density throughout) and uses the following approximations in Eq.~(\ref{eq:DiagonalLimit}):
\beq
\frac{\I_{34}}{\I_{44}} \rightarrow \frac{1}{\bar{n}_\textup{survey}};\ \ \  \frac{\I_{24}}{\I_{44}}\rightarrow \left(\frac{1}{\bar{n}_\textup{survey}}\right)^2;\ \ \frac{\I^2_{22}}{\I_{44}}\rightarrow \textup{V}_\textup{survey}\, ,
\label{eq:ForecastApprox}\eeq
 one gets the forecast approximation expression in Eq.~(\ref{eq:FisherCova}).
It is important to note that the approximations in Eq.~(\ref{eq:ForecastApprox}) can be \emph{particularly inaccurate for  realistic galaxy surveys}. For example, the NGC high-$z$ chunk gives the following values for the shot noise-like terms:
\be  \begin{split} &\{1/\bar{n}_\textup{survey}\, ,\, \I_{34}/\I_{44}\, ,\, \sqrt{\I_{24}/\I_{44}}\}
\\& = \{6.38,4.17,5.21\}\times 10^3 (\kMpc)^3
\end{split}
\ee 
and for the survey volume-like terms (\cite{AlaAtaBai1709,ReiHoPad16}): 
\be
 \{\textup{V}_\textup{survey}\, ,\, \I^2_{22}/\I_{44}\}=\{2.78, 1.91\}\quad (\text{Gpc}/h)^3
\ee
In the case when  the redshift distribution of the survey is known, but one has no knowledge of the survey geometry, we recommend the use of the diagonal limit expressions in Eq.~(\ref{eq:DiagonalLimit}) instead of Eq.~(\ref{eq:FisherCova}). Note that terms of the form $\I_{ij}$ given in Eq.~(\ref{eq:Iij}) are straightforward to calculate: one only needs to perform a one-dimensional integral by using the survey redshift distribution $\bar{n}(z)$. If one has information of the survey geometry, we recommend using the full Gaussian covariance in Eq.~(\ref{eq:GaussCova}). We also show a comparison of the forecast approximation case in Fig.~\ref{fig:ElementsCov_DiagLim}.



\begin{figure*}
\includegraphics[scale=0.5]{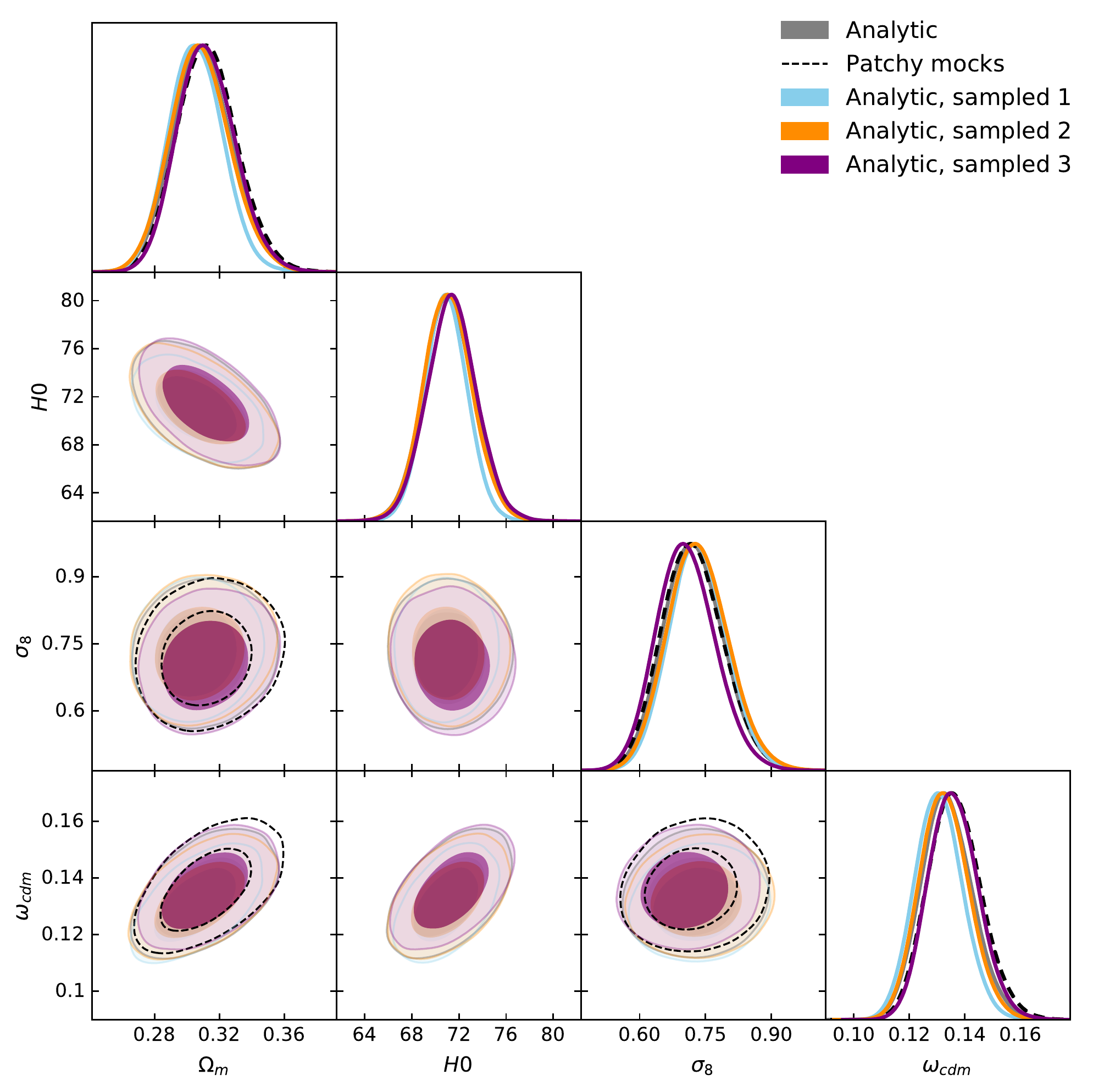}
\caption{Same as Fig.~\ref{fig:covar_ngc3} but also including three realizations of synthetic sample covariance matrices, each made by sampling 2048 power spectra from the analytic covariance. 
The samples were selected in a set of 50 based on their best-fit $\chi^2$ values; see the text for details.
The tabulated values are in Table~\ref{tab:ngcz3_samples}. The sampling noise in covariance leads to stochastic shifts in the parameter means and the error bars are both inflated and deflated in some cases. The most noticeable changes are for `Sample 1' where the $H_0$ errorbar contracts by $\sim 18$\% and mean of $\omega_\textup{cdm}$ shifts by 0.3$\sigma$.}
\label{fig:SamplingEffect}
\end{figure*}

\begin{table*}[th!]
  \begin{tabular}{|c||c|c|c|c|} \hline
    \diagbox{Parameter}{Covariance}   
     &  
     ~~~~
     Analytic
      ~~~~
      &  ~~~~
      Sample 1 
     ~~~~
     & ~~~~  Sample 2 ~~~~
      & ~~~~  Sample 3 ~~~~
      \\ [0.2cm]
      \hline 
$H_0$    
&  $71.15_{-2.3}^{+2.2}$
& $70.92_{-1.9}^{+1.8}$ 
& $71.09_{-2.2}^{+2}$
& $71.41_{-2.2}^{+2.1}$
\\ \hline
$A^{1/2}$   
& $0.8276_{-0.094}^{+0.078}$
& $0.8505_{-0.09}^{+0.076}$ 
& $0.8418_{-0.096}^{+0.08}$ 
& $0.799_{-0.091}^{+0.075}$
\\ 
\hline
     $\omega_\textup{cdm}$  
  & $0.1336_{-0.01}^{+0.0089}$ 
  & $0.1307_{-0.0091}^{+0.0083}$
  & $0.1327_{-0.0097}^{+0.0085}$ &
  $0.1359_{-0.0097}^{+0.0087}$
   \\ \hline \hline 
$\Omega_m$  
& $0.3098_{-0.02}^{+0.018}$
& $0.3058_{-0.018}^{+0.017}$ 
& $0.3084_{-0.02}^{+0.018}$
&$0.312_{-0.019}^{+0.017}$
\\ \hline
$\sigma_8$   
& $0.723_{-0.072}^{+0.063}$
& $0.733_{-0.068}^{+0.062}$ &$0.733_{-0.072}^{+0.065}$ &$0.707_{-0.071}^{+0.061}$\\ 
\hline
\end{tabular}
\caption{Same as Table~\ref{tab:ngcz3} but for  synthetic sampled versions of the analytic covariance matrix.
}
\label{tab:ngcz3_samples}
\end{table*}

\subsection{Non-Gaussian covariance} \label{apx:NGcova}

We give a very brief introduction of the terms in the non-Gaussian covariance in this section and refer the reader to \citetalias{WadSco19} for further details.
Using their notation, let us start by writing the FKP estimator for the galaxy overdensity as 
\begin{equation}
\hat{\delta}^\textup{FKP}(\x) = \frac{1}{\sqrt{\I_{22}}} \frac{\delta_W (\x)}{(1+\dng)^{1/2}}\, ,
\label{eq:delta_FKP}
\end{equation}
where $\dng$ is the long-wavelength fluctuation in the number of galaxies in the survey $(\dng \equiv [\int_{\x} \delta(\x) W_{10}(\x)]/\I_{10})$.
Ignoring the constant pre-factors, the covariance of the 3D power spectrum can be written as
\beq
 \left\langle \frac{|\delta_W(\k_1)|^2|\delta_W(\k_2)|^2)}{(1+\dng)^2}\right\rangle- \left\langle\frac{|\delta_W(\k_1)|^2}{(1+\dng)}\right\rangle \left\langle\frac{|\delta_W(\k_2)|^2}{(1+\dng)}\right\rangle  
\eeq
which can be decomposed into a Gaussian/disconnected part
\beq
\begin{split}
 \textbf{C}^\textup{G}(\k_1,\k_2) =& \langle \delta_W(\mathbf{k}_1) \delta_W(\mathbf{k}_2)\rangle \langle \delta_W(-\mathbf{k}_1) \delta_W(-\mathbf{k}_2)\rangle\\
&+\langle \delta_W(\mathbf{k}_1) \delta_W(-\mathbf{k}_2)\rangle \langle \delta_W(-\mathbf{k}_1) \delta_W(\mathbf{k}_2)\rangle 
\end{split}
\eeq
and all the remaining terms make up the non-Gaussian part:
\beq
\begin{split}
\textbf{C}^\textup{NG}(\k_1,\k_2)=&\langle|\delta_W(\k_1)|^2 |\delta_W(\k_2)|^2\rangle_c\\
&-\langle|\delta_W(\k_1)|^2 \dng\rangle \langle|\delta_W(\k_2)|^2 \rangle\\
&- \langle|\delta_W(\k_1)|^2 \rangle \langle|\delta_W(\k_2)|^2 \dng\rangle\\
&+\langle|\delta_W(\k_1)|^2\rangle \langle|\delta_W(\k_2)|^2\rangle \langle\dng^2\rangle
\end{split}\label{eq:A12}
\eeq
If we break the connected four-point function into a regular trispectrum part ($\textup{T}_0$), and a beat-coupling (BC) part which includes the contribution of the long modes to the trispectrum \cite{HamRimSco0609},
the NG part can be written as
\beq
\textbf{C}^\textup{NG}=\textbf{C}^{\textup{T}_0}+\textbf{C}^\textup{BC}+\textbf{C}^\textup{LA}
\eeq
where the three terms in Eq.~\ref{eq:A12} containing the $\dng$ variables are given the label `local average' (LA) \cite{PutWagMen1204}. Note that the $\textbf{C}^\textup{LA}$ expressions we use correspond to the FKP estimator and are different from those of Ref.~\cite{PutWagMen1204}. 
The total contribution of the super-survey modes is often referred to as the super-sample covariance (SSC) in the literature:
$\textbf{C}^\textup{SSC}
=\, \textbf{C}^\textup{BC}+\textbf{C}^\textup{LA}$. Due to the modification to the $\textbf{C}^\textup{LA}$ terms as mentioned earlier, the SSC effect becomes stronger than previously assumed for spectroscopic surveys in the literature (five times stronger for the case of real-space matter covariance \cite{WadSco19}).
 It is straightforward to account for the super survey modes in the analytic calculation as they are typically in the linear regime. We compute the trispectrum terms upto tree order in PT in this paper.
Finally, it is important to note that there is a significant shot noise contribution to the terms $\textbf{C}^{\textup{T}_0}$ and $\textbf{C}^\textup{LA}$ at high-$k$ and we have used the corresponding terms in the Poisson approximation using the formulae of \citetalias{WadSco19}.

Having discussed all the components of the analytic covariance, let us now continue our discussion on  reevaluation of the covariance for the best-fit cosmology from Sec.~\ref{sec:Results_BestFitupdate}.
The simulations of mocks is started before the data collection in the survey is complete and the mocks are therefore simulated for a fiducial set of cosmological parameters. Parameters corresponding to the bias and velocity dispersion are later adjusted to fit the two and three point clustering measurements of the survey data \cite{BauChu18}. However, in the Patchy mocks, there is some deviation in the mean power spectrum from mocks and the BOSS data as seen in the middle and lower panels of Fig.~\ref{fig:Ckk_fiducial}. We also compare the best-fit power spectrum from our model. We show in the top panel the change in auto-covariance diagonal elements on using the best-fit output cosmology as compared to the fiducial cosmology. We had however found in Fig.~\ref{fig:covar_ngc3} that this change in the covariance has a quite minor effect on the parameter constraints.

\section{Tests for noise in sample covariance matrices}
\label{AppB}

\begin{figure*}
\centering
\includegraphics[scale=0.4,keepaspectratio=true]{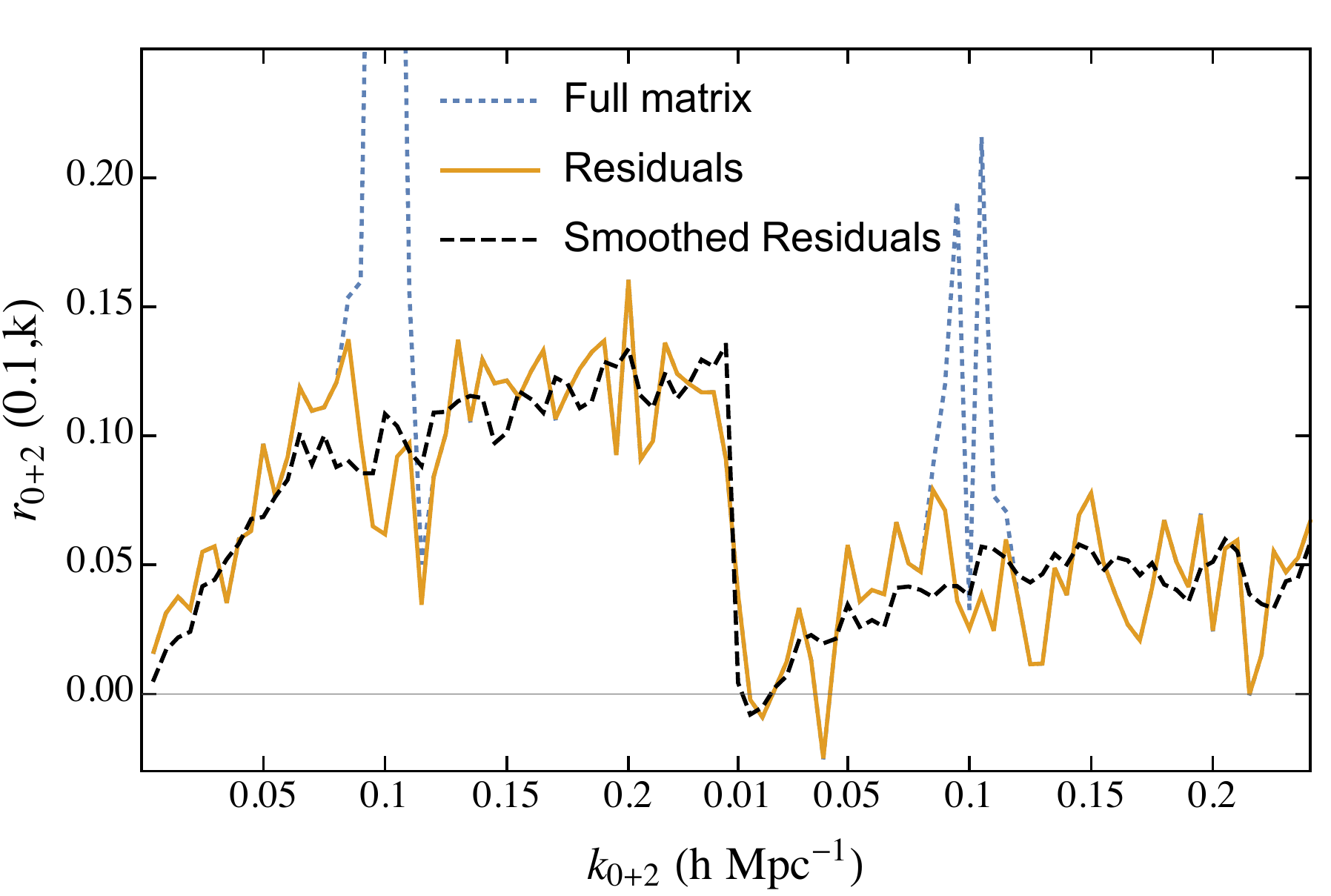}
\includegraphics[scale=0.53,keepaspectratio=true]{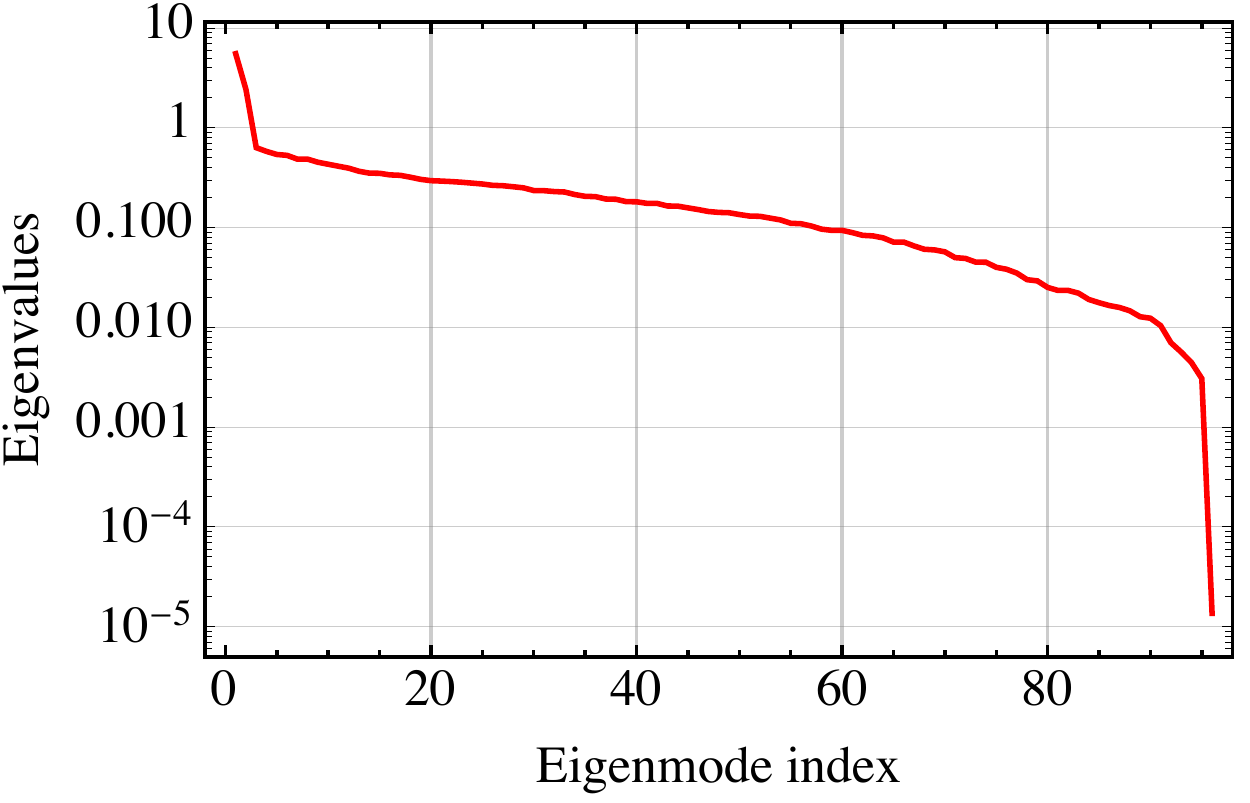}
\includegraphics[scale=0.53,keepaspectratio=true]{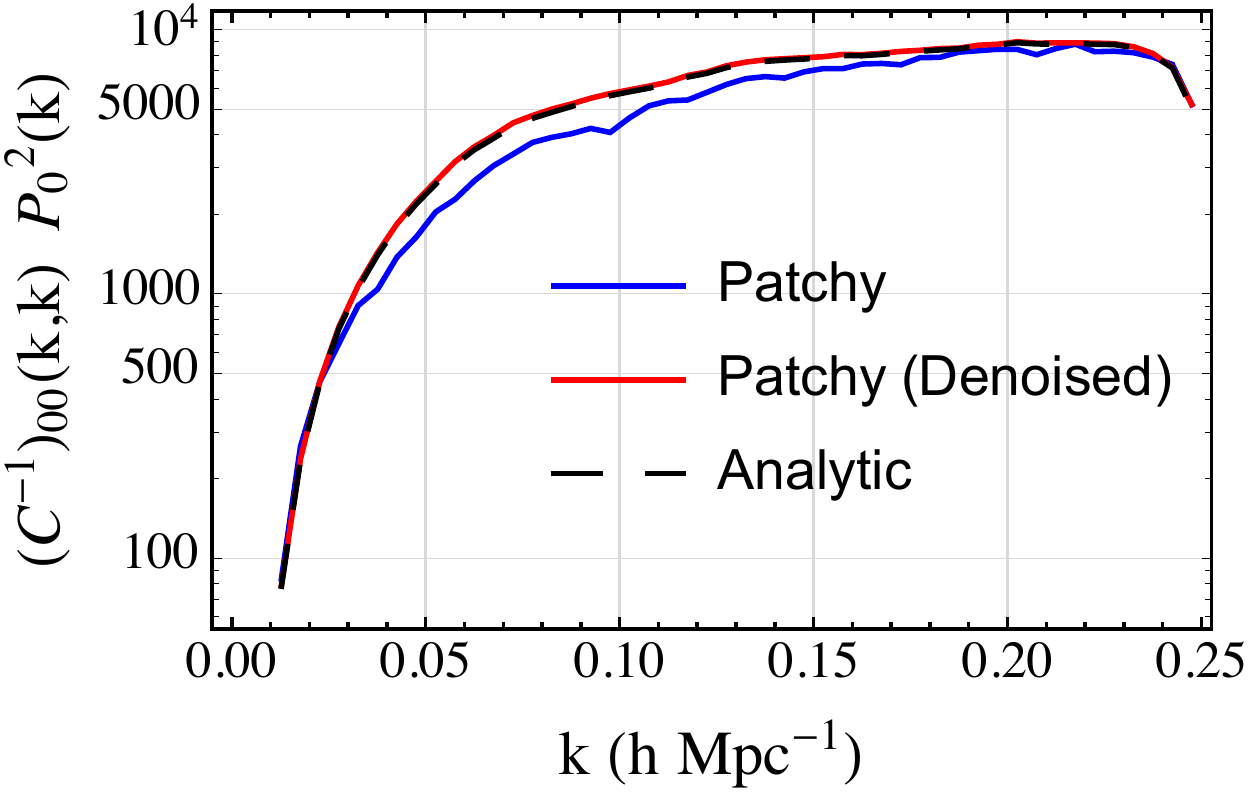}
\includegraphics[scale=0.53,keepaspectratio=true]{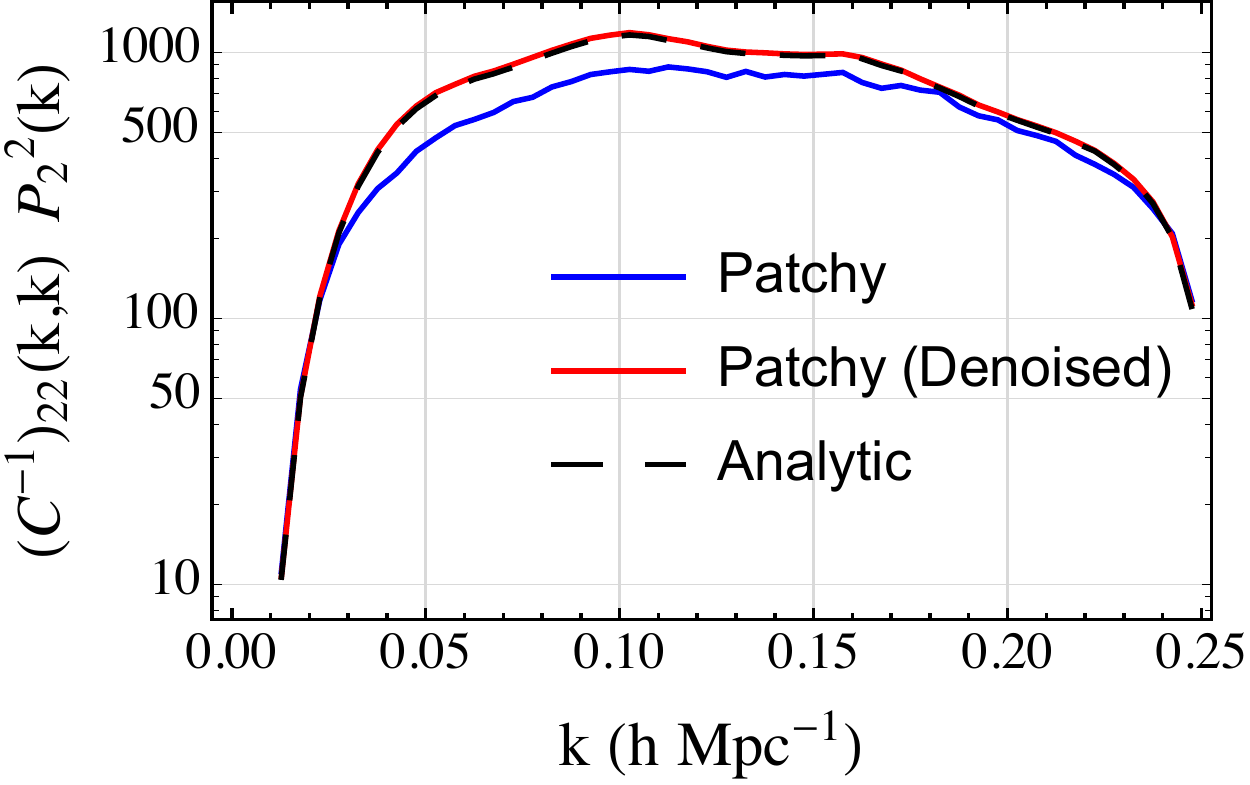}
\caption{Upper left panel: A row of the residuals of the Patchy multipole covariance matrix as defined in Eq.~\ref{eq:CovResiduals}. Upper right panel: Eigenvalues of the residuals matrix. Only including the eigenmodes with eigenvalues larger than $\sim 0.5$ gives the smoothed residuals in black the upper-left panel.
Bottom panels compare diagonals of the precision matrix where the denoised version of the Patchy covariance is constructed using the smoothed residuals by inverting Eq.~\ref{eq:CovResiduals} and it agrees well with the analytic covariance.}
\label{fig:Eigenmodes_SN}
\end{figure*}

In our results in Sec.~\ref{sec:Results}, we found $\sim 0.2\sigma$  shifts in parameter constraints between the analyses based on the analytic and Patchy mock covariance matrices. In this appendix we show that these shifts result from the noise in the sample covariance on the basis of two tests: showing the variation in parameter constraints upon using synthetic sample covariance matrices and undoing the tension caused due to parameter shifts by denoising the covariance matrix from Patchy mocks.

\subsection{Generating synthetic sample covariance matrices 
}
\label{apx:GeneratingSamples}

In this section we
verify that the sampling noise in a covariance matrix constructed from $\Nm=2048$ mocks can indeed cause $\sim 0.2\sigma$ 
shifts in parameter posteriors.
To this end, we generate synthetic sample covariance matrices using the method outlined below.
We want to sample 2048 power spectra from a Gaussian distribution with a given population mean $P^\textup{true}$ and covariance matrix $\C^\textup{true}$.
If the covariance matrix was diagonal, one could simply sample a realization of the power spectra $\hat{P}_i(k)$ from a Gaussian with mean $P^\textup{true}(k)$ and variance $\C^\textup{true}(k,k)$. In the general case of a non-diagonal covariance matrix, we first need to perform the Cholesky decomposition of the population covariance matrix $\C$ into a lower triangular matrix $\mathbb{L}$ as
\beq
\C^\textup{true} = \mathbb{L}\, \mathbb{L}^T
\eeq

We can use this to sample individual power spectrum vectors as 
\beq
\hat{P}_i = \mathbb{L}\, \textbf{z}_i + P^\textup{true}\,,
\eeq
where $\textbf{z}_i$ is a $d$ dimensional column vector with each element being a standard normal vector (mean=0, variance=1) and $d=96$ is the total number of $k$-bins in our analysis. 
We can then create a sample covariance matrix $\hat{\C}$ using the standard estimator in Eq.~(\ref{eq:CovEstimator}). One can verify that $\hat{\C}\rightarrow \C$ in the large $\Nm$ limit using the relation  $\langle \textbf{z}_i\cdot\textbf{z}_j^T\rangle = \delta^K_{ij}\, \mathbb{I}$, where $\mathbb{I}$ is the $d \times d$ identity vector.

Adopting $P^\textup{true}$ to be our best-fit NGC high-$z$ power spectrum and $C^\textup{true}$ to be the analytic covariance matrix for the NGC high-$z$ case,
we generate fifty realizations of the sample covariance matrix. We then calculate the $\chi^2$ using our best-fit power spectrum model and
label the realization with the largest (smallest) best-fit $\chi^2$ as `Sample 1' (`Sample 3') and the realization with the best-fit $\chi^2$ similar to the analytic matrix case as `Sample 2'. We then perform our likelihood analyses of the 
high-$z$ NGC BOSS data on the three samples.
The corner plot is shown in Fig.~\ref{fig:SamplingEffect}, whereas the
1d marginalized limits are given
in Table~\ref{tab:ngcz3_samples}.  For compactness, we show only the cosmological parameters.
The posterior distributions clearly perform a random walk
compatible with the 
expected stochasticity due to 
sampling noise. Both the mean values 
and the size of error bars of the inferred parameters are affected. In particular, the `Sample 3' exhibits a shift in $\omega_\textup{cdm}$ which is similar to the one observed when using Patchy mocks in Fig.~\ref{fig:covar_ngc3}.
It is important to note that the error-bars can also get spuriously smaller as a result of noise in the covariance, as seen in the case of $H_0$ for `Sample 1' where the errorbars shrink by $\sim$20\%. This hints at the dangers of sampling noise as it can lead to spurious tensions between different surveys.

Overall, the results we obtained in this section suggest that the errorbar rescaling factor $M_1$ (=1.02 in our case), which was derived for the ensemble-averaged case \cite{PerRosSan1404, DodSch1309}, can underestimate the effect of sampling noise in the case of a realistic parameter likelihood.
This gives strong motivation to use the analytic covaraince matrix.
We also see that $\sim 0.2\sigma$
shifts in cosmological 
parameters are likely to be observed when using $\sim 2000$ mock catalogs for calculating power spectrum covariances.


\subsection{Denoising the sample covariance matrix from mocks}
\label{apx:Denoising}
In this section we show that we can undo the tension caused due to parameter shifts if we denoise the covariance matrix using the procedure which we now outline.
We use the technique of singular-value decomposition (SVD) which is commonly used in the literature to denoise an estimated covariance matrix and has already been used in the analysis of both power spectra and bispectra \cite{Sco0012, GazSco0508, EisZal0101, TayJoaKit1306,BonJafKno98}. 


As we discussed in Sec.~\ref{sec:GaussCova}, the Gaussian part (G) of the covariance can be well modeled analytically. 
We will therefore use our knowledge of the analytic Gaussian covariance to help denoise the Patchy mock covariance matrix.
We perform the following transformation on the Patchy covariance matrix to get the residuals corresponding to the non-Gaussian part
\begin{equation}
r_{\ell,\ell'} (k,k') = \frac{C^\textup{\! mock}_{\ell,\ell'} (k,k')- C^{\rm G}_{\ell,\ell'} (k,k')}{\sqrt{C^{\rm G}_{\ell,\ell} (k,k) C^{\rm G}_{\ell',\ell'} (k',k')}}\, .
\label{eq:CovResiduals}
\end{equation}
Note that we also have normalized the covariance matrix by diagonals of the Gaussian part in order to remove the $k$-dependence. We show the residuals in the top-left panel of Fig.~\ref{fig:Eigenmodes_SN}.
We also show the normalized full matrix (which includes the Gaussian part) as the dotted blue line.
One can immediately see that the residuals are heavily affected by sampling noise, similar to what we have already seen in Fig.~\ref{fig:Error2D}.
We therefore perform a SVD decomposition of the residuals and the corresponding eigenvalues are shown in top-right of Fig.~\ref{fig:Eigenmodes_SN}.
The eigenmodes with low eigenvalues are expected to be the most affected by sampling noise and upon discarding the modes with eigenvalues smaller than 0.6, we obtain the result shown as the black dashed line in upper-left panel of Fig.~\ref{fig:Eigenmodes_SN}, which is relatively less noisy. 
We then reverse the transform in Eq.~(\ref{eq:CovResiduals}) on the smoothed residuals to obtain a denoised version of the estimated covariance matrix, the diagonals of which are shown in the lower panels of Fig.~\ref{fig:Eigenmodes_SN}.

It is worth commenting on some of the assumptions used in the aforementioned procedure. The first is the choice of cutoff in the eigenvalues of the residual matrix. The noise in the individual elements of the residual matrix is $\sim 1/\sqrt{\Nm}$, but it is not immediately clear how to translate that into the cutoff in the eigenvalues. 
Secondly, we have assumed an accurate theoretical Gaussian part and it is not clear if the denoising procedure would work in case the Gaussian covariance model is inaccurate.

\begin{figure*}
\includegraphics[scale=0.4]{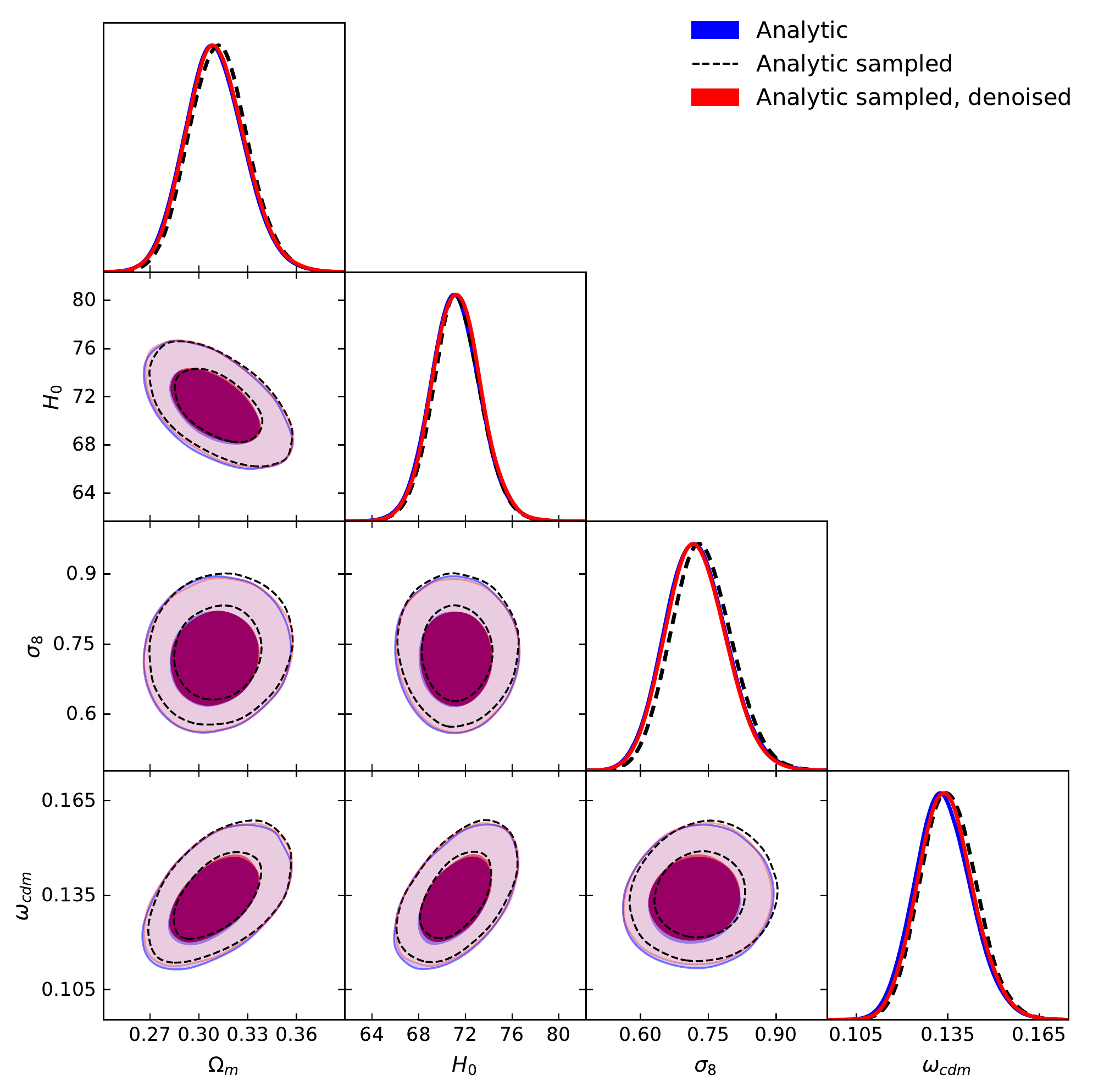}
\includegraphics[scale=0.4]{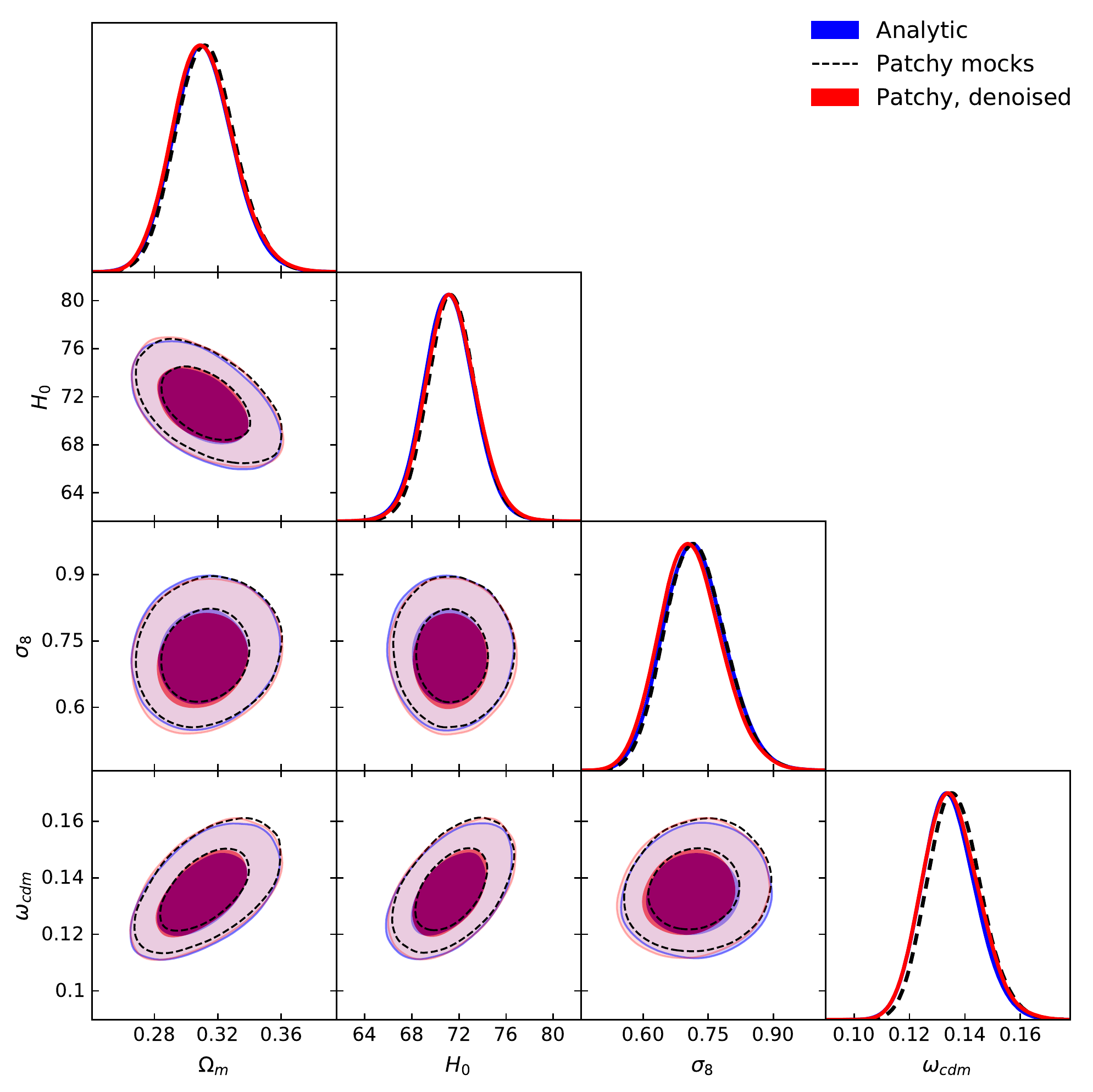}
\caption{Same as Fig.~\ref{fig:covar_ngc3} but also comparing the denoised versions of two cases of sample covariance matrices. Left panel shows a validation test of our denoising procedure, where we first construct a synthetic sample covariance and then denoise it which indeed undoes the small tension caused due to sampling noise. In the right panel, we apply denoising procedure to the covariance from Patchy mocks,
which eliminates 
the small tension with
the analytic covariance. Corresponding tabulated values are in Tables~\ref{tab:ngcz3_whitening} \&~\ref{tab:ngcz3_whitened_patchy}.}
    \label{fig:denoise}
\end{figure*}

We try to validate our procedure and also gauge the value of the eigenvalue cutoff by using our procedure on the synthetic sampled covariance matrices described in Appendix~\ref{apx:GeneratingSamples}, where the corresponding true covariance matrix is known. 
We have chosen one sample realization
of the covariance matrix that yields  $\approx 0.2\sigma$ shift of $\Omega_m$
just like what we have found in the analysis
based on the Patchy covariance.
Moreover, the noise in the chosen realization of the covariance matrix leads to systematically underestimated errorbars on $H_0$
and $\Omega_m$.
Then, we
have 
applied our denoising procedure on the chosen sampled covariance 
and reran the analysis. We find that using the eigenvalue cutoff of 0.6 helps reducing the shifts such that the new results are not in tension with the full analytic calculation as shown in the left panel of Fig.~\ref{fig:denoise} and in Table~\ref{tab:ngcz3_whitening} (we also show the case of the full analytic covariance for ease of comparison).
The denoised covariance matrix although slightly inflates the errorbars
and we believe that our denoising 
procedure can be improved even further to make the results based on the sample covariance agree better with the analytic case and also a theoretically motivated estimate of the cutoff in the eigenvalue space can in principle be derived. However, the current implementation is already enough for our goal which is to reduce the tension between the two results. 

As a next step, we applied the 
denoising procedure to the 
Patchy mock covariance. 
The results of all these analyses, along with the one based on the full analytic covariance computed for the fiducial Patchy cosmology, are shown in the right panel of Fig.~\ref{fig:denoise} and the parameter limits are presented in Table~\ref{tab:ngcz3_whitened_patchy}
(for compactness, we show only the cosmological parameters).
One can see that the denoising procedure leads to an increase in the errorbars, such that the new probability distributions enclose
the one from the analytic covariance. 
Overall, the upshot of this Appendix is that the (small) tension between the results based on the Patchy and the analytic covariance can be removed by denoising the Patchy covariance. Finally, it is worth mentioning that another way to dramatically reduce the noise in a sample covariance matrix is by projecting the power spectra to a lower-dimensional sub-space constructed using SVD, as is shown by Ref.~\cite{PhiIva20inprep}.

\begin{table*}[th!]
  \begin{tabular}{|c||c|c|c|} \hline
    \diagbox{Parameter}{Covariance}   
     &  
     ~~~~
     Analytic
      ~~~~
     &  Sampled analytic 
      & Sampled analytic denoised
      \\ [0.2cm]
      \hline 
$H_0$    
& $71.15_{-2.3}^{+2.2}$ 
& $71.27_{-2.1}^{+2}$
& $71.25_{-2.3}^{+2.1}$
\\ \hline
$A^{1/2}$   
& $0.8276_{-0.094}^{+0.078}$
& $0.8339_{-0.092}^{+0.076}$
& $0.824_{-0.092}^{+0.076}$
\\ 
\hline
     $\omega_\textup{cdm}$  
  & $0.1336_{-0.01}^{+0.0089}$ 
  &$0.1354_{-0.0098}^{+0.0087}$
  &$0.1344_{-0.0099}^{+0.0086}$ 
   \\ \hline \hline 
$\Omega_m$  
& $0.3098_{-0.02}^{+0.018}$
&$0.3121_{-0.019}^{+0.018}$
& $0.3105_{-0.02}^{+0.018}$
\\ \hline
$\sigma_8$   
& $0.723_{-0.072}^{+0.065}$
& $0.736_{-0.070}^{+0.062}$ &$0.723_{-0.072}^{+0.062}$\\ 
\hline
\end{tabular}
\caption{Same as Table~\ref{tab:ngcz3} but for the various choices of
the covariance matrix: analytic, synthetic sample covariance constructed from realizations of the data with the analytic covariance, and its denoised version.
}
\label{tab:ngcz3_whitening}
\end{table*}

\begin{table*}[th!]
  \begin{tabular}{|c||c|c|c|} \hline
    \diagbox{Parameter}{Covariance}   
     &  
     ~~~~
     Analytic (fiducial cosmo.)
      ~~~~
     &  ~~~~ Patchy  ~~~~
      & Patchy, denoised
      \\ [0.2cm]
      \hline 
$H_0$    
& $71.19_{-2.3}^{+2.1}$
& $71.44_{-2.2}^{+2.0}$
& $71.33_{-2.3}^{+2.1}$
\\ \hline
$A^{1/2}$   
& $0.8194_{-0.098}^{+0.08}$
& $0.8135_{-0.093}^{+0.077}$
&  $0.8069_{-0.097}^{+0.078}$
\\ 
\hline
     $\omega_\textup{cdm}$  
  & $0.1345_{-0.011}^{+0.0092}$  
  & $0.1364_{-0.01}^{+0.0091}$ 
  & $0.1351_{-0.011}^{+0.0094}$
   \\ \hline \hline 
$\Omega_m$  
& $0.3111_{-0.021}^{+0.019}$
& $0.3126_{-0.02}^{+0.018}$ 
& $0.3110_{-0.021}^{+0.018}$
\\ \hline
$\sigma_8$   
& $0.719_{-0.076}^{+0.065}$
& $0.721_{-0.074}^{+0.064}$ &$0.710_{-0.076}^{+0.065}$\\ 
\hline
\end{tabular}
\caption{Same as Table~\ref{tab:ngcz3} but for the various choices of
the covariance matrix: analytic covariance computed for the fiducial Patchy cosmology, the sample covariance of Patchy mocks, and its denoised version.
}
\label{tab:ngcz3_whitened_patchy}
\end{table*}

\section{Tests of the cubic bias treatement}
\label{sec:bg3}

In our baseline analysis we have set $b_{\Gamma_3}=0$
and varied only $b_{\mathcal{G}_2}$ in our MCMC chains. In this section we present additional tests to show that this choice does not bias our cosmological constraints.
We focus on the high-z NGC sample. We use the best-fit analytic covariance in all analyses presented in this Section. 

First, we have run the analysis having reset 
$b_{\Gamma_3}$ to the prediction of the local lagrangian approximation (LLA) within the coevolution model \cite{AbiBal1807},
\be 
b_{\Gamma_3}=b^{(LLA)}_{\Gamma_3}=\frac{23}{42}\left(b_1-1\right)\,.
\ee 

\begin{figure*}
\includegraphics[scale=0.45]{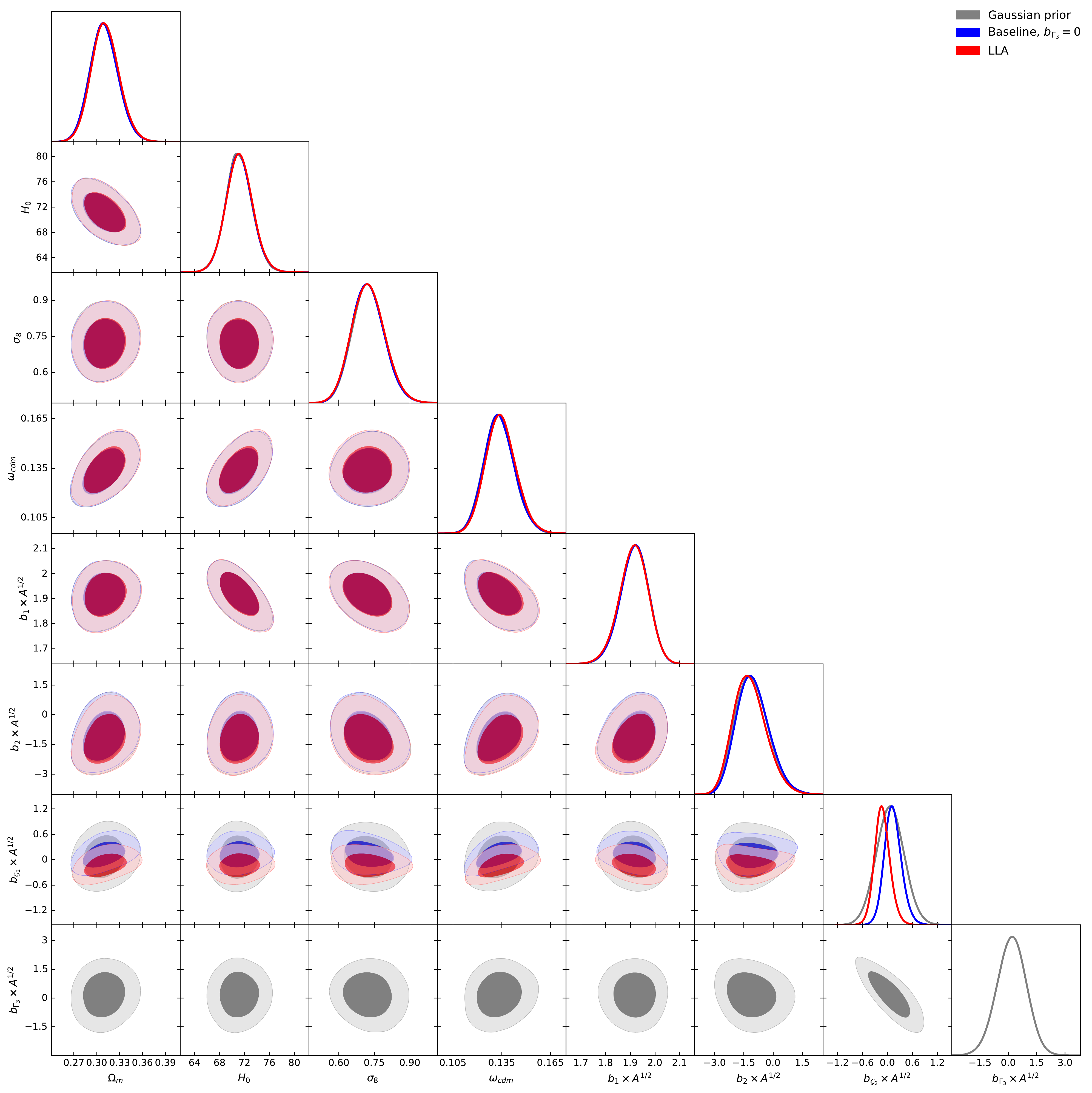}
\caption{Cosmological parameters inferred from the high-$z$ NGC data chunk with three different priors on the cubic bias:
the Gaussian prior centered at the prediction of the 
local Largangian approximation $b_{\Gamma_3}=\frac{23}{42}(b_1-1)$,
as well as infinitely strong priors 
$b_{\Gamma_3}=0$ (our baseline choice),
and $b_{\Gamma_3}=\frac{23}{42}(b_1-1)$. Corresponding tabulated values are in Table~\ref{tab:ngcz3_bG3}.
}
\label{fig:bg3}
\end{figure*}

As a second test, we marginalized over $b_{\Gamma_3}$ using the following Gaussian prior
with the mean equal to the LLA prediction evaluated for the best-fit $b_1$,
\be b_{\Gamma_3}\Big|_{\text{high-z NGC}}=0.71
\ee 
and variance equal to $1$,
\be 
b_{\Gamma_3}\sim \mathcal{N}(b^{(LLA)}_{\Gamma_3},1^2)\,.
\ee 
The results of our analyses are shown in Fig.~\ref{fig:bg3} and Table~\ref{tab:ngcz3_bG3}. 
For illustration purposes we also show the
posteriors for bias parameters $b_1\times A^{1/2}$, $b_{\mathcal{G}_2}\times A^{1/2}$
and $b_2\times A^{1/2}$, although the posterior for the latter
is not appreciably narrower than the prior.
These are the parameters mostly affected 
by the $b_{\Gamma_3}$ prior.
We can see that the prior on $b_{\Gamma_3}$
has noticeable impact only on the $b_{\mathcal{G}_2}$
posterior. The posterior distributions of other bias and cosmological parameters are largely unaffected.

\begin{table*}[t!]
  \begin{tabular}{|c||c|c|c|} \hline
    \diagbox{Parameter}{Prior}   &  ~~~~~~~~Gaussian prior~~~~~~~~ & ~~~~~~~~$b_{\Gamma_3}=\frac{23}{42}(b_1-1)$~~~~~~~~    &  ~~~~~~~~$b_{\Gamma_3}=0$  ~~~~~~~~
      \\ [0.2cm]
      \hline 
$H_0$ (km/s/Mpc)   & $71.1_{-2.3}^{+2.2}$
& $71.17_{-2.3}^{+2.2}$
& $71.15_{-2.3}^{+2.2}$ \\ \hline
$A^{1/2}$   
& $0.8316_{-0.096}^{+0.079}$ 
& $0.8261_{-0.096}^{+0.081}$
& $0.8276_{-0.094}^{+0.078}$
\\ 
\hline
     $\omega_\textup{cdm}$  & $0.1335_{-0.01}^{+0.0089}$
  & $0.1345_{-0.01}^{+0.0089}$
  & $0.1336_{-0.01}^{+0.0089}$ 
  \\ \hline
   $b_1 A^{1/2}$ & $1.916_{-0.055}^{+0.062}$
   & $1.913_{-0.056}^{+0.064}$
   & $1.916_{-0.055}^{+0.062}$
   \\ \hline
   $b_{\mathcal{G}_2} A^{1/2}$ 
   &  $0.06862_{-0.35}^{+0.34}$
   &  $-0.1217_{-0.2}^{+0.18}$
   &  $0.1345_{-0.23}^{+0.19}$
 \\  
 \hline
    $b_{\Gamma_3}$ 
   & $0.1716_{-0.78}^{+0.81}$ 
   &  $-$
   & $-$
 \\  
 \hline
   $b_2 A^{1/2}$ 
   & $-1.082_{-0.94}^{+0.77}$ 
   & $-1.178_{-0.96}^{+0.74}$
   & $-1.03_{-0.94}^{+0.75}$
   \\ \hline \hline
$\Omega_m$   & $0.3099_{-0.02}^{+0.018}$
& $0.3112_{-0.02}^{+0.018}$
& $0.3098_{-0.02}^{+0.018}$  \\ \hline
$\sigma_8$   & $0.726_{-0.072}^{+0.064}$
&$0.725_{-0.073}^{+0.065}$
& $0.724_{-0.072}^{+0.063}$ \\ 
\hline
\end{tabular}
\caption{Mean values and 68\% CL minimum credible
intervals for the parameters of the base $\Lambda$CDM model fitted to
the high-$z$ NGC chunk of the 
BOSS data. 
The upper part of the table displays the parameters that were sampled directly.
The lower group lists derived parameters. 
}
\label{tab:ngcz3_bG3}
\end{table*}
%

\bibliographystyle{apsrev4-1}
\bibliography{Cova}

\begin{thebibliography}{113}%
\makeatletter
\providecommand \@ifxundefined [1]{%
 \@ifx{#1\undefined}
}%
\providecommand \@ifnum [1]{%
 \ifnum #1\expandafter \@firstoftwo
 \else \expandafter \@secondoftwo
 \fi
}%
\providecommand \@ifx [1]{%
 \ifx #1\expandafter \@firstoftwo
 \else \expandafter \@secondoftwo
 \fi
}%
\providecommand \natexlab [1]{#1}%
\providecommand \enquote  [1]{``#1''}%
\providecommand \bibnamefont  [1]{#1}%
\providecommand \bibfnamefont [1]{#1}%
\providecommand \citenamefont [1]{#1}%
\providecommand \href@noop [0]{\@secondoftwo}%
\providecommand \href [0]{\begingroup \@sanitize@url \@href}%
\providecommand \@href[1]{\@@startlink{#1}\@@href}%
\providecommand \@@href[1]{\endgroup#1\@@endlink}%
\providecommand \@sanitize@url [0]{\catcode `\\12\catcode `\$12\catcode
  `\&12\catcode `\#12\catcode `\^12\catcode `\_12\catcode `\%12\relax}%
\providecommand \@@startlink[1]{}%
\providecommand \@@endlink[0]{}%
\providecommand \url  [0]{\begingroup\@sanitize@url \@url }%
\providecommand \@url [1]{\endgroup\@href {#1}{\urlprefix }}%
\providecommand \urlprefix  [0]{URL }%
\providecommand \Eprint [0]{\href }%
\providecommand \doibase [0]{http://dx.doi.org/}%
\providecommand \selectlanguage [0]{\@gobble}%
\providecommand \bibinfo  [0]{\@secondoftwo}%
\providecommand \bibfield  [0]{\@secondoftwo}%
\providecommand \translation [1]{[#1]}%
\providecommand \BibitemOpen [0]{}%
\providecommand \bibitemStop [0]{}%
\providecommand \bibitemNoStop [0]{.\EOS\space}%
\providecommand \EOS [0]{\spacefactor3000\relax}%
\providecommand \BibitemShut  [1]{\csname bibitem#1\endcsname}%
\let\auto@bib@innerbib\@empty
\bibitem [{\citenamefont {{Wadekar}}\ and\ \citenamefont
  {{Scoccimarro}}(2019)}]{WadSco19}%
  \BibitemOpen
  \bibfield  {author} {\bibinfo {author} {\bibfnamefont {D.}~\bibnamefont
  {{Wadekar}}}\ and\ \bibinfo {author} {\bibfnamefont {R.}~\bibnamefont
  {{Scoccimarro}}},\ }\href@noop {} {\bibfield  {journal} {\bibinfo  {journal}
  {arXiv e-prints}\ ,\ \bibinfo {eid} {arXiv:1910.02914}} (\bibinfo {year}
  {2019})},\ \bibinfo {note} {{(WS19)}},\ \Eprint
  {http://arxiv.org/abs/1910.02914} {arXiv:1910.02914 [astro-ph.CO]}
  \BibitemShut {NoStop}%
\bibitem [{\citenamefont {{Bond}}\ and\ \citenamefont
  {{Myers}}(1996)}]{BonMye9603}%
  \BibitemOpen
  \bibfield  {author} {\bibinfo {author} {\bibfnamefont {J.~R.}\ \bibnamefont
  {{Bond}}}\ and\ \bibinfo {author} {\bibfnamefont {S.~T.}\ \bibnamefont
  {{Myers}}},\ }\href {\doibase 10.1086/192267} {\bibfield  {journal} {\bibinfo
   {journal} {\apjs}\ }\textbf {\bibinfo {volume} {103}},\ \bibinfo {pages} {1}
  (\bibinfo {year} {1996})}\BibitemShut {NoStop}%
\bibitem [{\citenamefont {{Scoccimarro}}\ and\ \citenamefont
  {{Sheth}}(2002)}]{ScoShe02}%
  \BibitemOpen
  \bibfield  {author} {\bibinfo {author} {\bibfnamefont {R.}~\bibnamefont
  {{Scoccimarro}}}\ and\ \bibinfo {author} {\bibfnamefont {R.~K.}\ \bibnamefont
  {{Sheth}}},\ }\href {\doibase 10.1046/j.1365-8711.2002.04999.x} {\bibfield
  {journal} {\bibinfo  {journal} {\mnras}\ }\textbf {\bibinfo {volume} {329}},\
  \bibinfo {pages} {629} (\bibinfo {year} {2002})},\ \Eprint
  {http://arxiv.org/abs/astro-ph/0106120} {arXiv:astro-ph/0106120 [astro-ph]}
  \BibitemShut {NoStop}%
\bibitem [{\citenamefont {Manera}\ \emph {et~al.}(2013)\citenamefont {Manera}
  \emph {et~al.}}]{ManScoPer1301}%
  \BibitemOpen
  \bibfield  {author} {\bibinfo {author} {\bibfnamefont {M.}~\bibnamefont
  {Manera}} \emph {et~al.},\ }\href {\doibase 10.1093/mnras/sts084} {\bibfield
  {journal} {\bibinfo  {journal} {\mnras}\ }\textbf {\bibinfo {volume} {428}},\
  \bibinfo {pages} {1036} (\bibinfo {year} {2013})},\ \Eprint
  {http://arxiv.org/abs/1203.6609} {arXiv:1203.6609 [astro-ph.CO]} \BibitemShut
  {NoStop}%
\bibitem [{\citenamefont {{Tassev}}\ \emph {et~al.}(2013)\citenamefont
  {{Tassev}}, \citenamefont {{Zaldarriaga}},\ and\ \citenamefont
  {{Eisenstein}}}]{TasZalEis1306}%
  \BibitemOpen
  \bibfield  {author} {\bibinfo {author} {\bibfnamefont {S.}~\bibnamefont
  {{Tassev}}}, \bibinfo {author} {\bibfnamefont {M.}~\bibnamefont
  {{Zaldarriaga}}}, \ and\ \bibinfo {author} {\bibfnamefont {D.~J.}\
  \bibnamefont {{Eisenstein}}},\ }\href {\doibase
  10.1088/1475-7516/2013/06/036} {\bibfield  {journal} {\bibinfo  {journal}
  {\jcap}\ }\textbf {\bibinfo {volume} {2013}},\ \bibinfo {eid} {036} (\bibinfo
  {year} {2013})},\ \Eprint {http://arxiv.org/abs/1301.0322} {arXiv:1301.0322
  [astro-ph.CO]} \BibitemShut {NoStop}%
\bibitem [{\citenamefont {{White}}\ \emph {et~al.}(2014)\citenamefont
  {{White}}, \citenamefont {{Tinker}},\ and\ \citenamefont
  {{McBride}}}]{WhiTinMcB1401}%
  \BibitemOpen
  \bibfield  {author} {\bibinfo {author} {\bibfnamefont {M.}~\bibnamefont
  {{White}}}, \bibinfo {author} {\bibfnamefont {J.~L.}\ \bibnamefont
  {{Tinker}}}, \ and\ \bibinfo {author} {\bibfnamefont {C.~K.}\ \bibnamefont
  {{McBride}}},\ }\href {\doibase 10.1093/mnras/stt2071} {\bibfield  {journal}
  {\bibinfo  {journal} {\mnras}\ }\textbf {\bibinfo {volume} {437}},\ \bibinfo
  {pages} {2594} (\bibinfo {year} {2014})},\ \Eprint
  {http://arxiv.org/abs/1309.5532} {arXiv:1309.5532 [astro-ph.CO]} \BibitemShut
  {NoStop}%
\bibitem [{\citenamefont {{Kitaura}}\ \emph {et~al.}(2014)\citenamefont
  {{Kitaura}}, \citenamefont {{Yepes}},\ and\ \citenamefont
  {{Prada}}}]{KitYepPra14}%
  \BibitemOpen
  \bibfield  {author} {\bibinfo {author} {\bibfnamefont {F.~S.}\ \bibnamefont
  {{Kitaura}}}, \bibinfo {author} {\bibfnamefont {G.}~\bibnamefont {{Yepes}}},
  \ and\ \bibinfo {author} {\bibfnamefont {F.}~\bibnamefont {{Prada}}},\ }\href
  {\doibase 10.1093/mnrasl/slt172} {\bibfield  {journal} {\bibinfo  {journal}
  {\mnras}\ }\textbf {\bibinfo {volume} {439}},\ \bibinfo {pages} {L21}
  (\bibinfo {year} {2014})},\ \Eprint {http://arxiv.org/abs/1307.3285}
  {arXiv:1307.3285 [astro-ph.CO]} \BibitemShut {NoStop}%
\bibitem [{\citenamefont {{Chuang}}\ \emph {et~al.}(2015)\citenamefont
  {{Chuang}}, \citenamefont {{Kitaura}}, \citenamefont {{Prada}}, \citenamefont
  {{Zhao}},\ and\ \citenamefont {{Yepes}}}]{ChuKitPra15}%
  \BibitemOpen
  \bibfield  {author} {\bibinfo {author} {\bibfnamefont {C.-H.}\ \bibnamefont
  {{Chuang}}}, \bibinfo {author} {\bibfnamefont {F.-S.}\ \bibnamefont
  {{Kitaura}}}, \bibinfo {author} {\bibfnamefont {F.}~\bibnamefont {{Prada}}},
  \bibinfo {author} {\bibfnamefont {C.}~\bibnamefont {{Zhao}}}, \ and\ \bibinfo
  {author} {\bibfnamefont {G.}~\bibnamefont {{Yepes}}},\ }\href {\doibase
  10.1093/mnras/stu2301} {\bibfield  {journal} {\bibinfo  {journal} {\mnras}\
  }\textbf {\bibinfo {volume} {446}},\ \bibinfo {pages} {2621} (\bibinfo {year}
  {2015})},\ \Eprint {http://arxiv.org/abs/1409.1124} {arXiv:1409.1124
  [astro-ph.CO]} \BibitemShut {NoStop}%
\bibitem [{\citenamefont {{Izard}}\ \emph {et~al.}(2016)\citenamefont
  {{Izard}}, \citenamefont {{Crocce}},\ and\ \citenamefont
  {{Fosalba}}}]{IzaCroFos1607}%
  \BibitemOpen
  \bibfield  {author} {\bibinfo {author} {\bibfnamefont {A.}~\bibnamefont
  {{Izard}}}, \bibinfo {author} {\bibfnamefont {M.}~\bibnamefont {{Crocce}}}, \
  and\ \bibinfo {author} {\bibfnamefont {P.}~\bibnamefont {{Fosalba}}},\ }\href
  {\doibase 10.1093/mnras/stw797} {\bibfield  {journal} {\bibinfo  {journal}
  {\mnras}\ }\textbf {\bibinfo {volume} {459}},\ \bibinfo {pages} {2327}
  (\bibinfo {year} {2016})},\ \Eprint {http://arxiv.org/abs/1509.04685}
  {arXiv:1509.04685 [astro-ph.CO]} \BibitemShut {NoStop}%
\bibitem [{\citenamefont {{Lippich}}\ \emph {et~al.}(2019)\citenamefont
  {{Lippich}} \emph {et~al.}}]{LipSanCol1810}%
  \BibitemOpen
  \bibfield  {author} {\bibinfo {author} {\bibfnamefont {M.}~\bibnamefont
  {{Lippich}}} \emph {et~al.},\ }\href {\doibase 10.1093/mnras/sty2757}
  {\bibfield  {journal} {\bibinfo  {journal} {\mnras}\ }\textbf {\bibinfo
  {volume} {482}},\ \bibinfo {pages} {1786} (\bibinfo {year} {2019})},\ \Eprint
  {http://arxiv.org/abs/1806.09477} {arXiv:1806.09477 [astro-ph.CO]}
  \BibitemShut {NoStop}%
\bibitem [{\citenamefont {Blot}\ \emph {et~al.}(2019)\citenamefont {Blot} \emph
  {et~al.}}]{BloCroSef1905}%
  \BibitemOpen
  \bibfield  {author} {\bibinfo {author} {\bibfnamefont {L.}~\bibnamefont
  {Blot}} \emph {et~al.},\ }\href {\doibase 10.1093/mnras/stz507} {\bibfield
  {journal} {\bibinfo  {journal} {Mon. Not. Roy. Astron. Soc.}\ }\textbf
  {\bibinfo {volume} {485}},\ \bibinfo {pages} {2806} (\bibinfo {year}
  {2019})},\ \Eprint {http://arxiv.org/abs/1806.09497} {arXiv:1806.09497
  [astro-ph.CO]} \BibitemShut {NoStop}%
\bibitem [{\citenamefont {Colavincenzo}\ \emph {et~al.}(2019)\citenamefont
  {Colavincenzo} \emph {et~al.}}]{ColSefMon1811}%
  \BibitemOpen
  \bibfield  {author} {\bibinfo {author} {\bibfnamefont {M.}~\bibnamefont
  {Colavincenzo}} \emph {et~al.},\ }\href {\doibase 10.1093/mnras/sty2964}
  {\bibfield  {journal} {\bibinfo  {journal} {\mnras}\ }\textbf {\bibinfo
  {volume} {482}},\ \bibinfo {pages} {4883} (\bibinfo {year} {2019})},\ \Eprint
  {http://arxiv.org/abs/1806.09499} {arXiv:1806.09499 [astro-ph.CO]}
  \BibitemShut {NoStop}%
\bibitem [{\citenamefont {Kitaura}\ \emph {et~al.}(2016)\citenamefont {Kitaura}
  \emph {et~al.}}]{KitRodChu1603}%
  \BibitemOpen
  \bibfield  {author} {\bibinfo {author} {\bibfnamefont {F.-S.}\ \bibnamefont
  {Kitaura}} \emph {et~al.},\ }\href {\doibase 10.1093/mnras/stv2826}
  {\bibfield  {journal} {\bibinfo  {journal} {Mon. Not. Roy. Astron. Soc.}\
  }\textbf {\bibinfo {volume} {456}},\ \bibinfo {pages} {4156} (\bibinfo {year}
  {2016})},\ \Eprint {http://arxiv.org/abs/1509.06400} {arXiv:1509.06400
  [astro-ph.CO]} \BibitemShut {NoStop}%
\bibitem [{\citenamefont {Zhao}\ \emph {et~al.}(2020)\citenamefont {Zhao} \emph
  {et~al.}}]{Zha20}%
  \BibitemOpen
  \bibfield  {author} {\bibinfo {author} {\bibfnamefont {C.}~\bibnamefont
  {Zhao}} \emph {et~al.},\ }\href@noop {} {\  (\bibinfo {year} {2020})},\
  \Eprint {http://arxiv.org/abs/2007.08997} {arXiv:2007.08997 [astro-ph.CO]}
  \BibitemShut {NoStop}%
\bibitem [{\citenamefont {{Howlett}}\ and\ \citenamefont
  {{Percival}}(2017)}]{2017MNRAS.472.4935H}%
  \BibitemOpen
  \bibfield  {author} {\bibinfo {author} {\bibfnamefont {C.}~\bibnamefont
  {{Howlett}}}\ and\ \bibinfo {author} {\bibfnamefont {W.~J.}\ \bibnamefont
  {{Percival}}},\ }\href {\doibase 10.1093/mnras/stx2342} {\bibfield  {journal}
  {\bibinfo  {journal} {\mnras}\ }\textbf {\bibinfo {volume} {472}},\ \bibinfo
  {pages} {4935} (\bibinfo {year} {2017})},\ \Eprint
  {http://arxiv.org/abs/1709.03057} {arXiv:1709.03057 [astro-ph.CO]}
  \BibitemShut {NoStop}%
\bibitem [{\citenamefont {{Klypin}}\ and\ \citenamefont
  {{Prada}}(2018)}]{2018MNRAS.478.4602K}%
  \BibitemOpen
  \bibfield  {author} {\bibinfo {author} {\bibfnamefont {A.}~\bibnamefont
  {{Klypin}}}\ and\ \bibinfo {author} {\bibfnamefont {F.}~\bibnamefont
  {{Prada}}},\ }\href {\doibase 10.1093/mnras/sty1340} {\bibfield  {journal}
  {\bibinfo  {journal} {\mnras}\ }\textbf {\bibinfo {volume} {478}},\ \bibinfo
  {pages} {4602} (\bibinfo {year} {2018})},\ \Eprint
  {http://arxiv.org/abs/1701.05690} {arXiv:1701.05690 [astro-ph.CO]}
  \BibitemShut {NoStop}%
\bibitem [{\citenamefont {{O'Connell}}\ \emph {et~al.}(2016)\citenamefont
  {{O'Connell}}, \citenamefont {{Eisenstein}}, \citenamefont {{Vargas}},
  \citenamefont {{Ho}},\ and\ \citenamefont {{Padmanabhan}}}]{OcoEisVar1611}%
  \BibitemOpen
  \bibfield  {author} {\bibinfo {author} {\bibfnamefont {R.}~\bibnamefont
  {{O'Connell}}}, \bibinfo {author} {\bibfnamefont {D.}~\bibnamefont
  {{Eisenstein}}}, \bibinfo {author} {\bibfnamefont {M.}~\bibnamefont
  {{Vargas}}}, \bibinfo {author} {\bibfnamefont {S.}~\bibnamefont {{Ho}}}, \
  and\ \bibinfo {author} {\bibfnamefont {N.}~\bibnamefont {{Padmanabhan}}},\
  }\href {\doibase 10.1093/mnras/stw1821} {\bibfield  {journal} {\bibinfo
  {journal} {\mnras}\ }\textbf {\bibinfo {volume} {462}},\ \bibinfo {pages}
  {2681} (\bibinfo {year} {2016})}\BibitemShut {NoStop}%
\bibitem [{\citenamefont {{O'Connell}}\ and\ \citenamefont
  {{Eisenstein}}(2018)}]{OcoEis18}%
  \BibitemOpen
  \bibfield  {author} {\bibinfo {author} {\bibfnamefont {R.}~\bibnamefont
  {{O'Connell}}}\ and\ \bibinfo {author} {\bibfnamefont {D.~J.}\ \bibnamefont
  {{Eisenstein}}},\ }\href@noop {} {\bibfield  {journal} {\bibinfo  {journal}
  {arXiv e-prints}\ ,\ \bibinfo {eid} {arXiv:1808.05978}} (\bibinfo {year}
  {2018})},\ \Eprint {http://arxiv.org/abs/1808.05978} {arXiv:1808.05978
  [astro-ph.CO]} \BibitemShut {NoStop}%
\bibitem [{\citenamefont {{Philcox}}\ \emph
  {et~al.}(2020{\natexlab{a}})\citenamefont {{Philcox}}, \citenamefont
  {{Eisenstein}}, \citenamefont {{O'Connell}},\ and\ \citenamefont
  {{Wiegand}}}]{2020MNRAS.491.3290P}%
  \BibitemOpen
  \bibfield  {author} {\bibinfo {author} {\bibfnamefont {O.~H.~E.}\
  \bibnamefont {{Philcox}}}, \bibinfo {author} {\bibfnamefont {D.~J.}\
  \bibnamefont {{Eisenstein}}}, \bibinfo {author} {\bibfnamefont
  {R.}~\bibnamefont {{O'Connell}}}, \ and\ \bibinfo {author} {\bibfnamefont
  {A.}~\bibnamefont {{Wiegand}}},\ }\href {\doibase 10.1093/mnras/stz3218}
  {\bibfield  {journal} {\bibinfo  {journal} {\mnras}\ }\textbf {\bibinfo
  {volume} {491}},\ \bibinfo {pages} {3290} (\bibinfo {year}
  {2020}{\natexlab{a}})},\ \Eprint {http://arxiv.org/abs/1904.11070}
  {arXiv:1904.11070 [astro-ph.CO]} \BibitemShut {NoStop}%
\bibitem [{\citenamefont {{Tegmark}}(1997)}]{Teg9711}%
  \BibitemOpen
  \bibfield  {author} {\bibinfo {author} {\bibfnamefont {M.}~\bibnamefont
  {{Tegmark}}},\ }\href {\doibase 10.1103/PhysRevLett.79.3806} {\bibfield
  {journal} {\bibinfo  {journal} {\prl}\ }\textbf {\bibinfo {volume} {79}},\
  \bibinfo {pages} {3806} (\bibinfo {year} {1997})},\ \Eprint
  {http://arxiv.org/abs/astro-ph/9706198} {arXiv:astro-ph/9706198 [astro-ph]}
  \BibitemShut {NoStop}%
\bibitem [{\citenamefont {Tegmark}(1997)}]{Tegmark:1996qt}%
  \BibitemOpen
  \bibfield  {author} {\bibinfo {author} {\bibfnamefont {M.}~\bibnamefont
  {Tegmark}},\ }\href {\doibase 10.1103/PhysRevD.55.5895} {\bibfield  {journal}
  {\bibinfo  {journal} {Phys. Rev. D}\ }\textbf {\bibinfo {volume} {55}},\
  \bibinfo {pages} {5895} (\bibinfo {year} {1997})},\ \Eprint
  {http://arxiv.org/abs/astro-ph/9611174} {arXiv:astro-ph/9611174} \BibitemShut
  {NoStop}%
\bibitem [{\citenamefont {{Eifler}}\ \emph {et~al.}(2009)\citenamefont
  {{Eifler}}, \citenamefont {{Schneider}},\ and\ \citenamefont
  {{Hartlap}}}]{EifSchHar09}%
  \BibitemOpen
  \bibfield  {author} {\bibinfo {author} {\bibfnamefont {T.}~\bibnamefont
  {{Eifler}}}, \bibinfo {author} {\bibfnamefont {P.}~\bibnamefont
  {{Schneider}}}, \ and\ \bibinfo {author} {\bibfnamefont {J.}~\bibnamefont
  {{Hartlap}}},\ }\href {\doibase 10.1051/0004-6361/200811276} {\bibfield
  {journal} {\bibinfo  {journal} {\aap}\ }\textbf {\bibinfo {volume} {502}},\
  \bibinfo {pages} {721} (\bibinfo {year} {2009})},\ \Eprint
  {http://arxiv.org/abs/0810.4254} {arXiv:0810.4254 [astro-ph]} \BibitemShut
  {NoStop}%
\bibitem [{\citenamefont {{White}}\ and\ \citenamefont
  {{Padmanabhan}}(2015)}]{WhiPad15}%
  \BibitemOpen
  \bibfield  {author} {\bibinfo {author} {\bibfnamefont {M.}~\bibnamefont
  {{White}}}\ and\ \bibinfo {author} {\bibfnamefont {N.}~\bibnamefont
  {{Padmanabhan}}},\ }\href {\doibase 10.1088/1475-7516/2015/12/058} {\bibfield
   {journal} {\bibinfo  {journal} {\jcap}\ }\textbf {\bibinfo {volume}
  {2015}},\ \bibinfo {eid} {058} (\bibinfo {year} {2015})},\ \Eprint
  {http://arxiv.org/abs/1508.00566} {arXiv:1508.00566 [astro-ph.CO]}
  \BibitemShut {NoStop}%
\bibitem [{\citenamefont {{Morrison}}\ and\ \citenamefont
  {{Schneider}}(2013)}]{MorSch13}%
  \BibitemOpen
  \bibfield  {author} {\bibinfo {author} {\bibfnamefont {C.~B.}\ \bibnamefont
  {{Morrison}}}\ and\ \bibinfo {author} {\bibfnamefont {M.~D.}\ \bibnamefont
  {{Schneider}}},\ }\href {\doibase 10.1088/1475-7516/2013/11/009} {\bibfield
  {journal} {\bibinfo  {journal} {\jcap}\ }\textbf {\bibinfo {volume} {2013}},\
  \bibinfo {eid} {009} (\bibinfo {year} {2013})},\ \Eprint
  {http://arxiv.org/abs/1304.7789} {arXiv:1304.7789 [astro-ph.CO]} \BibitemShut
  {NoStop}%
\bibitem [{\citenamefont {{Hartlap}}\ \emph {et~al.}(2007)\citenamefont
  {{Hartlap}}, \citenamefont {{Simon}},\ and\ \citenamefont
  {{Schneider}}}]{HarSimSch0703}%
  \BibitemOpen
  \bibfield  {author} {\bibinfo {author} {\bibfnamefont {J.}~\bibnamefont
  {{Hartlap}}}, \bibinfo {author} {\bibfnamefont {P.}~\bibnamefont {{Simon}}},
  \ and\ \bibinfo {author} {\bibfnamefont {P.}~\bibnamefont {{Schneider}}},\
  }\href {\doibase 10.1051/0004-6361:20066170} {\bibfield  {journal} {\bibinfo
  {journal} {\aap}\ }\textbf {\bibinfo {volume} {464}},\ \bibinfo {pages} {399}
  (\bibinfo {year} {2007})},\ \Eprint {http://arxiv.org/abs/astro-ph/0608064}
  {arXiv:astro-ph/0608064 [astro-ph]} \BibitemShut {NoStop}%
\bibitem [{\citenamefont {{Taylor}}\ \emph {et~al.}(2013)\citenamefont
  {{Taylor}}, \citenamefont {{Joachimi}},\ and\ \citenamefont
  {{Kitching}}}]{TayJoaKit1306}%
  \BibitemOpen
  \bibfield  {author} {\bibinfo {author} {\bibfnamefont {A.}~\bibnamefont
  {{Taylor}}}, \bibinfo {author} {\bibfnamefont {B.}~\bibnamefont
  {{Joachimi}}}, \ and\ \bibinfo {author} {\bibfnamefont {T.}~\bibnamefont
  {{Kitching}}},\ }\href {\doibase 10.1093/mnras/stt270} {\bibfield  {journal}
  {\bibinfo  {journal} {\mnras}\ }\textbf {\bibinfo {volume} {432}},\ \bibinfo
  {pages} {1928} (\bibinfo {year} {2013})},\ \Eprint
  {http://arxiv.org/abs/1212.4359} {arXiv:1212.4359 [astro-ph.CO]} \BibitemShut
  {NoStop}%
\bibitem [{\citenamefont {{Dodelson}}\ and\ \citenamefont
  {{Schneider}}(2013)}]{DodSch1309}%
  \BibitemOpen
  \bibfield  {author} {\bibinfo {author} {\bibfnamefont {S.}~\bibnamefont
  {{Dodelson}}}\ and\ \bibinfo {author} {\bibfnamefont {M.~D.}\ \bibnamefont
  {{Schneider}}},\ }\href {\doibase 10.1103/PhysRevD.88.063537} {\bibfield
  {journal} {\bibinfo  {journal} {\prd}\ }\textbf {\bibinfo {volume} {88}},\
  \bibinfo {eid} {063537} (\bibinfo {year} {2013})},\ \Eprint
  {http://arxiv.org/abs/1304.2593} {arXiv:1304.2593 [astro-ph.CO]} \BibitemShut
  {NoStop}%
\bibitem [{\citenamefont {Percival}\ \emph {et~al.}(2014)\citenamefont
  {Percival} \emph {et~al.}}]{PerRosSan1404}%
  \BibitemOpen
  \bibfield  {author} {\bibinfo {author} {\bibfnamefont {W.~J.}\ \bibnamefont
  {Percival}} \emph {et~al.},\ }\href {\doibase 10.1093/mnras/stu112}
  {\bibfield  {journal} {\bibinfo  {journal} {Mon. Not. Roy. Astron. Soc.}\
  }\textbf {\bibinfo {volume} {439}},\ \bibinfo {pages} {2531} (\bibinfo {year}
  {2014})},\ \Eprint {http://arxiv.org/abs/1312.4841} {arXiv:1312.4841
  [astro-ph.CO]} \BibitemShut {NoStop}%
\bibitem [{\citenamefont {{Taylor}}\ and\ \citenamefont
  {{Joachimi}}(2014)}]{TayJoa14}%
  \BibitemOpen
  \bibfield  {author} {\bibinfo {author} {\bibfnamefont {A.}~\bibnamefont
  {{Taylor}}}\ and\ \bibinfo {author} {\bibfnamefont {B.}~\bibnamefont
  {{Joachimi}}},\ }\href {\doibase 10.1093/mnras/stu996} {\bibfield  {journal}
  {\bibinfo  {journal} {\mnras}\ }\textbf {\bibinfo {volume} {442}},\ \bibinfo
  {pages} {2728} (\bibinfo {year} {2014})},\ \Eprint
  {http://arxiv.org/abs/1402.6983} {arXiv:1402.6983 [astro-ph.CO]} \BibitemShut
  {NoStop}%
\bibitem [{\citenamefont {{Sellentin}}\ and\ \citenamefont
  {{Heavens}}(2016)}]{SelHea1602}%
  \BibitemOpen
  \bibfield  {author} {\bibinfo {author} {\bibfnamefont {E.}~\bibnamefont
  {{Sellentin}}}\ and\ \bibinfo {author} {\bibfnamefont {A.~F.}\ \bibnamefont
  {{Heavens}}},\ }\href {\doibase 10.1093/mnrasl/slv190} {\bibfield  {journal}
  {\bibinfo  {journal} {\mnras}\ }\textbf {\bibinfo {volume} {456}},\ \bibinfo
  {pages} {L132} (\bibinfo {year} {2016})},\ \Eprint
  {http://arxiv.org/abs/1511.05969} {arXiv:1511.05969 [astro-ph.CO]}
  \BibitemShut {NoStop}%
\bibitem [{\citenamefont {{Paz}}\ and\ \citenamefont
  {{S{\'a}nchez}}(2015)}]{2015MNRAS.454.4326P}%
  \BibitemOpen
  \bibfield  {author} {\bibinfo {author} {\bibfnamefont {D.~J.}\ \bibnamefont
  {{Paz}}}\ and\ \bibinfo {author} {\bibfnamefont {A.~G.}\ \bibnamefont
  {{S{\'a}nchez}}},\ }\href {\doibase 10.1093/mnras/stv2259} {\bibfield
  {journal} {\bibinfo  {journal} {\mnras}\ }\textbf {\bibinfo {volume} {454}},\
  \bibinfo {pages} {4326} (\bibinfo {year} {2015})},\ \Eprint
  {http://arxiv.org/abs/1508.03162} {arXiv:1508.03162 [astro-ph.CO]}
  \BibitemShut {NoStop}%
\bibitem [{\citenamefont {{Joachimi}}(2017)}]{2017MNRAS.466L..83J}%
  \BibitemOpen
  \bibfield  {author} {\bibinfo {author} {\bibfnamefont {B.}~\bibnamefont
  {{Joachimi}}},\ }\href {\doibase 10.1093/mnrasl/slw240} {\bibfield  {journal}
  {\bibinfo  {journal} {\mnras}\ }\textbf {\bibinfo {volume} {466}},\ \bibinfo
  {pages} {L83} (\bibinfo {year} {2017})},\ \Eprint
  {http://arxiv.org/abs/1612.00752} {arXiv:1612.00752 [astro-ph.IM]}
  \BibitemShut {NoStop}%
\bibitem [{\citenamefont {{Padmanabhan}}\ \emph {et~al.}(2016)\citenamefont
  {{Padmanabhan}}, \citenamefont {{White}}, \citenamefont {{Zhou}},\ and\
  \citenamefont {{O'Connell}}}]{2016MNRAS.460.1567P}%
  \BibitemOpen
  \bibfield  {author} {\bibinfo {author} {\bibfnamefont {N.}~\bibnamefont
  {{Padmanabhan}}}, \bibinfo {author} {\bibfnamefont {M.}~\bibnamefont
  {{White}}}, \bibinfo {author} {\bibfnamefont {H.~H.}\ \bibnamefont {{Zhou}}},
  \ and\ \bibinfo {author} {\bibfnamefont {R.}~\bibnamefont {{O'Connell}}},\
  }\href {\doibase 10.1093/mnras/stw1042} {\bibfield  {journal} {\bibinfo
  {journal} {\mnras}\ }\textbf {\bibinfo {volume} {460}},\ \bibinfo {pages}
  {1567} (\bibinfo {year} {2016})},\ \Eprint {http://arxiv.org/abs/1512.01241}
  {arXiv:1512.01241 [astro-ph.IM]} \BibitemShut {NoStop}%
\bibitem [{\citenamefont {{Gazta{\~n}aga}}\ and\ \citenamefont
  {{Scoccimarro}}(2005)}]{GazSco0508}%
  \BibitemOpen
  \bibfield  {author} {\bibinfo {author} {\bibfnamefont {E.}~\bibnamefont
  {{Gazta{\~n}aga}}}\ and\ \bibinfo {author} {\bibfnamefont {R.}~\bibnamefont
  {{Scoccimarro}}},\ }\href {\doibase 10.1111/j.1365-2966.2005.09234.x}
  {\bibfield  {journal} {\bibinfo  {journal} {\mnras}\ }\textbf {\bibinfo
  {volume} {361}},\ \bibinfo {pages} {824} (\bibinfo {year} {2005})},\ \Eprint
  {http://arxiv.org/abs/astro-ph/0501637} {arXiv:astro-ph/0501637 [astro-ph]}
  \BibitemShut {NoStop}%
\bibitem [{\citenamefont {{Philcox}}\ \emph
  {et~al.}(2020{\natexlab{b}})\citenamefont {{Philcox}}, \citenamefont
  {{Ivanov}}, \citenamefont {{Zaldarriaga}}, \citenamefont {{Simonovic}},\ and\
  \citenamefont {{Schmittfull}}}]{PhiIva20inprep}%
  \BibitemOpen
  \bibfield  {author} {\bibinfo {author} {\bibfnamefont {O.~H.~E.}\
  \bibnamefont {{Philcox}}}, \bibinfo {author} {\bibfnamefont {M.~M.}\
  \bibnamefont {{Ivanov}}}, \bibinfo {author} {\bibfnamefont {M.}~\bibnamefont
  {{Zaldarriaga}}}, \bibinfo {author} {\bibfnamefont {M.}~\bibnamefont
  {{Simonovic}}}, \ and\ \bibinfo {author} {\bibfnamefont {M.}~\bibnamefont
  {{Schmittfull}}},\ }\href@noop {} {\bibfield  {journal} {\bibinfo  {journal}
  {arXiv e-prints}\ ,\ \bibinfo {eid} {arXiv:2009.03311}} (\bibinfo {year}
  {2020}{\natexlab{b}})},\ \Eprint {http://arxiv.org/abs/2009.03311}
  {arXiv:2009.03311 [astro-ph.CO]} \BibitemShut {NoStop}%
\bibitem [{\citenamefont {{Scoccimarro}}(2000)}]{Sco0012}%
  \BibitemOpen
  \bibfield  {author} {\bibinfo {author} {\bibfnamefont {R.}~\bibnamefont
  {{Scoccimarro}}},\ }\href {\doibase 10.1086/317248} {\bibfield  {journal}
  {\bibinfo  {journal} {\apj}\ }\textbf {\bibinfo {volume} {544}},\ \bibinfo
  {pages} {597} (\bibinfo {year} {2000})},\ \Eprint
  {http://arxiv.org/abs/astro-ph/0004086} {arXiv:astro-ph/0004086 [astro-ph]}
  \BibitemShut {NoStop}%
\bibitem [{\citenamefont {{Eisenstein}}\ and\ \citenamefont
  {{Zaldarriaga}}(2001)}]{EisZal0101}%
  \BibitemOpen
  \bibfield  {author} {\bibinfo {author} {\bibfnamefont {D.~J.}\ \bibnamefont
  {{Eisenstein}}}\ and\ \bibinfo {author} {\bibfnamefont {M.}~\bibnamefont
  {{Zaldarriaga}}},\ }\href {\doibase 10.1086/318226} {\bibfield  {journal}
  {\bibinfo  {journal} {\apj}\ }\textbf {\bibinfo {volume} {546}},\ \bibinfo
  {pages} {2} (\bibinfo {year} {2001})},\ \Eprint
  {http://arxiv.org/abs/astro-ph/9912149} {arXiv:astro-ph/9912149 [astro-ph]}
  \BibitemShut {NoStop}%
\bibitem [{\citenamefont {{Friedrich}}\ and\ \citenamefont
  {{Eifler}}(2018)}]{2018MNRAS.473.4150F}%
  \BibitemOpen
  \bibfield  {author} {\bibinfo {author} {\bibfnamefont {O.}~\bibnamefont
  {{Friedrich}}}\ and\ \bibinfo {author} {\bibfnamefont {T.}~\bibnamefont
  {{Eifler}}},\ }\href {\doibase 10.1093/mnras/stx2566} {\bibfield  {journal}
  {\bibinfo  {journal} {\mnras}\ }\textbf {\bibinfo {volume} {473}},\ \bibinfo
  {pages} {4150} (\bibinfo {year} {2018})},\ \Eprint
  {http://arxiv.org/abs/1703.07786} {arXiv:1703.07786 [astro-ph.IM]}
  \BibitemShut {NoStop}%
\bibitem [{\citenamefont {{Ivanov}}\ \emph {et~al.}(2019)\citenamefont
  {{Ivanov}}, \citenamefont {{Simonovi{\'c}}},\ and\ \citenamefont
  {{Zaldarriaga}}}]{IvaSimZal19}%
  \BibitemOpen
  \bibfield  {author} {\bibinfo {author} {\bibfnamefont {M.~M.}\ \bibnamefont
  {{Ivanov}}}, \bibinfo {author} {\bibfnamefont {M.}~\bibnamefont
  {{Simonovi{\'c}}}}, \ and\ \bibinfo {author} {\bibfnamefont {M.}~\bibnamefont
  {{Zaldarriaga}}},\ }\href@noop {} {\bibfield  {journal} {\bibinfo  {journal}
  {arXiv e-prints}\ ,\ \bibinfo {eid} {arXiv:1909.05277}} (\bibinfo {year}
  {2019})},\ \Eprint {http://arxiv.org/abs/1909.05277} {arXiv:1909.05277
  [astro-ph.CO]} \BibitemShut {NoStop}%
\bibitem [{\citenamefont {D'Amico}\ \emph {et~al.}(2020)\citenamefont
  {D'Amico}, \citenamefont {Gleyzes}, \citenamefont {Kokron}, \citenamefont
  {Markovic}, \citenamefont {Senatore}, \citenamefont {Zhang}, \citenamefont
  {Beutler},\ and\ \citenamefont {Gil-MarÃ­n}}]{DAmico:2019fhj}%
  \BibitemOpen
  \bibfield  {author} {\bibinfo {author} {\bibfnamefont {G.}~\bibnamefont
  {D'Amico}}, \bibinfo {author} {\bibfnamefont {J.}~\bibnamefont {Gleyzes}},
  \bibinfo {author} {\bibfnamefont {N.}~\bibnamefont {Kokron}}, \bibinfo
  {author} {\bibfnamefont {D.}~\bibnamefont {Markovic}}, \bibinfo {author}
  {\bibfnamefont {L.}~\bibnamefont {Senatore}}, \bibinfo {author}
  {\bibfnamefont {P.}~\bibnamefont {Zhang}}, \bibinfo {author} {\bibfnamefont
  {F.}~\bibnamefont {Beutler}}, \ and\ \bibinfo {author} {\bibfnamefont
  {H.}~\bibnamefont {Gil-MarÃ­n}},\ }\href {\doibase
  10.1088/1475-7516/2020/05/005} {\bibfield  {journal} {\bibinfo  {journal}
  {JCAP}\ }\textbf {\bibinfo {volume} {05}},\ \bibinfo {pages} {005} (\bibinfo
  {year} {2020})},\ \Eprint {http://arxiv.org/abs/1909.05271} {arXiv:1909.05271
  [astro-ph.CO]} \BibitemShut {NoStop}%
\bibitem [{\citenamefont {Ivanov}\ \emph
  {et~al.}(2020{\natexlab{a}})\citenamefont {Ivanov}, \citenamefont
  {Simonovi\'c},\ and\ \citenamefont {Zaldarriaga}}]{Ivanov:2019hqk}%
  \BibitemOpen
  \bibfield  {author} {\bibinfo {author} {\bibfnamefont {M.~M.}\ \bibnamefont
  {Ivanov}}, \bibinfo {author} {\bibfnamefont {M.}~\bibnamefont {Simonovi\'c}},
  \ and\ \bibinfo {author} {\bibfnamefont {M.}~\bibnamefont {Zaldarriaga}},\
  }\href {\doibase 10.1103/PhysRevD.101.083504} {\bibfield  {journal} {\bibinfo
   {journal} {Phys. Rev. D}\ }\textbf {\bibinfo {volume} {101}},\ \bibinfo
  {pages} {083504} (\bibinfo {year} {2020}{\natexlab{a}})},\ \Eprint
  {http://arxiv.org/abs/1912.08208} {arXiv:1912.08208 [astro-ph.CO]}
  \BibitemShut {NoStop}%
\bibitem [{\citenamefont {{S{\'a}nchez}}\ \emph {et~al.}(2017)\citenamefont
  {{S{\'a}nchez}}, \citenamefont {{Scoccimarro}}, \citenamefont {{Crocce}},
  \citenamefont {{Grieb}}, \citenamefont {{Salazar-Albornoz}}, \citenamefont
  {{Dalla Vecchia}}, \citenamefont {{Lippich}}, \citenamefont {{Beutler}},
  \citenamefont {{Brownstein}}, \citenamefont {{Chuang}}, \citenamefont
  {{Eisenstein}}, \citenamefont {{Kitaura}}, \citenamefont {{Olmstead}},
  \citenamefont {{Percival}}, \citenamefont {{Prada}}, \citenamefont
  {{Rodr{\'{\i}}guez-Torres}}, \citenamefont {{Ross}}, \citenamefont
  {{Samushia}}, \citenamefont {{Seo}}, \citenamefont {{Tinker}}, \citenamefont
  {{Tojeiro}}, \citenamefont {{Vargas-Maga{\~n}a}}, \citenamefont {{Wang}},\
  and\ \citenamefont {{Zhao}}}]{SanScoCro1701}%
  \BibitemOpen
  \bibfield  {author} {\bibinfo {author} {\bibfnamefont {A.~G.}\ \bibnamefont
  {{S{\'a}nchez}}}, \bibinfo {author} {\bibfnamefont {R.}~\bibnamefont
  {{Scoccimarro}}}, \bibinfo {author} {\bibfnamefont {M.}~\bibnamefont
  {{Crocce}}}, \bibinfo {author} {\bibfnamefont {J.~N.}\ \bibnamefont
  {{Grieb}}}, \bibinfo {author} {\bibfnamefont {S.}~\bibnamefont
  {{Salazar-Albornoz}}}, \bibinfo {author} {\bibfnamefont {C.}~\bibnamefont
  {{Dalla Vecchia}}}, \bibinfo {author} {\bibfnamefont {M.}~\bibnamefont
  {{Lippich}}}, \bibinfo {author} {\bibfnamefont {F.}~\bibnamefont
  {{Beutler}}}, \bibinfo {author} {\bibfnamefont {J.~R.}\ \bibnamefont
  {{Brownstein}}}, \bibinfo {author} {\bibfnamefont {C.-H.}\ \bibnamefont
  {{Chuang}}}, \bibinfo {author} {\bibfnamefont {D.~J.}\ \bibnamefont
  {{Eisenstein}}}, \bibinfo {author} {\bibfnamefont {F.-S.}\ \bibnamefont
  {{Kitaura}}}, \bibinfo {author} {\bibfnamefont {M.~D.}\ \bibnamefont
  {{Olmstead}}}, \bibinfo {author} {\bibfnamefont {W.~J.}\ \bibnamefont
  {{Percival}}}, \bibinfo {author} {\bibfnamefont {F.}~\bibnamefont {{Prada}}},
  \bibinfo {author} {\bibfnamefont {S.}~\bibnamefont
  {{Rodr{\'{\i}}guez-Torres}}}, \bibinfo {author} {\bibfnamefont {A.~J.}\
  \bibnamefont {{Ross}}}, \bibinfo {author} {\bibfnamefont {L.}~\bibnamefont
  {{Samushia}}}, \bibinfo {author} {\bibfnamefont {H.-J.}\ \bibnamefont
  {{Seo}}}, \bibinfo {author} {\bibfnamefont {J.}~\bibnamefont {{Tinker}}},
  \bibinfo {author} {\bibfnamefont {R.}~\bibnamefont {{Tojeiro}}}, \bibinfo
  {author} {\bibfnamefont {M.}~\bibnamefont {{Vargas-Maga{\~n}a}}}, \bibinfo
  {author} {\bibfnamefont {Y.}~\bibnamefont {{Wang}}}, \ and\ \bibinfo {author}
  {\bibfnamefont {G.-B.}\ \bibnamefont {{Zhao}}},\ }\href {\doibase
  10.1093/mnras/stw2443} {\bibfield  {journal} {\bibinfo  {journal} {\mnras}\
  }\textbf {\bibinfo {volume} {464}},\ \bibinfo {pages} {1640} (\bibinfo {year}
  {2017})},\ \Eprint {http://arxiv.org/abs/1607.03147} {arXiv:1607.03147}
  \BibitemShut {NoStop}%
\bibitem [{\citenamefont {{Gil-Mar{\'\i}n}}\ \emph {et~al.}(2017)\citenamefont
  {{Gil-Mar{\'\i}n}}, \citenamefont {{Percival}}, \citenamefont {{Verde}},
  \citenamefont {{Brownstein}}, \citenamefont {{Chuang}}, \citenamefont
  {{Kitaura}}, \citenamefont {{Rodr{\'\i}guez-Torres}},\ and\ \citenamefont
  {{Olmstead}}}]{GilPerVer1702}%
  \BibitemOpen
  \bibfield  {author} {\bibinfo {author} {\bibfnamefont {H.}~\bibnamefont
  {{Gil-Mar{\'\i}n}}}, \bibinfo {author} {\bibfnamefont {W.~J.}\ \bibnamefont
  {{Percival}}}, \bibinfo {author} {\bibfnamefont {L.}~\bibnamefont {{Verde}}},
  \bibinfo {author} {\bibfnamefont {J.~R.}\ \bibnamefont {{Brownstein}}},
  \bibinfo {author} {\bibfnamefont {C.-H.}\ \bibnamefont {{Chuang}}}, \bibinfo
  {author} {\bibfnamefont {F.-S.}\ \bibnamefont {{Kitaura}}}, \bibinfo {author}
  {\bibfnamefont {S.~A.}\ \bibnamefont {{Rodr{\'\i}guez-Torres}}}, \ and\
  \bibinfo {author} {\bibfnamefont {M.~D.}\ \bibnamefont {{Olmstead}}},\ }\href
  {\doibase 10.1093/mnras/stw2679} {\bibfield  {journal} {\bibinfo  {journal}
  {\mnras}\ }\textbf {\bibinfo {volume} {465}},\ \bibinfo {pages} {1757}
  (\bibinfo {year} {2017})},\ \Eprint {http://arxiv.org/abs/1606.00439}
  {arXiv:1606.00439 [astro-ph.CO]} \BibitemShut {NoStop}%
\bibitem [{\citenamefont {Beutler}\ \emph
  {et~al.}(2017{\natexlab{a}})\citenamefont {Beutler} \emph
  {et~al.}}]{BeuSeoSai1704}%
  \BibitemOpen
  \bibfield  {author} {\bibinfo {author} {\bibfnamefont {F.}~\bibnamefont
  {Beutler}} \emph {et~al.} (\bibinfo {collaboration} {BOSS}),\ }\href
  {\doibase 10.1093/mnras/stw3298} {\bibfield  {journal} {\bibinfo  {journal}
  {Mon. Not. Roy. Astron. Soc.}\ }\textbf {\bibinfo {volume} {466}},\ \bibinfo
  {pages} {2242} (\bibinfo {year} {2017}{\natexlab{a}})},\ \Eprint
  {http://arxiv.org/abs/1607.03150} {arXiv:1607.03150 [astro-ph.CO]}
  \BibitemShut {NoStop}%
\bibitem [{\citenamefont {{Grieb}}\ \emph {et~al.}(2017)\citenamefont
  {{Grieb}}, \citenamefont {{S{\'a}nchez}}, \citenamefont {{Salazar-Albornoz}},
  \citenamefont {{Scoccimarro}}, \citenamefont {{Crocce}}, \citenamefont
  {{Dalla Vecchia}}, \citenamefont {{Montesano}}, \citenamefont
  {{Gil-Mar{\'{\i}}n}}, \citenamefont {{Ross}}, \citenamefont {{Beutler}},
  \citenamefont {{Rodr{\'{\i}}guez-Torres}}, \citenamefont {{Chuang}},
  \citenamefont {{Prada}}, \citenamefont {{Kitaura}}, \citenamefont {{Cuesta}},
  \citenamefont {{Eisenstein}}, \citenamefont {{Percival}}, \citenamefont
  {{Vargas-Maga{\~n}a}}, \citenamefont {{Tinker}}, \citenamefont {{Tojeiro}},
  \citenamefont {{Brownstein}}, \citenamefont {{Maraston}}, \citenamefont
  {{Nichol}}, \citenamefont {{Olmstead}}, \citenamefont {{Samushia}},
  \citenamefont {{Seo}}, \citenamefont {{Streblyanska}},\ and\ \citenamefont
  {{Zhao}}}]{GriSanSal1705}%
  \BibitemOpen
  \bibfield  {author} {\bibinfo {author} {\bibfnamefont {J.~N.}\ \bibnamefont
  {{Grieb}}}, \bibinfo {author} {\bibfnamefont {A.~G.}\ \bibnamefont
  {{S{\'a}nchez}}}, \bibinfo {author} {\bibfnamefont {S.}~\bibnamefont
  {{Salazar-Albornoz}}}, \bibinfo {author} {\bibfnamefont {R.}~\bibnamefont
  {{Scoccimarro}}}, \bibinfo {author} {\bibfnamefont {M.}~\bibnamefont
  {{Crocce}}}, \bibinfo {author} {\bibfnamefont {C.}~\bibnamefont {{Dalla
  Vecchia}}}, \bibinfo {author} {\bibfnamefont {F.}~\bibnamefont
  {{Montesano}}}, \bibinfo {author} {\bibfnamefont {H.}~\bibnamefont
  {{Gil-Mar{\'{\i}}n}}}, \bibinfo {author} {\bibfnamefont {A.~J.}\ \bibnamefont
  {{Ross}}}, \bibinfo {author} {\bibfnamefont {F.}~\bibnamefont {{Beutler}}},
  \bibinfo {author} {\bibfnamefont {S.}~\bibnamefont
  {{Rodr{\'{\i}}guez-Torres}}}, \bibinfo {author} {\bibfnamefont {C.-H.}\
  \bibnamefont {{Chuang}}}, \bibinfo {author} {\bibfnamefont {F.}~\bibnamefont
  {{Prada}}}, \bibinfo {author} {\bibfnamefont {F.-S.}\ \bibnamefont
  {{Kitaura}}}, \bibinfo {author} {\bibfnamefont {A.~J.}\ \bibnamefont
  {{Cuesta}}}, \bibinfo {author} {\bibfnamefont {D.~J.}\ \bibnamefont
  {{Eisenstein}}}, \bibinfo {author} {\bibfnamefont {W.~J.}\ \bibnamefont
  {{Percival}}}, \bibinfo {author} {\bibfnamefont {M.}~\bibnamefont
  {{Vargas-Maga{\~n}a}}}, \bibinfo {author} {\bibfnamefont {J.~L.}\
  \bibnamefont {{Tinker}}}, \bibinfo {author} {\bibfnamefont {R.}~\bibnamefont
  {{Tojeiro}}}, \bibinfo {author} {\bibfnamefont {J.~R.}\ \bibnamefont
  {{Brownstein}}}, \bibinfo {author} {\bibfnamefont {C.}~\bibnamefont
  {{Maraston}}}, \bibinfo {author} {\bibfnamefont {R.~C.}\ \bibnamefont
  {{Nichol}}}, \bibinfo {author} {\bibfnamefont {M.~D.}\ \bibnamefont
  {{Olmstead}}}, \bibinfo {author} {\bibfnamefont {L.}~\bibnamefont
  {{Samushia}}}, \bibinfo {author} {\bibfnamefont {H.-J.}\ \bibnamefont
  {{Seo}}}, \bibinfo {author} {\bibfnamefont {A.}~\bibnamefont
  {{Streblyanska}}}, \ and\ \bibinfo {author} {\bibfnamefont {G.-b.}\
  \bibnamefont {{Zhao}}},\ }\href {\doibase 10.1093/mnras/stw3384} {\bibfield
  {journal} {\bibinfo  {journal} {\mnras}\ }\textbf {\bibinfo {volume} {467}},\
  \bibinfo {pages} {2085} (\bibinfo {year} {2017})},\ \Eprint
  {http://arxiv.org/abs/1607.03143} {arXiv:1607.03143} \BibitemShut {NoStop}%
\bibitem [{\citenamefont {Alam}\ \emph {et~al.}(2017)\citenamefont {Alam} \emph
  {et~al.}}]{AlaAtaBai1709}%
  \BibitemOpen
  \bibfield  {author} {\bibinfo {author} {\bibfnamefont {S.}~\bibnamefont
  {Alam}} \emph {et~al.} (\bibinfo {collaboration} {BOSS}),\ }\href {\doibase
  10.1093/mnras/stx721} {\bibfield  {journal} {\bibinfo  {journal} {Mon. Not.
  Roy. Astron. Soc.}\ }\textbf {\bibinfo {volume} {470}},\ \bibinfo {pages}
  {2617} (\bibinfo {year} {2017})},\ \Eprint {http://arxiv.org/abs/1607.03155}
  {arXiv:1607.03155 [astro-ph.CO]} \BibitemShut {NoStop}%
\bibitem [{\citenamefont {Abbott}\ \emph {et~al.}(2018)\citenamefont {Abbott}
  \emph {et~al.}}]{Abbott:2017wau}%
  \BibitemOpen
  \bibfield  {author} {\bibinfo {author} {\bibfnamefont {T.}~\bibnamefont
  {Abbott}} \emph {et~al.} (\bibinfo {collaboration} {DES}),\ }\href {\doibase
  10.1103/PhysRevD.98.043526} {\bibfield  {journal} {\bibinfo  {journal} {Phys.
  Rev. D}\ }\textbf {\bibinfo {volume} {98}},\ \bibinfo {pages} {043526}
  (\bibinfo {year} {2018})},\ \Eprint {http://arxiv.org/abs/1708.01530}
  {arXiv:1708.01530 [astro-ph.CO]} \BibitemShut {NoStop}%
\bibitem [{\citenamefont {Aghanim}\ \emph {et~al.}(2019)\citenamefont {Aghanim}
  \emph {et~al.}}]{Aghanim:2019ame}%
  \BibitemOpen
  \bibfield  {author} {\bibinfo {author} {\bibfnamefont {N.}~\bibnamefont
  {Aghanim}} \emph {et~al.} (\bibinfo {collaboration} {Planck}),\ }\href@noop
  {} {\  (\bibinfo {year} {2019})},\ \Eprint {http://arxiv.org/abs/1907.12875}
  {arXiv:1907.12875 [astro-ph.CO]} \BibitemShut {NoStop}%
\bibitem [{\citenamefont {Spergel}\ \emph {et~al.}(2003)\citenamefont {Spergel}
  \emph {et~al.}}]{Spe03}%
  \BibitemOpen
  \bibfield  {author} {\bibinfo {author} {\bibfnamefont {D.}~\bibnamefont
  {Spergel}} \emph {et~al.} (\bibinfo {collaboration} {WMAP}),\ }\href
  {\doibase 10.1086/377226} {\bibfield  {journal} {\bibinfo  {journal}
  {Astrophys. J. Suppl.}\ }\textbf {\bibinfo {volume} {148}},\ \bibinfo {pages}
  {175} (\bibinfo {year} {2003})},\ \Eprint
  {http://arxiv.org/abs/astro-ph/0302209} {arXiv:astro-ph/0302209} \BibitemShut
  {NoStop}%
\bibitem [{\citenamefont {Hikage}\ \emph {et~al.}(2019)\citenamefont {Hikage}
  \emph {et~al.}}]{HikOguHam1903}%
  \BibitemOpen
  \bibfield  {author} {\bibinfo {author} {\bibfnamefont {C.}~\bibnamefont
  {Hikage}} \emph {et~al.} (\bibinfo {collaboration} {HSC}),\ }\href {\doibase
  10.1093/pasj/psz010} {\bibfield  {journal} {\bibinfo  {journal} {Publ.
  Astron. Soc. Jap.}\ }\textbf {\bibinfo {volume} {71}},\ \bibinfo {pages}
  {Publications of the Astronomical Society of Japan, Volume 71, Issue 2, April
  2019, 43, https://doi.org/10.1093/pasj/psz010} (\bibinfo {year} {2019})},\
  \Eprint {http://arxiv.org/abs/1809.09148} {arXiv:1809.09148 [astro-ph.CO]}
  \BibitemShut {NoStop}%
\bibitem [{\citenamefont {Joachimi}\ \emph {et~al.}(2020)\citenamefont
  {Joachimi} \emph {et~al.}}]{JoaLimAsg20}%
  \BibitemOpen
  \bibfield  {author} {\bibinfo {author} {\bibfnamefont {B.}~\bibnamefont
  {Joachimi}} \emph {et~al.},\ }\href@noop {} {\  (\bibinfo {year} {2020})},\
  \Eprint {http://arxiv.org/abs/2007.01844} {arXiv:2007.01844 [astro-ph.CO]}
  \BibitemShut {NoStop}%
\bibitem [{\citenamefont {{Scoccimarro}}\ \emph {et~al.}(1999)\citenamefont
  {{Scoccimarro}}, \citenamefont {{Zaldarriaga}},\ and\ \citenamefont
  {{Hui}}}]{ScoZalHui9912}%
  \BibitemOpen
  \bibfield  {author} {\bibinfo {author} {\bibfnamefont {R.}~\bibnamefont
  {{Scoccimarro}}}, \bibinfo {author} {\bibfnamefont {M.}~\bibnamefont
  {{Zaldarriaga}}}, \ and\ \bibinfo {author} {\bibfnamefont {L.}~\bibnamefont
  {{Hui}}},\ }\href {\doibase 10.1086/308059} {\bibfield  {journal} {\bibinfo
  {journal} {\apj}\ }\textbf {\bibinfo {volume} {527}},\ \bibinfo {pages} {1}
  (\bibinfo {year} {1999})},\ \Eprint {http://arxiv.org/abs/astro-ph/9901099}
  {astro-ph/9901099} \BibitemShut {NoStop}%
\bibitem [{\citenamefont {{Hamilton}}\ \emph {et~al.}(2006)\citenamefont
  {{Hamilton}}, \citenamefont {{Rimes}},\ and\ \citenamefont
  {{Scoccimarro}}}]{HamRimSco0609}%
  \BibitemOpen
  \bibfield  {author} {\bibinfo {author} {\bibfnamefont {A.~J.~S.}\
  \bibnamefont {{Hamilton}}}, \bibinfo {author} {\bibfnamefont {C.~D.}\
  \bibnamefont {{Rimes}}}, \ and\ \bibinfo {author} {\bibfnamefont
  {R.}~\bibnamefont {{Scoccimarro}}},\ }\href {\doibase
  10.1111/j.1365-2966.2006.10709.x} {\bibfield  {journal} {\bibinfo  {journal}
  {\mnras}\ }\textbf {\bibinfo {volume} {371}},\ \bibinfo {pages} {1188}
  (\bibinfo {year} {2006})},\ \Eprint {http://arxiv.org/abs/astro-ph/0511416}
  {astro-ph/0511416} \BibitemShut {NoStop}%
\bibitem [{\citenamefont {{Sefusatti}}\ and\ \citenamefont
  {{Scoccimarro}}(2005)}]{SefSco0503}%
  \BibitemOpen
  \bibfield  {author} {\bibinfo {author} {\bibfnamefont {E.}~\bibnamefont
  {{Sefusatti}}}\ and\ \bibinfo {author} {\bibfnamefont {R.}~\bibnamefont
  {{Scoccimarro}}},\ }\href {\doibase 10.1103/PhysRevD.71.063001} {\bibfield
  {journal} {\bibinfo  {journal} {\prd}\ }\textbf {\bibinfo {volume} {71}},\
  \bibinfo {eid} {063001} (\bibinfo {year} {2005})},\ \Eprint
  {http://arxiv.org/abs/astro-ph/0412626} {arXiv:astro-ph/0412626 [astro-ph]}
  \BibitemShut {NoStop}%
\bibitem [{\citenamefont {{Sefusatti}}\ \emph {et~al.}(2006)\citenamefont
  {{Sefusatti}}, \citenamefont {{Crocce}}, \citenamefont {{Pueblas}},\ and\
  \citenamefont {{Scoccimarro}}}]{SefCroPue0607}%
  \BibitemOpen
  \bibfield  {author} {\bibinfo {author} {\bibfnamefont {E.}~\bibnamefont
  {{Sefusatti}}}, \bibinfo {author} {\bibfnamefont {M.}~\bibnamefont
  {{Crocce}}}, \bibinfo {author} {\bibfnamefont {S.}~\bibnamefont {{Pueblas}}},
  \ and\ \bibinfo {author} {\bibfnamefont {R.}~\bibnamefont {{Scoccimarro}}},\
  }\href {\doibase 10.1103/PhysRevD.74.023522} {\bibfield  {journal} {\bibinfo
  {journal} {\prd}\ }\textbf {\bibinfo {volume} {74}},\ \bibinfo {eid} {023522}
  (\bibinfo {year} {2006})},\ \Eprint {http://arxiv.org/abs/astro-ph/0604505}
  {arXiv:astro-ph/0604505 [astro-ph]} \BibitemShut {NoStop}%
\bibitem [{\citenamefont {{de Putter}}\ \emph {et~al.}(2012)\citenamefont {{de
  Putter}}, \citenamefont {{Wagner}}, \citenamefont {{Mena}}, \citenamefont
  {{Verde}},\ and\ \citenamefont {{Percival}}}]{PutWagMen1204}%
  \BibitemOpen
  \bibfield  {author} {\bibinfo {author} {\bibfnamefont {R.}~\bibnamefont {{de
  Putter}}}, \bibinfo {author} {\bibfnamefont {C.}~\bibnamefont {{Wagner}}},
  \bibinfo {author} {\bibfnamefont {O.}~\bibnamefont {{Mena}}}, \bibinfo
  {author} {\bibfnamefont {L.}~\bibnamefont {{Verde}}}, \ and\ \bibinfo
  {author} {\bibfnamefont {W.~J.}\ \bibnamefont {{Percival}}},\ }\href
  {\doibase 10.1088/1475-7516/2012/04/019} {\bibfield  {journal} {\bibinfo
  {journal} {\jcap}\ }\textbf {\bibinfo {volume} {4}},\ \bibinfo {eid} {019}
  (\bibinfo {year} {2012})},\ \Eprint {http://arxiv.org/abs/1111.6596}
  {arXiv:1111.6596} \BibitemShut {NoStop}%
\bibitem [{\citenamefont {{Takada}}\ and\ \citenamefont
  {{Hu}}(2013)}]{TakHu1306}%
  \BibitemOpen
  \bibfield  {author} {\bibinfo {author} {\bibfnamefont {M.}~\bibnamefont
  {{Takada}}}\ and\ \bibinfo {author} {\bibfnamefont {W.}~\bibnamefont
  {{Hu}}},\ }\href {\doibase 10.1103/PhysRevD.87.123504} {\bibfield  {journal}
  {\bibinfo  {journal} {\prd}\ }\textbf {\bibinfo {volume} {87}},\ \bibinfo
  {eid} {123504} (\bibinfo {year} {2013})},\ \Eprint
  {http://arxiv.org/abs/1302.6994} {arXiv:1302.6994 [astro-ph.CO]} \BibitemShut
  {NoStop}%
\bibitem [{\citenamefont {{Li}}\ \emph
  {et~al.}(2014{\natexlab{a}})\citenamefont {{Li}}, \citenamefont {{Hu}},\ and\
  \citenamefont {{Takada}}}]{LiHuTak1404}%
  \BibitemOpen
  \bibfield  {author} {\bibinfo {author} {\bibfnamefont {Y.}~\bibnamefont
  {{Li}}}, \bibinfo {author} {\bibfnamefont {W.}~\bibnamefont {{Hu}}}, \ and\
  \bibinfo {author} {\bibfnamefont {M.}~\bibnamefont {{Takada}}},\ }\href
  {\doibase 10.1103/PhysRevD.89.083519} {\bibfield  {journal} {\bibinfo
  {journal} {\prd}\ }\textbf {\bibinfo {volume} {89}},\ \bibinfo {eid} {083519}
  (\bibinfo {year} {2014}{\natexlab{a}})},\ \Eprint
  {http://arxiv.org/abs/1401.0385} {arXiv:1401.0385} \BibitemShut {NoStop}%
\bibitem [{\citenamefont {{Li}}\ \emph
  {et~al.}(2014{\natexlab{b}})\citenamefont {{Li}}, \citenamefont {{Hu}},\ and\
  \citenamefont {{Takada}}}]{LiHuTak1411}%
  \BibitemOpen
  \bibfield  {author} {\bibinfo {author} {\bibfnamefont {Y.}~\bibnamefont
  {{Li}}}, \bibinfo {author} {\bibfnamefont {W.}~\bibnamefont {{Hu}}}, \ and\
  \bibinfo {author} {\bibfnamefont {M.}~\bibnamefont {{Takada}}},\ }\href
  {\doibase 10.1103/PhysRevD.90.103530} {\bibfield  {journal} {\bibinfo
  {journal} {\prd}\ }\textbf {\bibinfo {volume} {90}},\ \bibinfo {eid} {103530}
  (\bibinfo {year} {2014}{\natexlab{b}})},\ \Eprint
  {http://arxiv.org/abs/1408.1081} {arXiv:1408.1081} \BibitemShut {NoStop}%
\bibitem [{\citenamefont {{Akitsu}}\ \emph {et~al.}(2017)\citenamefont
  {{Akitsu}}, \citenamefont {{Takada}},\ and\ \citenamefont
  {{Li}}}]{AkiTakLi1704}%
  \BibitemOpen
  \bibfield  {author} {\bibinfo {author} {\bibfnamefont {K.}~\bibnamefont
  {{Akitsu}}}, \bibinfo {author} {\bibfnamefont {M.}~\bibnamefont {{Takada}}},
  \ and\ \bibinfo {author} {\bibfnamefont {Y.}~\bibnamefont {{Li}}},\ }\href
  {\doibase 10.1103/PhysRevD.95.083522} {\bibfield  {journal} {\bibinfo
  {journal} {\prd}\ }\textbf {\bibinfo {volume} {95}},\ \bibinfo {eid} {083522}
  (\bibinfo {year} {2017})},\ \Eprint {http://arxiv.org/abs/1611.04723}
  {arXiv:1611.04723 [astro-ph.CO]} \BibitemShut {NoStop}%
\bibitem [{\citenamefont {{Li}}\ \emph {et~al.}(2018)\citenamefont {{Li}},
  \citenamefont {{Schmittfull}},\ and\ \citenamefont
  {{Seljak}}}]{LiSchSel1802}%
  \BibitemOpen
  \bibfield  {author} {\bibinfo {author} {\bibfnamefont {Y.}~\bibnamefont
  {{Li}}}, \bibinfo {author} {\bibfnamefont {M.}~\bibnamefont {{Schmittfull}}},
  \ and\ \bibinfo {author} {\bibfnamefont {U.}~\bibnamefont {{Seljak}}},\
  }\href {\doibase 10.1088/1475-7516/2018/02/022} {\bibfield  {journal}
  {\bibinfo  {journal} {\jcap}\ }\textbf {\bibinfo {volume} {2}},\ \bibinfo
  {eid} {022} (\bibinfo {year} {2018})},\ \Eprint
  {http://arxiv.org/abs/1711.00018} {arXiv:1711.00018} \BibitemShut {NoStop}%
\bibitem [{\citenamefont {{Chan}}\ and\ \citenamefont
  {{Blot}}(2017)}]{ChaBlo17}%
  \BibitemOpen
  \bibfield  {author} {\bibinfo {author} {\bibfnamefont {K.~C.}\ \bibnamefont
  {{Chan}}}\ and\ \bibinfo {author} {\bibfnamefont {L.}~\bibnamefont
  {{Blot}}},\ }\href {\doibase 10.1103/PhysRevD.96.023528} {\bibfield
  {journal} {\bibinfo  {journal} {\prd}\ }\textbf {\bibinfo {volume} {96}},\
  \bibinfo {eid} {023528} (\bibinfo {year} {2017})},\ \Eprint
  {http://arxiv.org/abs/1610.06585} {arXiv:1610.06585 [astro-ph.CO]}
  \BibitemShut {NoStop}%
\bibitem [{\citenamefont {{Taruya}}\ \emph {et~al.}(2020)\citenamefont
  {{Taruya}}, \citenamefont {{Nishimichi}},\ and\ \citenamefont
  {{Jeong}}}]{TarNisJeo20}%
  \BibitemOpen
  \bibfield  {author} {\bibinfo {author} {\bibfnamefont {A.}~\bibnamefont
  {{Taruya}}}, \bibinfo {author} {\bibfnamefont {T.}~\bibnamefont
  {{Nishimichi}}}, \ and\ \bibinfo {author} {\bibfnamefont {D.}~\bibnamefont
  {{Jeong}}},\ }\href@noop {} {\bibfield  {journal} {\bibinfo  {journal} {arXiv
  e-prints}\ ,\ \bibinfo {eid} {arXiv:2007.05504}} (\bibinfo {year} {2020})},\
  \Eprint {http://arxiv.org/abs/2007.05504} {arXiv:2007.05504 [astro-ph.CO]}
  \BibitemShut {NoStop}%
\bibitem [{\citenamefont {{Hikage}}\ \emph {et~al.}(2020)\citenamefont
  {{Hikage}}, \citenamefont {{Takahashi}},\ and\ \citenamefont
  {{Koyama}}}]{HikTakKoy20}%
  \BibitemOpen
  \bibfield  {author} {\bibinfo {author} {\bibfnamefont {C.}~\bibnamefont
  {{Hikage}}}, \bibinfo {author} {\bibfnamefont {R.}~\bibnamefont
  {{Takahashi}}}, \ and\ \bibinfo {author} {\bibfnamefont {K.}~\bibnamefont
  {{Koyama}}},\ }\href@noop {} {\bibfield  {journal} {\bibinfo  {journal}
  {arXiv e-prints}\ ,\ \bibinfo {eid} {arXiv:2007.13998}} (\bibinfo {year}
  {2020})},\ \Eprint {http://arxiv.org/abs/2007.13998} {arXiv:2007.13998
  [astro-ph.CO]} \BibitemShut {NoStop}%
\bibitem [{\citenamefont {{Sugiyama}}\ \emph {et~al.}(2019)\citenamefont
  {{Sugiyama}}, \citenamefont {{Saito}}, \citenamefont {{Beutler}},\ and\
  \citenamefont {{Seo}}}]{SugSaiBeu1908}%
  \BibitemOpen
  \bibfield  {author} {\bibinfo {author} {\bibfnamefont {N.~S.}\ \bibnamefont
  {{Sugiyama}}}, \bibinfo {author} {\bibfnamefont {S.}~\bibnamefont {{Saito}}},
  \bibinfo {author} {\bibfnamefont {F.}~\bibnamefont {{Beutler}}}, \ and\
  \bibinfo {author} {\bibfnamefont {H.-J.}\ \bibnamefont {{Seo}}},\ }\href@noop
  {} {\bibfield  {journal} {\bibinfo  {journal} {arXiv e-prints}\ ,\ \bibinfo
  {eid} {arXiv:1908.06234}} (\bibinfo {year} {2019})},\ \Eprint
  {http://arxiv.org/abs/1908.06234} {arXiv:1908.06234 [astro-ph.CO]}
  \BibitemShut {NoStop}%
\bibitem [{\citenamefont {Yamamoto}\ \emph {et~al.}(2006)\citenamefont
  {Yamamoto}, \citenamefont {Nakamichi}, \citenamefont {Kamino}, \citenamefont
  {Bassett},\ and\ \citenamefont {Nishioka}}]{Yamamoto:2005dz}%
  \BibitemOpen
  \bibfield  {author} {\bibinfo {author} {\bibfnamefont {K.}~\bibnamefont
  {Yamamoto}}, \bibinfo {author} {\bibfnamefont {M.}~\bibnamefont {Nakamichi}},
  \bibinfo {author} {\bibfnamefont {A.}~\bibnamefont {Kamino}}, \bibinfo
  {author} {\bibfnamefont {B.~A.}\ \bibnamefont {Bassett}}, \ and\ \bibinfo
  {author} {\bibfnamefont {H.}~\bibnamefont {Nishioka}},\ }\href {\doibase
  10.1093/pasj/58.1.93} {\bibfield  {journal} {\bibinfo  {journal} {Publ.
  Astron. Soc. Jap.}\ }\textbf {\bibinfo {volume} {58}},\ \bibinfo {pages} {93}
  (\bibinfo {year} {2006})},\ \Eprint {http://arxiv.org/abs/astro-ph/0505115}
  {arXiv:astro-ph/0505115} \BibitemShut {NoStop}%
\bibitem [{\citenamefont {Grieb}\ \emph {et~al.}(2016)\citenamefont {Grieb},
  \citenamefont {SÃ¡nchez}, \citenamefont {Salazar-Albornoz},\ and\
  \citenamefont {Dalla~Vecchia}}]{Grieb:2015bia}%
  \BibitemOpen
  \bibfield  {author} {\bibinfo {author} {\bibfnamefont {J.~N.}\ \bibnamefont
  {Grieb}}, \bibinfo {author} {\bibfnamefont {A.~G.}\ \bibnamefont
  {SÃ¡nchez}}, \bibinfo {author} {\bibfnamefont {S.}~\bibnamefont
  {Salazar-Albornoz}}, \ and\ \bibinfo {author} {\bibfnamefont
  {C.}~\bibnamefont {Dalla~Vecchia}},\ }\href {\doibase 10.1093/mnras/stw065}
  {\bibfield  {journal} {\bibinfo  {journal} {Mon. Not. Roy. Astron. Soc.}\
  }\textbf {\bibinfo {volume} {457}},\ \bibinfo {pages} {1577} (\bibinfo {year}
  {2016})},\ \Eprint {http://arxiv.org/abs/1509.04293} {arXiv:1509.04293
  [astro-ph.CO]} \BibitemShut {NoStop}%
\bibitem [{\citenamefont {{Blake}}\ \emph {et~al.}(2018)\citenamefont
  {{Blake}}, \citenamefont {{Carter}},\ and\ \citenamefont
  {{Koda}}}]{BlaCarKod1810}%
  \BibitemOpen
  \bibfield  {author} {\bibinfo {author} {\bibfnamefont {C.}~\bibnamefont
  {{Blake}}}, \bibinfo {author} {\bibfnamefont {P.}~\bibnamefont {{Carter}}}, \
  and\ \bibinfo {author} {\bibfnamefont {J.}~\bibnamefont {{Koda}}},\ }\href
  {\doibase 10.1093/mnras/sty1814} {\bibfield  {journal} {\bibinfo  {journal}
  {\mnras}\ }\textbf {\bibinfo {volume} {479}},\ \bibinfo {pages} {5168}
  (\bibinfo {year} {2018})},\ \Eprint {http://arxiv.org/abs/1801.04969}
  {arXiv:1801.04969 [astro-ph.CO]} \BibitemShut {NoStop}%
\bibitem [{\citenamefont {{Philcox}}\ and\ \citenamefont
  {{Eisenstein}}(2019{\natexlab{a}})}]{2019MNRAS.490.5931P}%
  \BibitemOpen
  \bibfield  {author} {\bibinfo {author} {\bibfnamefont {O.~H.~E.}\
  \bibnamefont {{Philcox}}}\ and\ \bibinfo {author} {\bibfnamefont {D.~J.}\
  \bibnamefont {{Eisenstein}}},\ }\href {\doibase 10.1093/mnras/stz2896}
  {\bibfield  {journal} {\bibinfo  {journal} {\mnras}\ }\textbf {\bibinfo
  {volume} {490}},\ \bibinfo {pages} {5931} (\bibinfo {year}
  {2019}{\natexlab{a}})},\ \Eprint {http://arxiv.org/abs/1910.04764}
  {arXiv:1910.04764 [astro-ph.CO]} \BibitemShut {NoStop}%
\bibitem [{\citenamefont {{Li}}\ \emph {et~al.}(2019)\citenamefont {{Li}},
  \citenamefont {{Singh}}, \citenamefont {{Yu}}, \citenamefont {{Feng}},\ and\
  \citenamefont {{Seljak}}}]{LiSinYu1811}%
  \BibitemOpen
  \bibfield  {author} {\bibinfo {author} {\bibfnamefont {Y.}~\bibnamefont
  {{Li}}}, \bibinfo {author} {\bibfnamefont {S.}~\bibnamefont {{Singh}}},
  \bibinfo {author} {\bibfnamefont {B.}~\bibnamefont {{Yu}}}, \bibinfo {author}
  {\bibfnamefont {Y.}~\bibnamefont {{Feng}}}, \ and\ \bibinfo {author}
  {\bibfnamefont {U.}~\bibnamefont {{Seljak}}},\ }\href {\doibase
  10.1088/1475-7516/2019/01/016} {\bibfield  {journal} {\bibinfo  {journal}
  {Journal of Cosmology and Astro-Particle Physics}\ }\textbf {\bibinfo
  {volume} {2019}},\ \bibinfo {eid} {016} (\bibinfo {year} {2019})},\ \Eprint
  {http://arxiv.org/abs/1811.05714} {arXiv:1811.05714 [astro-ph.CO]}
  \BibitemShut {NoStop}%
\bibitem [{\citenamefont {{Scoccimarro}}(2015)}]{Sco1510}%
  \BibitemOpen
  \bibfield  {author} {\bibinfo {author} {\bibfnamefont {R.}~\bibnamefont
  {{Scoccimarro}}},\ }\href {\doibase 10.1103/PhysRevD.92.083532} {\bibfield
  {journal} {\bibinfo  {journal} {\prd}\ }\textbf {\bibinfo {volume} {92}},\
  \bibinfo {eid} {083532} (\bibinfo {year} {2015})},\ \Eprint
  {http://arxiv.org/abs/1506.02729} {arXiv:1506.02729} \BibitemShut {NoStop}%
\bibitem [{\citenamefont {Hand}\ \emph {et~al.}(2018)\citenamefont {Hand},
  \citenamefont {Feng}, \citenamefont {Beutler}, \citenamefont {Li},
  \citenamefont {Modi}, \citenamefont {Seljak},\ and\ \citenamefont
  {Slepian}}]{Hand:2017pqn}%
  \BibitemOpen
  \bibfield  {author} {\bibinfo {author} {\bibfnamefont {N.}~\bibnamefont
  {Hand}}, \bibinfo {author} {\bibfnamefont {Y.}~\bibnamefont {Feng}}, \bibinfo
  {author} {\bibfnamefont {F.}~\bibnamefont {Beutler}}, \bibinfo {author}
  {\bibfnamefont {Y.}~\bibnamefont {Li}}, \bibinfo {author} {\bibfnamefont
  {C.}~\bibnamefont {Modi}}, \bibinfo {author} {\bibfnamefont {U.}~\bibnamefont
  {Seljak}}, \ and\ \bibinfo {author} {\bibfnamefont {Z.}~\bibnamefont
  {Slepian}},\ }\href {\doibase 10.3847/1538-3881/aadae0} {\bibfield  {journal}
  {\bibinfo  {journal} {Astron. J.}\ }\textbf {\bibinfo {volume} {156}},\
  \bibinfo {pages} {160} (\bibinfo {year} {2018})},\ \Eprint
  {http://arxiv.org/abs/1712.05834} {arXiv:1712.05834 [astro-ph.IM]}
  \BibitemShut {NoStop}%
\bibitem [{\citenamefont {Blas}\ \emph {et~al.}(2016)\citenamefont {Blas},
  \citenamefont {Garny}, \citenamefont {Ivanov},\ and\ \citenamefont
  {Sibiryakov}}]{Blas:2016sfa}%
  \BibitemOpen
  \bibfield  {author} {\bibinfo {author} {\bibfnamefont {D.}~\bibnamefont
  {Blas}}, \bibinfo {author} {\bibfnamefont {M.}~\bibnamefont {Garny}},
  \bibinfo {author} {\bibfnamefont {M.~M.}\ \bibnamefont {Ivanov}}, \ and\
  \bibinfo {author} {\bibfnamefont {S.}~\bibnamefont {Sibiryakov}},\ }\href
  {\doibase 10.1088/1475-7516/2016/07/028} {\bibfield  {journal} {\bibinfo
  {journal} {JCAP}\ }\textbf {\bibinfo {volume} {07}},\ \bibinfo {pages} {028}
  (\bibinfo {year} {2016})},\ \Eprint {http://arxiv.org/abs/1605.02149}
  {arXiv:1605.02149 [astro-ph.CO]} \BibitemShut {NoStop}%
\bibitem [{\citenamefont {Ivanov}\ and\ \citenamefont
  {Sibiryakov}(2018)}]{Ivanov:2018gjr}%
  \BibitemOpen
  \bibfield  {author} {\bibinfo {author} {\bibfnamefont {M.~M.}\ \bibnamefont
  {Ivanov}}\ and\ \bibinfo {author} {\bibfnamefont {S.}~\bibnamefont
  {Sibiryakov}},\ }\href {\doibase 10.1088/1475-7516/2018/07/053} {\bibfield
  {journal} {\bibinfo  {journal} {JCAP}\ }\textbf {\bibinfo {volume} {07}},\
  \bibinfo {pages} {053} (\bibinfo {year} {2018})},\ \Eprint
  {http://arxiv.org/abs/1804.05080} {arXiv:1804.05080 [astro-ph.CO]}
  \BibitemShut {NoStop}%
\bibitem [{\citenamefont {Nishimichi}\ \emph {et~al.}(2020)\citenamefont
  {Nishimichi}, \citenamefont {D'Amico}, \citenamefont {Ivanov}, \citenamefont
  {Senatore}, \citenamefont {Simonovi\'c}, \citenamefont {Takada},
  \citenamefont {Zaldarriaga},\ and\ \citenamefont
  {Zhang}}]{Nishimichi:2020tvu}%
  \BibitemOpen
  \bibfield  {author} {\bibinfo {author} {\bibfnamefont {T.}~\bibnamefont
  {Nishimichi}}, \bibinfo {author} {\bibfnamefont {G.}~\bibnamefont {D'Amico}},
  \bibinfo {author} {\bibfnamefont {M.~M.}\ \bibnamefont {Ivanov}}, \bibinfo
  {author} {\bibfnamefont {L.}~\bibnamefont {Senatore}}, \bibinfo {author}
  {\bibfnamefont {M.}~\bibnamefont {Simonovi\'c}}, \bibinfo {author}
  {\bibfnamefont {M.}~\bibnamefont {Takada}}, \bibinfo {author} {\bibfnamefont
  {M.}~\bibnamefont {Zaldarriaga}}, \ and\ \bibinfo {author} {\bibfnamefont
  {P.}~\bibnamefont {Zhang}},\ }\href@noop {} {\  (\bibinfo {year} {2020})},\
  \Eprint {http://arxiv.org/abs/2003.08277} {arXiv:2003.08277 [astro-ph.CO]}
  \BibitemShut {NoStop}%
\bibitem [{\citenamefont {{Chudaykin}}\ \emph {et~al.}(2020)\citenamefont
  {{Chudaykin}}, \citenamefont {{Ivanov}},\ and\ \citenamefont
  {{Simonovi{\'c}}}}]{ChuIvaSim20}%
  \BibitemOpen
  \bibfield  {author} {\bibinfo {author} {\bibfnamefont {A.}~\bibnamefont
  {{Chudaykin}}}, \bibinfo {author} {\bibfnamefont {M.~M.}\ \bibnamefont
  {{Ivanov}}}, \ and\ \bibinfo {author} {\bibfnamefont {M.}~\bibnamefont
  {{Simonovi{\'c}}}},\ }\href@noop {} {\bibfield  {journal} {\bibinfo
  {journal} {arXiv e-prints}\ ,\ \bibinfo {eid} {arXiv:2004.10607}} (\bibinfo
  {year} {2020})},\ \Eprint {http://arxiv.org/abs/2004.10607} {arXiv:2004.10607
  [astro-ph.CO]} \BibitemShut {NoStop}%
\bibitem [{\citenamefont {Simonovi\'c}\ \emph {et~al.}(2018)\citenamefont
  {Simonovi\'c}, \citenamefont {Baldauf}, \citenamefont {Zaldarriaga},
  \citenamefont {Carrasco},\ and\ \citenamefont
  {Kollmeier}}]{Simonovic:2017mhp}%
  \BibitemOpen
  \bibfield  {author} {\bibinfo {author} {\bibfnamefont {M.}~\bibnamefont
  {Simonovi\'c}}, \bibinfo {author} {\bibfnamefont {T.}~\bibnamefont
  {Baldauf}}, \bibinfo {author} {\bibfnamefont {M.}~\bibnamefont
  {Zaldarriaga}}, \bibinfo {author} {\bibfnamefont {J.~J.}\ \bibnamefont
  {Carrasco}}, \ and\ \bibinfo {author} {\bibfnamefont {J.~A.}\ \bibnamefont
  {Kollmeier}},\ }\href {\doibase 10.1088/1475-7516/2018/04/030} {\bibfield
  {journal} {\bibinfo  {journal} {JCAP}\ }\textbf {\bibinfo {volume} {04}},\
  \bibinfo {pages} {030} (\bibinfo {year} {2018})},\ \Eprint
  {http://arxiv.org/abs/1708.08130} {arXiv:1708.08130 [astro-ph.CO]}
  \BibitemShut {NoStop}%
\bibitem [{\citenamefont {Aghanim}\ \emph {et~al.}(2018)\citenamefont {Aghanim}
  \emph {et~al.}}]{Aghanim:2018eyx}%
  \BibitemOpen
  \bibfield  {author} {\bibinfo {author} {\bibfnamefont {N.}~\bibnamefont
  {Aghanim}} \emph {et~al.} (\bibinfo {collaboration} {Planck}),\ }\href@noop
  {} {\  (\bibinfo {year} {2018})},\ \Eprint {http://arxiv.org/abs/1807.06209}
  {arXiv:1807.06209 [astro-ph.CO]} \BibitemShut {NoStop}%
\bibitem [{\citenamefont {{Feldman}}\ \emph {et~al.}(1994)\citenamefont
  {{Feldman}}, \citenamefont {{Kaiser}},\ and\ \citenamefont
  {{Peacock}}}]{FelKaiPea9405}%
  \BibitemOpen
  \bibfield  {author} {\bibinfo {author} {\bibfnamefont {H.~A.}\ \bibnamefont
  {{Feldman}}}, \bibinfo {author} {\bibfnamefont {N.}~\bibnamefont {{Kaiser}}},
  \ and\ \bibinfo {author} {\bibfnamefont {J.~A.}\ \bibnamefont {{Peacock}}},\
  }\href {\doibase 10.1086/174036} {\bibfield  {journal} {\bibinfo  {journal}
  {\apj}\ }\textbf {\bibinfo {volume} {426}},\ \bibinfo {pages} {23} (\bibinfo
  {year} {1994})},\ \Eprint {http://arxiv.org/abs/astro-ph/9304022}
  {astro-ph/9304022} \BibitemShut {NoStop}%
\bibitem [{\citenamefont {Audren}\ \emph {et~al.}(2013)\citenamefont {Audren},
  \citenamefont {Lesgourgues}, \citenamefont {Bird}, \citenamefont {Haehnelt},\
  and\ \citenamefont {Viel}}]{Audren:2012vy}%
  \BibitemOpen
  \bibfield  {author} {\bibinfo {author} {\bibfnamefont {B.}~\bibnamefont
  {Audren}}, \bibinfo {author} {\bibfnamefont {J.}~\bibnamefont {Lesgourgues}},
  \bibinfo {author} {\bibfnamefont {S.}~\bibnamefont {Bird}}, \bibinfo {author}
  {\bibfnamefont {M.~G.}\ \bibnamefont {Haehnelt}}, \ and\ \bibinfo {author}
  {\bibfnamefont {M.}~\bibnamefont {Viel}},\ }\href {\doibase
  10.1088/1475-7516/2013/01/026} {\bibfield  {journal} {\bibinfo  {journal}
  {JCAP}\ }\textbf {\bibinfo {volume} {01}},\ \bibinfo {pages} {026} (\bibinfo
  {year} {2013})},\ \Eprint {http://arxiv.org/abs/1210.2194} {arXiv:1210.2194
  [astro-ph.CO]} \BibitemShut {NoStop}%
\bibitem [{\citenamefont {Brinckmann}\ and\ \citenamefont
  {Lesgourgues}(2018)}]{Brinckmann:2018cvx}%
  \BibitemOpen
  \bibfield  {author} {\bibinfo {author} {\bibfnamefont {T.}~\bibnamefont
  {Brinckmann}}\ and\ \bibinfo {author} {\bibfnamefont {J.}~\bibnamefont
  {Lesgourgues}},\ }\href@noop {} {\  (\bibinfo {year} {2018})},\ \Eprint
  {http://arxiv.org/abs/1804.07261} {arXiv:1804.07261 [astro-ph.CO]}
  \BibitemShut {NoStop}%
\bibitem [{\citenamefont {Gelman}\ and\ \citenamefont
  {Rubin}(1992)}]{Gelman:1992zz}%
  \BibitemOpen
  \bibfield  {author} {\bibinfo {author} {\bibfnamefont {A.}~\bibnamefont
  {Gelman}}\ and\ \bibinfo {author} {\bibfnamefont {D.~B.}\ \bibnamefont
  {Rubin}},\ }\href {\doibase 10.1214/ss/1177011136} {\bibfield  {journal}
  {\bibinfo  {journal} {Statist. Sci.}\ }\textbf {\bibinfo {volume} {7}},\
  \bibinfo {pages} {457} (\bibinfo {year} {1992})}\BibitemShut {NoStop}%
\bibitem [{\citenamefont {Brooks}\ and\ \citenamefont
  {Gelman}(1997)}]{Brooks:1997me}%
  \BibitemOpen
  \bibfield  {author} {\bibinfo {author} {\bibfnamefont {S.~P.}\ \bibnamefont
  {Brooks}}\ and\ \bibinfo {author} {\bibfnamefont {A.}~\bibnamefont
  {Gelman}},\ }\href {\doibase 10.1080/10618600.1998.10474787} {\bibfield
  {journal} {\bibinfo  {journal} {J. Comp. Graph. Stat.}\ }\textbf {\bibinfo
  {volume} {7}},\ \bibinfo {pages} {434} (\bibinfo {year} {1997})}\BibitemShut
  {NoStop}%
\bibitem [{\citenamefont {Lewis}(2019)}]{Lewis:2019xzd}%
  \BibitemOpen
  \bibfield  {author} {\bibinfo {author} {\bibfnamefont {A.}~\bibnamefont
  {Lewis}},\ }\href@noop {} {\  (\bibinfo {year} {2019})},\ \Eprint
  {http://arxiv.org/abs/1910.13970} {arXiv:1910.13970 [astro-ph.IM]}
  \BibitemShut {NoStop}%
\bibitem [{\citenamefont {{Philcox}}\ and\ \citenamefont
  {{Eisenstein}}(2019{\natexlab{b}})}]{PhiEis19}%
  \BibitemOpen
  \bibfield  {author} {\bibinfo {author} {\bibfnamefont {O.~H.~E.}\
  \bibnamefont {{Philcox}}}\ and\ \bibinfo {author} {\bibfnamefont {D.~J.}\
  \bibnamefont {{Eisenstein}}},\ }\href {\doibase 10.1093/mnras/stz2896}
  {\bibfield  {journal} {\bibinfo  {journal} {\mnras}\ ,\ \bibinfo {pages}
  {2492}} (\bibinfo {year} {2019}{\natexlab{b}})},\ \Eprint
  {http://arxiv.org/abs/1910.04764} {arXiv:1910.04764 [astro-ph.CO]}
  \BibitemShut {NoStop}%
\bibitem [{\citenamefont {{Harnois-D{\'e}raps}}\ and\ \citenamefont
  {{Pen}}(2012)}]{HarPen12}%
  \BibitemOpen
  \bibfield  {author} {\bibinfo {author} {\bibfnamefont {J.}~\bibnamefont
  {{Harnois-D{\'e}raps}}}\ and\ \bibinfo {author} {\bibfnamefont {U.-L.}\
  \bibnamefont {{Pen}}},\ }\href {\doibase 10.1111/j.1365-2966.2012.21039.x}
  {\bibfield  {journal} {\bibinfo  {journal} {\mnras}\ }\textbf {\bibinfo
  {volume} {423}},\ \bibinfo {pages} {2288} (\bibinfo {year} {2012})},\ \Eprint
  {http://arxiv.org/abs/1109.5746} {arXiv:1109.5746 [astro-ph.CO]} \BibitemShut
  {NoStop}%
\bibitem [{\citenamefont {{Harnois-D{\'e}raps}}\ and\ \citenamefont
  {{Pen}}(2013)}]{HarPen13}%
  \BibitemOpen
  \bibfield  {author} {\bibinfo {author} {\bibfnamefont {J.}~\bibnamefont
  {{Harnois-D{\'e}raps}}}\ and\ \bibinfo {author} {\bibfnamefont {U.-L.}\
  \bibnamefont {{Pen}}},\ }\href {\doibase 10.1093/mnras/stt413} {\bibfield
  {journal} {\bibinfo  {journal} {\mnras}\ }\textbf {\bibinfo {volume} {431}},\
  \bibinfo {pages} {3349} (\bibinfo {year} {2013})},\ \Eprint
  {http://arxiv.org/abs/1211.6213} {arXiv:1211.6213 [astro-ph.CO]} \BibitemShut
  {NoStop}%
\bibitem [{\citenamefont {Bertolini}\ \emph {et~al.}(2016)\citenamefont
  {Bertolini}, \citenamefont {Schutz}, \citenamefont {Solon}, \citenamefont
  {Walsh},\ and\ \citenamefont {Zurek}}]{Bertolini:2015fya}%
  \BibitemOpen
  \bibfield  {author} {\bibinfo {author} {\bibfnamefont {D.}~\bibnamefont
  {Bertolini}}, \bibinfo {author} {\bibfnamefont {K.}~\bibnamefont {Schutz}},
  \bibinfo {author} {\bibfnamefont {M.~P.}\ \bibnamefont {Solon}}, \bibinfo
  {author} {\bibfnamefont {J.~R.}\ \bibnamefont {Walsh}}, \ and\ \bibinfo
  {author} {\bibfnamefont {K.~M.}\ \bibnamefont {Zurek}},\ }\href {\doibase
  10.1103/PhysRevD.93.123505} {\bibfield  {journal} {\bibinfo  {journal} {Phys.
  Rev. D}\ }\textbf {\bibinfo {volume} {93}},\ \bibinfo {pages} {123505}
  (\bibinfo {year} {2016})},\ \Eprint {http://arxiv.org/abs/1512.07630}
  {arXiv:1512.07630 [astro-ph.CO]} \BibitemShut {NoStop}%
\bibitem [{\citenamefont {{Mohammed}}\ \emph {et~al.}(2017)\citenamefont
  {{Mohammed}}, \citenamefont {{Seljak}},\ and\ \citenamefont
  {{Vlah}}}]{MohSelVla1704}%
  \BibitemOpen
  \bibfield  {author} {\bibinfo {author} {\bibfnamefont {I.}~\bibnamefont
  {{Mohammed}}}, \bibinfo {author} {\bibfnamefont {U.}~\bibnamefont
  {{Seljak}}}, \ and\ \bibinfo {author} {\bibfnamefont {Z.}~\bibnamefont
  {{Vlah}}},\ }\href {\doibase 10.1093/mnras/stw3196} {\bibfield  {journal}
  {\bibinfo  {journal} {\mnras}\ }\textbf {\bibinfo {volume} {466}},\ \bibinfo
  {pages} {780} (\bibinfo {year} {2017})},\ \Eprint
  {http://arxiv.org/abs/1607.00043} {arXiv:1607.00043} \BibitemShut {NoStop}%
\bibitem [{\citenamefont {{Kobayashi}}\ \emph {et~al.}(2020)\citenamefont
  {{Kobayashi}}, \citenamefont {{Nishimichi}}, \citenamefont {{Takada}},\ and\
  \citenamefont {{Takahashi}}}]{KobNisTak20}%
  \BibitemOpen
  \bibfield  {author} {\bibinfo {author} {\bibfnamefont {Y.}~\bibnamefont
  {{Kobayashi}}}, \bibinfo {author} {\bibfnamefont {T.}~\bibnamefont
  {{Nishimichi}}}, \bibinfo {author} {\bibfnamefont {M.}~\bibnamefont
  {{Takada}}}, \ and\ \bibinfo {author} {\bibfnamefont {R.}~\bibnamefont
  {{Takahashi}}},\ }\href {\doibase 10.1103/PhysRevD.101.023510} {\bibfield
  {journal} {\bibinfo  {journal} {\prd}\ }\textbf {\bibinfo {volume} {101}},\
  \bibinfo {eid} {023510} (\bibinfo {year} {2020})},\ \Eprint
  {http://arxiv.org/abs/1907.08515} {arXiv:1907.08515 [astro-ph.CO]}
  \BibitemShut {NoStop}%
\bibitem [{\citenamefont {{Yu}}\ \emph {et~al.}(2020)\citenamefont {{Yu}},
  \citenamefont {{Li}}, \citenamefont {{Singh}},\ and\ \citenamefont
  {{Seljak}}}]{YuSel20inprep}%
  \BibitemOpen
  \bibfield  {author} {\bibinfo {author} {\bibfnamefont {B.}~\bibnamefont
  {{Yu}}}, \bibinfo {author} {\bibfnamefont {Y.}~\bibnamefont {{Li}}}, \bibinfo
  {author} {\bibfnamefont {S.}~\bibnamefont {{Singh}}}, \ and\ \bibinfo
  {author} {\bibfnamefont {U.}~\bibnamefont {{Seljak}}},\ }\href@noop {}
  {\bibfield  {journal} {\bibinfo  {journal} {in preparation}\ } (\bibinfo
  {year} {2020})}\BibitemShut {NoStop}%
\bibitem [{\citenamefont {{Klypin}}\ \emph {et~al.}(2016)\citenamefont
  {{Klypin}}, \citenamefont {{Yepes}}, \citenamefont {{Gottl{\"o}ber}},
  \citenamefont {{Prada}},\ and\ \citenamefont {{He{\ss}}}}]{KlyYep1604}%
  \BibitemOpen
  \bibfield  {author} {\bibinfo {author} {\bibfnamefont {A.}~\bibnamefont
  {{Klypin}}}, \bibinfo {author} {\bibfnamefont {G.}~\bibnamefont {{Yepes}}},
  \bibinfo {author} {\bibfnamefont {S.}~\bibnamefont {{Gottl{\"o}ber}}},
  \bibinfo {author} {\bibfnamefont {F.}~\bibnamefont {{Prada}}}, \ and\
  \bibinfo {author} {\bibfnamefont {S.}~\bibnamefont {{He{\ss}}}},\ }\href
  {\doibase 10.1093/mnras/stw248} {\bibfield  {journal} {\bibinfo  {journal}
  {\mnras}\ }\textbf {\bibinfo {volume} {457}},\ \bibinfo {pages} {4340}
  (\bibinfo {year} {2016})},\ \Eprint {http://arxiv.org/abs/1411.4001}
  {arXiv:1411.4001 [astro-ph.CO]} \BibitemShut {NoStop}%
\bibitem [{\citenamefont {Rodr{\'{\i}}guez-Torres}\ \emph
  {et~al.}(2016)\citenamefont {Rodr{\'{\i}}guez-Torres} \emph
  {et~al.}}]{RodChuPra1610}%
  \BibitemOpen
  \bibfield  {author} {\bibinfo {author} {\bibfnamefont {S.~A.}\ \bibnamefont
  {Rodr{\'{\i}}guez-Torres}} \emph {et~al.},\ }\href {\doibase
  10.1093/mnras/stw1014} {\bibfield  {journal} {\bibinfo  {journal} {Mon. Not.
  Roy. Astron. Soc.}\ }\textbf {\bibinfo {volume} {460}},\ \bibinfo {pages}
  {1173} (\bibinfo {year} {2016})},\ \Eprint {http://arxiv.org/abs/1509.06404}
  {arXiv:1509.06404 [astro-ph.CO]} \BibitemShut {NoStop}%
\bibitem [{\citenamefont {{Hahn}}\ \emph {et~al.}(2017)\citenamefont {{Hahn}},
  \citenamefont {{Scoccimarro}}, \citenamefont {{Blanton}}, \citenamefont
  {{Tinker}},\ and\ \citenamefont {{Rodr{\'\i}guez-Torres}}}]{HahScoBla17}%
  \BibitemOpen
  \bibfield  {author} {\bibinfo {author} {\bibfnamefont {C.}~\bibnamefont
  {{Hahn}}}, \bibinfo {author} {\bibfnamefont {R.}~\bibnamefont
  {{Scoccimarro}}}, \bibinfo {author} {\bibfnamefont {M.~R.}\ \bibnamefont
  {{Blanton}}}, \bibinfo {author} {\bibfnamefont {J.~L.}\ \bibnamefont
  {{Tinker}}}, \ and\ \bibinfo {author} {\bibfnamefont {S.~A.}\ \bibnamefont
  {{Rodr{\'\i}guez-Torres}}},\ }\href {\doibase 10.1093/mnras/stx185}
  {\bibfield  {journal} {\bibinfo  {journal} {\mnras}\ }\textbf {\bibinfo
  {volume} {467}},\ \bibinfo {pages} {1940} (\bibinfo {year} {2017})},\ \Eprint
  {http://arxiv.org/abs/1609.01714} {arXiv:1609.01714 [astro-ph.CO]}
  \BibitemShut {NoStop}%
\bibitem [{\citenamefont {{Anderson}}(2003)}]{And03}%
  \BibitemOpen
  \bibfield  {author} {\bibinfo {author} {\bibfnamefont {T.}~\bibnamefont
  {{Anderson}}},\ }\href@noop {} {\emph {\bibinfo {title} {{An introduction to
  multivariate statistical analysis, 3rd edn.}}}}\ (\bibinfo  {publisher}
  {Wiley-Interscience},\ \bibinfo {year} {2003})\BibitemShut {NoStop}%
\bibitem [{\citenamefont {{Wishart}}(1928)}]{Wishart28}%
  \BibitemOpen
  \bibfield  {author} {\bibinfo {author} {\bibfnamefont {J.}~\bibnamefont
  {{Wishart}}},\ }\href@noop {} {\bibfield  {journal} {\bibinfo  {journal}
  {Biometrika}\ }\textbf {\bibinfo {volume} {20A}},\ \bibinfo {pages} {32}
  (\bibinfo {year} {1928})}\BibitemShut {NoStop}%
\bibitem [{\citenamefont {Beutler}\ \emph
  {et~al.}(2017{\natexlab{b}})\citenamefont {Beutler} \emph
  {et~al.}}]{BeuSeoRos1701}%
  \BibitemOpen
  \bibfield  {author} {\bibinfo {author} {\bibfnamefont {F.}~\bibnamefont
  {Beutler}} \emph {et~al.} (\bibinfo {collaboration} {BOSS}),\ }\href
  {\doibase 10.1093/mnras/stw2373} {\bibfield  {journal} {\bibinfo  {journal}
  {Mon. Not. Roy. Astron. Soc.}\ }\textbf {\bibinfo {volume} {464}},\ \bibinfo
  {pages} {3409} (\bibinfo {year} {2017}{\natexlab{b}})},\ \Eprint
  {http://arxiv.org/abs/1607.03149} {arXiv:1607.03149 [astro-ph.CO]}
  \BibitemShut {NoStop}%
\bibitem [{\citenamefont {Alam}\ \emph {et~al.}(2020)\citenamefont {Alam} \emph
  {et~al.}}]{Ala20}%
  \BibitemOpen
  \bibfield  {author} {\bibinfo {author} {\bibfnamefont {S.}~\bibnamefont
  {Alam}} \emph {et~al.} (\bibinfo {collaboration} {eBOSS}),\ }\href@noop {} {\
   (\bibinfo {year} {2020})},\ \Eprint {http://arxiv.org/abs/2007.08991}
  {arXiv:2007.08991 [astro-ph.CO]} \BibitemShut {NoStop}%
\bibitem [{\citenamefont {{Barreira}}\ \emph {et~al.}(2018)\citenamefont
  {{Barreira}}, \citenamefont {{Krause}},\ and\ \citenamefont
  {{Schmidt}}}]{BarKraSch1806}%
  \BibitemOpen
  \bibfield  {author} {\bibinfo {author} {\bibfnamefont {A.}~\bibnamefont
  {{Barreira}}}, \bibinfo {author} {\bibfnamefont {E.}~\bibnamefont
  {{Krause}}}, \ and\ \bibinfo {author} {\bibfnamefont {F.}~\bibnamefont
  {{Schmidt}}},\ }\href {\doibase 10.1088/1475-7516/2018/06/015} {\bibfield
  {journal} {\bibinfo  {journal} {\jcap}\ }\textbf {\bibinfo {volume} {6}},\
  \bibinfo {eid} {015} (\bibinfo {year} {2018})},\ \Eprint
  {http://arxiv.org/abs/1711.07467} {arXiv:1711.07467} \BibitemShut {NoStop}%
\bibitem [{\citenamefont {{Lacasa}}\ and\ \citenamefont
  {{Grain}}(2019)}]{LacGra19}%
  \BibitemOpen
  \bibfield  {author} {\bibinfo {author} {\bibfnamefont {F.}~\bibnamefont
  {{Lacasa}}}\ and\ \bibinfo {author} {\bibfnamefont {J.}~\bibnamefont
  {{Grain}}},\ }\href {\doibase 10.1051/0004-6361/201834343} {\bibfield
  {journal} {\bibinfo  {journal} {\aap}\ }\textbf {\bibinfo {volume} {624}},\
  \bibinfo {eid} {A61} (\bibinfo {year} {2019})},\ \Eprint
  {http://arxiv.org/abs/1809.05437} {arXiv:1809.05437 [astro-ph.CO]}
  \BibitemShut {NoStop}%
\bibitem [{\citenamefont {Laureijs}\ \emph {et~al.}(2011)\citenamefont
  {Laureijs} \emph {et~al.}}]{Laureijs:2011gra}%
  \BibitemOpen
  \bibfield  {author} {\bibinfo {author} {\bibfnamefont {R.}~\bibnamefont
  {Laureijs}} \emph {et~al.} (\bibinfo {collaboration} {EUCLID}),\ }\href@noop
  {} {\  (\bibinfo {year} {2011})},\ \Eprint {http://arxiv.org/abs/1110.3193}
  {arXiv:1110.3193 [astro-ph.CO]} \BibitemShut {NoStop}%
\bibitem [{\citenamefont {Amendola}\ \emph {et~al.}(2018)\citenamefont
  {Amendola} \emph {et~al.}}]{Amendola:2016saw}%
  \BibitemOpen
  \bibfield  {author} {\bibinfo {author} {\bibfnamefont {L.}~\bibnamefont
  {Amendola}} \emph {et~al.},\ }\href {\doibase 10.1007/s41114-017-0010-3}
  {\bibfield  {journal} {\bibinfo  {journal} {Living Rev. Rel.}\ }\textbf
  {\bibinfo {volume} {21}},\ \bibinfo {pages} {2} (\bibinfo {year} {2018})},\
  \Eprint {http://arxiv.org/abs/1606.00180} {arXiv:1606.00180 [astro-ph.CO]}
  \BibitemShut {NoStop}%
\bibitem [{\citenamefont {Aghamousa}\ \emph {et~al.}(2016)\citenamefont
  {Aghamousa} \emph {et~al.}}]{Aghamousa:2016zmz}%
  \BibitemOpen
  \bibfield  {author} {\bibinfo {author} {\bibfnamefont {A.}~\bibnamefont
  {Aghamousa}} \emph {et~al.} (\bibinfo {collaboration} {DESI}),\ }\href@noop
  {} {\  (\bibinfo {year} {2016})},\ \Eprint {http://arxiv.org/abs/1611.00036}
  {arXiv:1611.00036 [astro-ph.IM]} \BibitemShut {NoStop}%
\bibitem [{\citenamefont {{Font-Ribera}}\ \emph {et~al.}(2014)\citenamefont
  {{Font-Ribera}}, \citenamefont {{McDonald}}, \citenamefont {{Mostek}},
  \citenamefont {{Reid}}, \citenamefont {{Seo}},\ and\ \citenamefont
  {{Slosar}}}]{FonMcDMos1405}%
  \BibitemOpen
  \bibfield  {author} {\bibinfo {author} {\bibfnamefont {A.}~\bibnamefont
  {{Font-Ribera}}}, \bibinfo {author} {\bibfnamefont {P.}~\bibnamefont
  {{McDonald}}}, \bibinfo {author} {\bibfnamefont {N.}~\bibnamefont
  {{Mostek}}}, \bibinfo {author} {\bibfnamefont {B.~A.}\ \bibnamefont
  {{Reid}}}, \bibinfo {author} {\bibfnamefont {H.-J.}\ \bibnamefont {{Seo}}}, \
  and\ \bibinfo {author} {\bibfnamefont {A.}~\bibnamefont {{Slosar}}},\ }\href
  {\doibase 10.1088/1475-7516/2014/05/023} {\bibfield  {journal} {\bibinfo
  {journal} {Journal of Cosmology and Astro-Particle Physics}\ }\textbf
  {\bibinfo {volume} {2014}},\ \bibinfo {eid} {023} (\bibinfo {year} {2014})},\
  \Eprint {http://arxiv.org/abs/1308.4164} {arXiv:1308.4164 [astro-ph.CO]}
  \BibitemShut {NoStop}%
\bibitem [{\citenamefont {Philcox}\ \emph {et~al.}(2020)\citenamefont
  {Philcox}, \citenamefont {Ivanov}, \citenamefont {Simonovi\'c},\ and\
  \citenamefont {Zaldarriaga}}]{Philcox:2020vvt}%
  \BibitemOpen
  \bibfield  {author} {\bibinfo {author} {\bibfnamefont {O.~H.}\ \bibnamefont
  {Philcox}}, \bibinfo {author} {\bibfnamefont {M.~M.}\ \bibnamefont {Ivanov}},
  \bibinfo {author} {\bibfnamefont {M.}~\bibnamefont {Simonovi\'c}}, \ and\
  \bibinfo {author} {\bibfnamefont {M.}~\bibnamefont {Zaldarriaga}},\ }\href
  {\doibase 10.1088/1475-7516/2020/05/032} {\bibfield  {journal} {\bibinfo
  {journal} {JCAP}\ }\textbf {\bibinfo {volume} {05}},\ \bibinfo {pages} {032}
  (\bibinfo {year} {2020})},\ \Eprint {http://arxiv.org/abs/2002.04035}
  {arXiv:2002.04035 [astro-ph.CO]} \BibitemShut {NoStop}%
\bibitem [{\citenamefont {Chudaykin}\ and\ \citenamefont
  {Ivanov}(2019)}]{Chudaykin:2019ock}%
  \BibitemOpen
  \bibfield  {author} {\bibinfo {author} {\bibfnamefont {A.}~\bibnamefont
  {Chudaykin}}\ and\ \bibinfo {author} {\bibfnamefont {M.~M.}\ \bibnamefont
  {Ivanov}},\ }\href {\doibase 10.1088/1475-7516/2019/11/034} {\bibfield
  {journal} {\bibinfo  {journal} {JCAP}\ }\textbf {\bibinfo {volume} {11}},\
  \bibinfo {pages} {034} (\bibinfo {year} {2019})},\ \Eprint
  {http://arxiv.org/abs/1907.06666} {arXiv:1907.06666 [astro-ph.CO]}
  \BibitemShut {NoStop}%
\bibitem [{\citenamefont {Blanchard}\ \emph {et~al.}(2019)\citenamefont
  {Blanchard} \emph {et~al.}}]{Blanchard:2019oqi}%
  \BibitemOpen
  \bibfield  {author} {\bibinfo {author} {\bibfnamefont {A.}~\bibnamefont
  {Blanchard}} \emph {et~al.} (\bibinfo {collaboration} {Euclid}),\ }\href@noop
  {} {\  (\bibinfo {year} {2019})},\ \Eprint {http://arxiv.org/abs/1910.09273}
  {arXiv:1910.09273 [astro-ph.CO]} \BibitemShut {NoStop}%
\bibitem [{\citenamefont {{Agarwal}}\ \emph {et~al.}(2020)\citenamefont
  {{Agarwal}}, \citenamefont {{Desjacques}}, \citenamefont {{Jeong}},\ and\
  \citenamefont {{Schmidt}}}]{Nis20}%
  \BibitemOpen
  \bibfield  {author} {\bibinfo {author} {\bibfnamefont {N.}~\bibnamefont
  {{Agarwal}}}, \bibinfo {author} {\bibfnamefont {V.}~\bibnamefont
  {{Desjacques}}}, \bibinfo {author} {\bibfnamefont {D.}~\bibnamefont
  {{Jeong}}}, \ and\ \bibinfo {author} {\bibfnamefont {F.}~\bibnamefont
  {{Schmidt}}},\ }\href@noop {} {\bibfield  {journal} {\bibinfo  {journal}
  {arXiv e-prints}\ ,\ \bibinfo {eid} {arXiv:2007.04340}} (\bibinfo {year}
  {2020})},\ \Eprint {http://arxiv.org/abs/2007.04340} {arXiv:2007.04340
  [astro-ph.CO]} \BibitemShut {NoStop}%
\bibitem [{\citenamefont {Ivanov}\ \emph
  {et~al.}(2020{\natexlab{b}})\citenamefont {Ivanov}, \citenamefont
  {McDonough}, \citenamefont {Hill}, \citenamefont {Simonovi\'c}, \citenamefont
  {Toomey}, \citenamefont {Alexander},\ and\ \citenamefont
  {Zaldarriaga}}]{Ivanov:2020ril}%
  \BibitemOpen
  \bibfield  {author} {\bibinfo {author} {\bibfnamefont {M.~M.}\ \bibnamefont
  {Ivanov}}, \bibinfo {author} {\bibfnamefont {E.}~\bibnamefont {McDonough}},
  \bibinfo {author} {\bibfnamefont {J.~C.}\ \bibnamefont {Hill}}, \bibinfo
  {author} {\bibfnamefont {M.}~\bibnamefont {Simonovi\'c}}, \bibinfo {author}
  {\bibfnamefont {M.~W.}\ \bibnamefont {Toomey}}, \bibinfo {author}
  {\bibfnamefont {S.}~\bibnamefont {Alexander}}, \ and\ \bibinfo {author}
  {\bibfnamefont {M.}~\bibnamefont {Zaldarriaga}},\ }\href@noop {} {\
  (\bibinfo {year} {2020}{\natexlab{b}})},\ \Eprint
  {http://arxiv.org/abs/2006.11235} {arXiv:2006.11235 [astro-ph.CO]}
  \BibitemShut {NoStop}%
\bibitem [{\citenamefont {{Reid}}\ \emph {et~al.}(2016)\citenamefont {{Reid}}
  \emph {et~al.}}]{ReiHoPad16}%
  \BibitemOpen
  \bibfield  {author} {\bibinfo {author} {\bibfnamefont {B.}~\bibnamefont
  {{Reid}}} \emph {et~al.},\ }\href {\doibase 10.1093/mnras/stv2382} {\bibfield
   {journal} {\bibinfo  {journal} {\mnras}\ }\textbf {\bibinfo {volume}
  {455}},\ \bibinfo {pages} {1553} (\bibinfo {year} {2016})},\ \Eprint
  {http://arxiv.org/abs/1509.06529} {arXiv:1509.06529 [astro-ph.CO]}
  \BibitemShut {NoStop}%
\bibitem [{\citenamefont {{Baumgarten}}\ and\ \citenamefont
  {{Chuang}}(2018)}]{BauChu18}%
  \BibitemOpen
  \bibfield  {author} {\bibinfo {author} {\bibfnamefont {F.}~\bibnamefont
  {{Baumgarten}}}\ and\ \bibinfo {author} {\bibfnamefont {C.-H.}\ \bibnamefont
  {{Chuang}}},\ }\href {\doibase 10.1093/mnras/sty1971} {\bibfield  {journal}
  {\bibinfo  {journal} {\mnras}\ }\textbf {\bibinfo {volume} {480}},\ \bibinfo
  {pages} {2535} (\bibinfo {year} {2018})},\ \Eprint
  {http://arxiv.org/abs/1802.04462} {arXiv:1802.04462 [astro-ph.CO]}
  \BibitemShut {NoStop}%
\bibitem [{\citenamefont {{Bond}}\ \emph {et~al.}(1998)\citenamefont {{Bond}},
  \citenamefont {{Jaffe}},\ and\ \citenamefont {{Knox}}}]{BonJafKno98}%
  \BibitemOpen
  \bibfield  {author} {\bibinfo {author} {\bibfnamefont {J.~R.}\ \bibnamefont
  {{Bond}}}, \bibinfo {author} {\bibfnamefont {A.~H.}\ \bibnamefont {{Jaffe}}},
  \ and\ \bibinfo {author} {\bibfnamefont {L.}~\bibnamefont {{Knox}}},\ }\href
  {\doibase 10.1103/PhysRevD.57.2117} {\bibfield  {journal} {\bibinfo
  {journal} {\prd}\ }\textbf {\bibinfo {volume} {57}},\ \bibinfo {pages} {2117}
  (\bibinfo {year} {1998})},\ \Eprint {http://arxiv.org/abs/astro-ph/9708203}
  {arXiv:astro-ph/9708203 [astro-ph]} \BibitemShut {NoStop}%
\bibitem [{\citenamefont {{Abidi}}\ and\ \citenamefont
  {{Baldauf}}(2018)}]{AbiBal1807}%
  \BibitemOpen
  \bibfield  {author} {\bibinfo {author} {\bibfnamefont {M.~M.}\ \bibnamefont
  {{Abidi}}}\ and\ \bibinfo {author} {\bibfnamefont {T.}~\bibnamefont
  {{Baldauf}}},\ }\href {\doibase 10.1088/1475-7516/2018/07/029} {\bibfield
  {journal} {\bibinfo  {journal} {\jcap}\ }\textbf {\bibinfo {volume} {2018}},\
  \bibinfo {eid} {029} (\bibinfo {year} {2018})},\ \Eprint
  {http://arxiv.org/abs/1802.07622} {arXiv:1802.07622 [astro-ph.CO]}
  \BibitemShut {NoStop}%
\end{thebibliography}%
\end{document}